%%%%%%%%%%%%%%%%%%%%%%%%%%%%%%%%%%%%%%%%%%%%%%%%%%%%%%%%%%%%%%%%%%%%%%%%%%
%                                                                        %
%  									 % 
%                by P. Di Francesco, O. Golinelli and E. Guitter         %
%                TEX file, using harvmac.tex macros  		         %
%									 %
%                    SPhT        hep-th/94....     			 %
%%%%%%%%%%%%%%%%%%%%%%%%%%%%%%%%%%%%%%%%%%%%%%%%%%%%%%%%%%%%%%%%%%%%%%%%%%
\input harvmac 
\input epsf.tex

\overfullrule=0mm
\newcount\figno
\figno=0
\def\fig#1#2#3{
\par\begingroup\parindent=0pt\leftskip=1cm\rightskip=1cm\parindent=0pt
\baselineskip=11pt
\global\advance\figno by 1
\midinsert
\epsfxsize=#3
\centerline{\epsfbox{#2}}
\vskip 12pt
{\bf Fig. \the\figno:} #1\par
\endinsert\endgroup\par
}
\def\figlabel#1{\xdef#1{\the\figno}}
\def\encadremath#1{\vbox{\hrule\hbox{\vrule\kern8pt\vbox{\kern8pt
\hbox{$\displaystyle #1$}\kern8pt}
\kern8pt\vrule}\hrule}}

%Macros 
%%%%%%%%%%%%%%%%%%%%%%%%%%%%%%%%%%%%%%%%%%%%%%%%%%%%%%%%%%%%%%%%%
\def\tvi{\vrule height 12pt depth 6pt width 0pt}
\def\tv{\tvi\vrule}

\def\IR{\relax{\rm I\kern-.18em R}}
\font\cmss=cmss10 \font\cmsss=cmss10 at 7pt
\def\IZ{\relax\ifmmode\mathchoice
{\hbox{\cmss Z\kern-.4em Z}}{\hbox{\cmss Z\kern-.4em Z}}
{\lower.9pt\hbox{\cmsss Z\kern-.4em Z}}
{\lower1.2pt\hbox{\cmsss Z\kern-.4em Z}}\else{\cmss Z\kern-.4em Z}\fi}
\def\buildrel#1\under#2{\mathrel{\mathop{\kern0pt #2}\limits_{#1}}}

%%%%%%%%%%%%%%%%%%%%%%%%%%%%%%%%%%%%%%%%%%%%%%%%%%%%%%%%%%%%%%%%%

\Title{SPhT/95-059}
{{\vbox {
%\centerline{}
\bigskip
\centerline{Meander, Folding and Arch Statistics}
}}}
\bigskip
\centerline{P. Di Francesco,}
\medskip
\centerline{O. Golinelli}
\medskip
\centerline{and} 
\medskip
\centerline{E. 
Guitter\footnote*{e-mails: philippe,golinel,guitter@amoco.saclay.cea.fr},}

\bigskip

\centerline{ \it Service de Physique Th\'eorique de Saclay,}
%{Laboratoire de la Direction des Sciences 
%de la Mati\`ere du Commissariat \`a l'Energie Atomique.}

\centerline{ \it F-91191 Gif sur Yvette Cedex, France}

\vskip .5in

The statistics of meander and 
related problems are studied as particular realizations of compact polymer chain foldings.  
This paper presents a general discussion of these topics, 
with a particular emphasis on three points: (i) the use of a direct recursive relation
for building (semi) meanders (ii) the equivalence with a random matrix model (iii)
the exact solution of simpler related problems, such as arch configurations or irreducible
meanders.

\noindent
\Date{Keywords: meanders, polymers, folding, matrix models.}

%\writetoc

%references
\nref\LZ{S. Lando and A. Zvonkin, {\it Plane and Projective Meanders}, Theor. Comp.
Science {\bf 117} (1993) 227 and {\it Meanders}, Selecta Math. Sov. {\bf 11} (1992) 117.}
\nref\TOU{J. Touchard, {\it Contributions \`a l'\'etude du probl\`eme des
timbres poste}, Canad. J. Math. {\bf 2} (1950) 385.}
\nref\LUN{W. Lunnon, {\it A map--folding problem}, Math. of Computation {\bf 22} 
(1968) 193.}
\nref\SLO{N. Sloane, {\it The on--line encyclopedia of integer sequences}, 
e-mail: sequences@research.att.com}
\nref\DGZ{E. Br\'ezin, C. Itzykson, G. Parisi and J.-B. Zuber,
{\it Planar Diagrams}, Commun. Math. Phys. {\bf 69} (1979) 147; 
see also P. Di Francesco, P. Ginsparg and J. Zinn--Justin, {\it 2D Gravity and
Random Matrices}, Phys. Rep. {\bf 254} (1995) 1, for a review and more references.}

%text

\newsec{Introduction}

The concept of folding has an important place in polymer physics.
Typically one considers the statistical model of a polymer chain made of say $n$
identical constituents, and which may be folded onto itself. 
The entropy of such a system
is obtained by counting the number of inequivalent ways of folding the chain.
The combinatorial problem of enumerating all the {\it compact} foldings 
of a closed polymer chain happens to be equivalent to another geometrical problem, 
that of enumerating meanders \LZ, i.e. configurations of a closed road crossing a river 
through $n$ bridges. 
To our knowledge, this is still an open problem, which indeed has been adressed
by very few authors.

In the present paper, we study various aspects of this meander problem.
In section 2, we define the meander problem itself, as well as
a somewhat simpler semi--meander problem corresponding to the
compact folding of an open polymer chain. We also gather in this
section a number of data on exact enumeration of some meanders, as well as conjectural 
analytic structure for some of these data.
A meander may be viewed as a particular gluing of two arch configurations, representing
the configuration of the road respectively above an below the river. 
Section 3 is devoted 
to the derivation of exact results for the statistical distribution
of arches in arch configurations. In section 4, we introduce an exact
recursion relation for constructing all semi--meanders, for which
we develop various approximations leading to estimates of the entropy of 
folding of open polymers. 
Section 5 presents an alternative description of the meander problems in terms
of random matrix models, and gathers a few consequent results. 
A number of sum rules and inequalities satisfied by meander and semi--meander numbers
are displayed in section 6, together with exact solutions for simpler meandric problems.
In section 7, we address the question of irreducible meanders \LZ, 
i.e. systems of several roads
crossing a river, which are interlocked in an irreducible way: in this case,
an exact solution is derived for both irreducible meander and semi--meander numbers.
A few technical details are gathered in appendix A.

\newsec{Definitions and generalities}

\subsec{Meanders}

\fig{The $M_2=2$ meanders of order $2$ (a), and the $M_2^{(2)}=2$ two--component 
meanders of order $2$. The infinite river is represented as a horizontal line.}{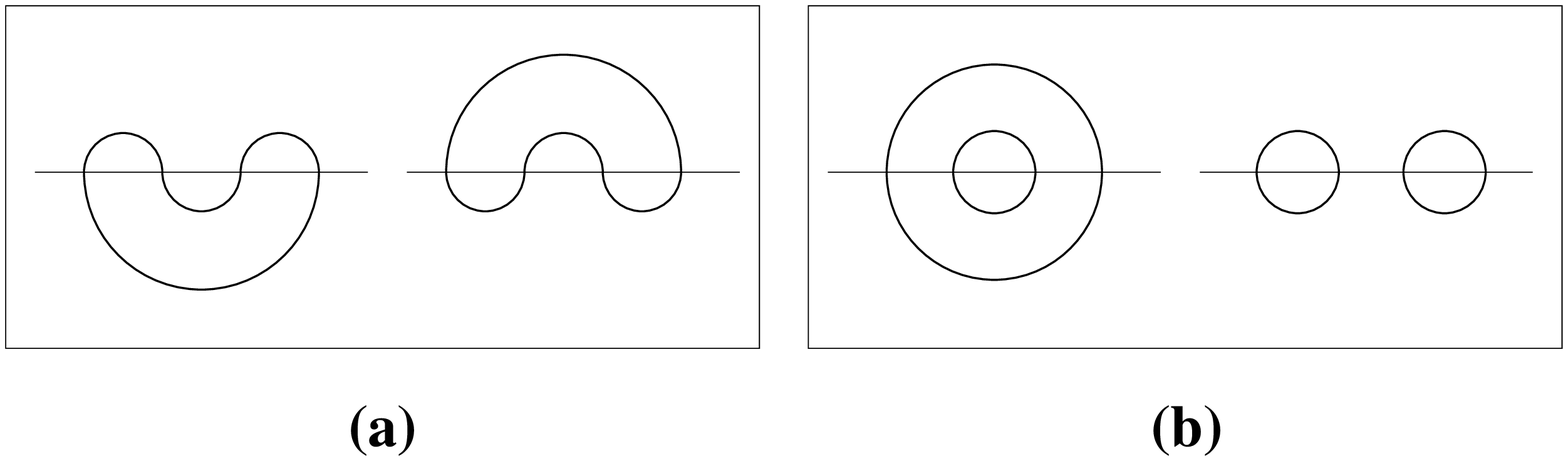}{8.5truecm}
\figlabel\exam

A {\bf meander} is defined as follows. Let us consider an infinite straight
line (river).
A meander of order $n$ is a closed self--avoiding connected loop (road)
which intersects the line through $2n$ points\foot{ The
number of intersections between a loop and an infinite line is necessarily
an even number.} 
(bridges). 
A meander of order $n$
can clearly be viewed as a compact folding configuration of a closed chain
of $2n$ constituents (in one--to--one correspondence with the $2n$ bridges),
by putting a hinge on each section of road between two bridges.
Note also that the road and the river play indeed similar roles, by appropriately
cutting the road and closing the river.

In the following, we
will study the number $M_n$ of inequivalent meanders of order $n$ (by inequivalent
we mean meanders which cannot be smoothly deformed into each other without
changing the order of the bridges).
We will also be interested in the numbers $M_n^{(k)}$ of inequivalent meanders
of order $n$ with $k$ connected components, i.e. made
of $k$ closed connected non--intersecting but possibly interlocking loops,
which cross the river through a total of $2n$ bridges.
Note that with this last definition,
$M_n=M_n^{(1)}$.
The $M_2=M_2^{(1)}=2$ meanders of order $2$ and the 
$M_2^{(2)}=2$ two--component meanders of order $2$ are depicted in 
Fig.\exam\ (a) and (b) respectively for illustration.

\subsec{Folding a strip of stamps, semi--meanders}

As mentioned in the introduction, the meander problem is equivalent to that of 
compact folding of a closed polymer chain. In this section, we will instead
consider the case of an {\it open} polymer chain, moreover attached by one of
its extremities. In another language, the problem is nothing but
that of {\bf folding a strip of stamps} \TOU\ \LUN,
and leads to a slightly different version of the meander problem,
the {\it semi--meander} problem, which we describe now.

\fig{The $4$ inequivalent foldings of a strip of $3$ stamps. The 
fixed stamp is indicated by the empty circle. The other circles correspond to the
edges of the stamps. The first stamp is fixed and attached to a support
(shaded area).}{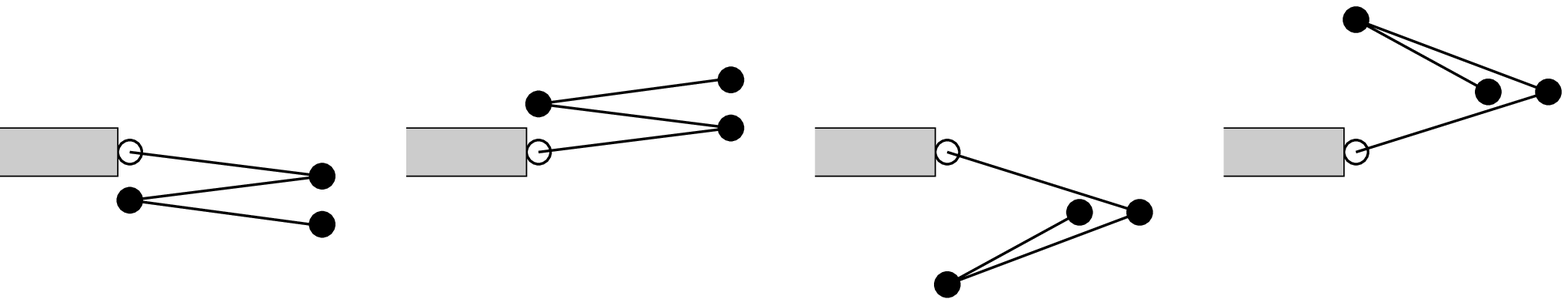}{9.5truecm}
\figlabel\stamp

The problem of 
{\bf folding a strip of stamps}
may be stated as follows. 
One considers a strip of $n-1$ stamps, the first
of which (say the leftmost one) is fixed, and supposedly attached to some support, 
preventing the strip from winding around the first stamp. 
A folding of the strip is a complete piling of its stamps, 
which preserves the (non--intersecting) stamps and their connections, 
and only affects the 
relative positions of any two adjacent stamps: 
each stamp is folded either on top of or below
the preceding one in the strip.

The number of inequivalent ways of folding a strip of $n-1$ stamps
is denoted by $S_n$.
In Fig.\stamp\ (a), we display 
the $S_4=4$ inequivalent foldings of a strip of $3$ stamps (stamps are represented in
side view, by segments, and also not completely folded to clearly indicate 
the succession of folds).

\fig{The ${\bar M}_3=2$ and ${\bar M}_4=4$ semi--meanders of order $3$ and $4$. 
The source of the corresponding semi--infinite river is
indicated by an asterisk.}{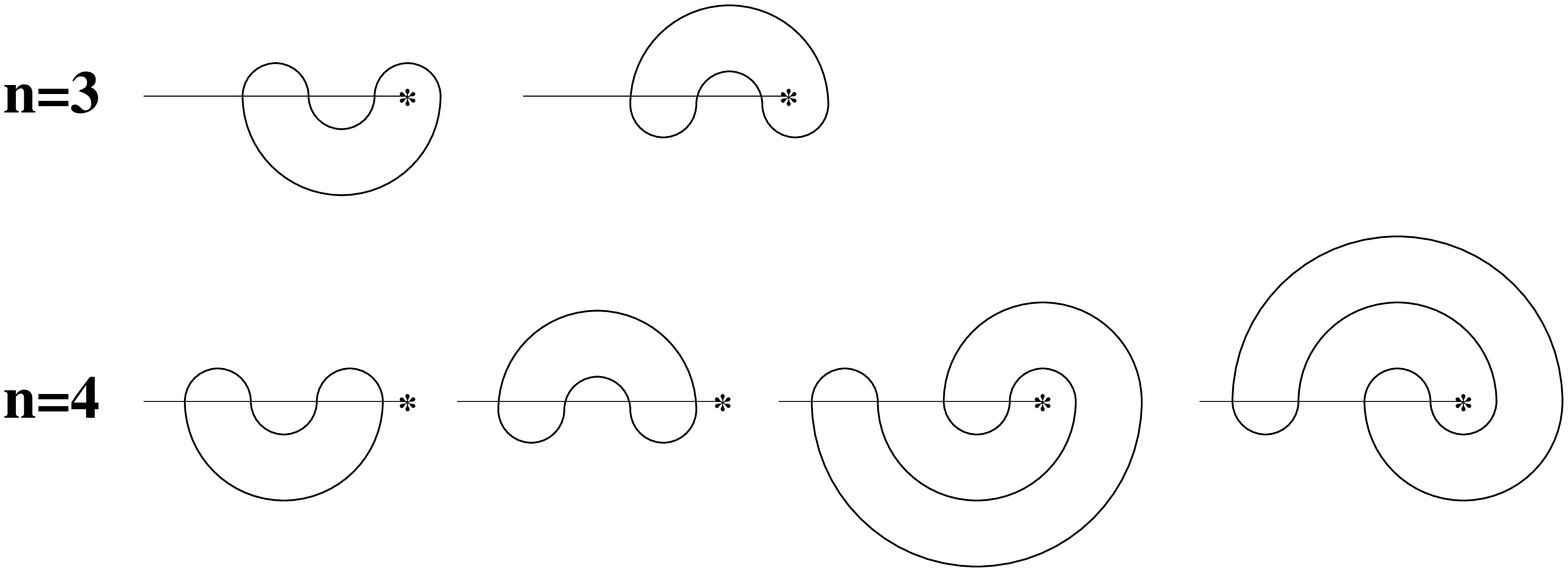}{10.truecm}
\figlabel\semime

This folding problem turns out to be equivalent to a particular meander problem,
which we will refer to as the {\bf semi--meander} problem.  Let us consider
a half (semi--infinite)  straight line (river) starting at a point (source).
A semi--meander of order $n$ is a closed self--avoiding
connected loop which intersects the
half--line through $n$ points\foot{The number $n$ of bridges for a semi--meander
need not be even, hence $n=1,2,3,...$} (bridges). 
Let us denote by ${\bar M}_n$ the number
of inequivalent semi--meanders of order $n$.   In analogy with the definition of 
meanders with $k$ connected components, we also denote by ${\bar M}_n^{(k)}$
the number of inequivalent semi--meanders of order $n$ with $k$ connected components.
As before, we have in particular ${\bar M}_n^{(1)}={\bar M}_n$.
The ${\bar M}_3=2$ and ${\bar M}_4=4$ semi--meanders of order $3$ and 
$4$ are depicted in Fig.\semime.

\fig{The mapping from semi--meanders to folded strips of stamps.
(i) cut the leftmost arch of the semi--meander along the river.
(ii) stretch the open circuit into a line. 
(iii) identify the segments of bent river with edges of stamps. 
(iv) draw the stamps according to the relative positions 
of crossings.}{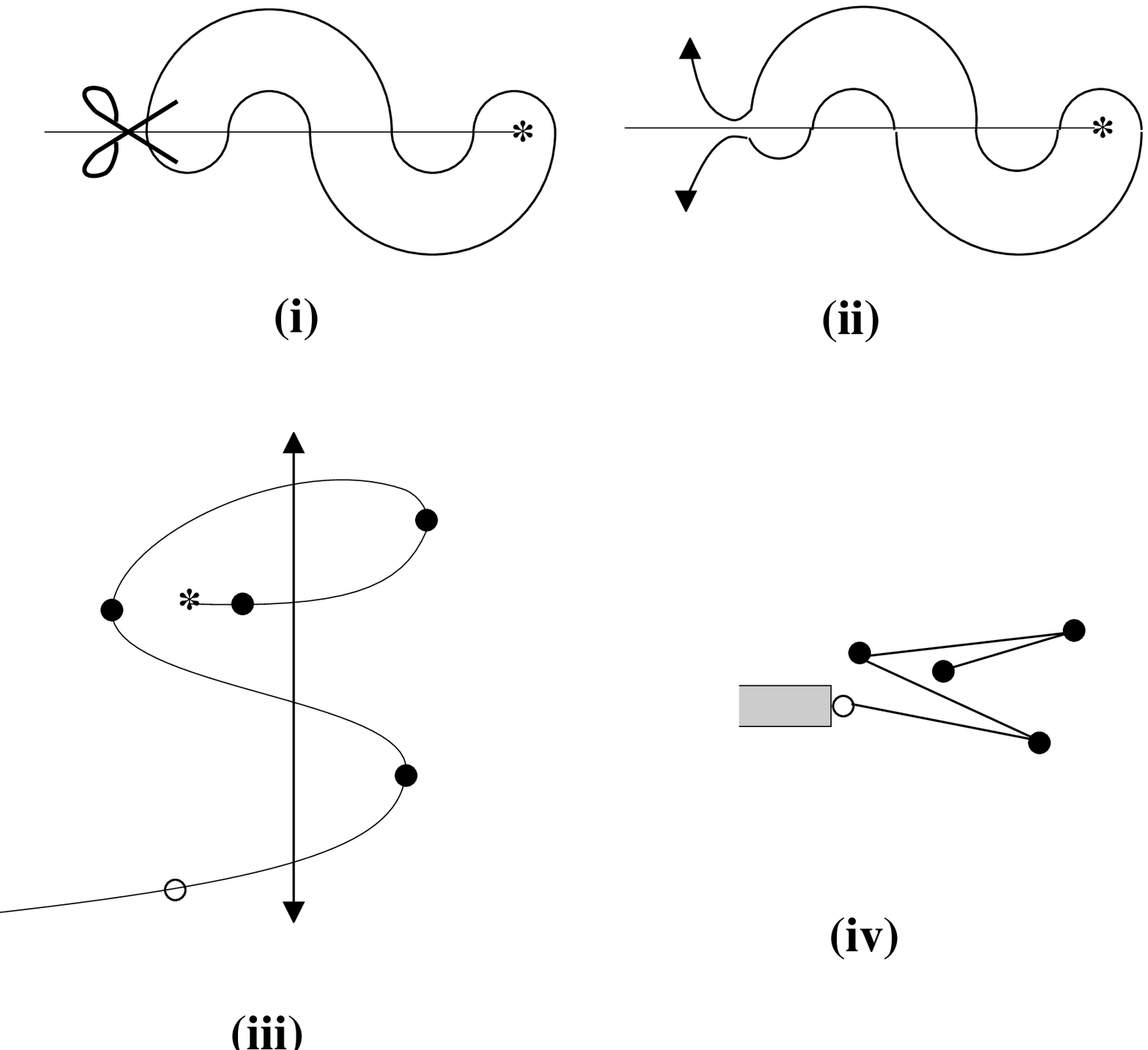}{8.truecm}
\figlabel\cutme

Let us prove that 
\eqn\equimest{ S_n~=~ {\bar M}_n ~.}

Starting from a semi--meander of order $n$, let us construct a folding
of the strip  of $n-1$ stamps as follows. Cut the leftmost arch of the semi--meander
along the river as indicated in Fig.\cutme\ (i), and stretch the (now open) circuit
into an infinite line (ii). The river has been bent in this process, 
but the structure of its
crossings with the line is preserved (except that the leftmost bridge has been erased). 
In a third step (iii), each segment of the river between
two crossings of the infinite line is identified with an edge of stamp.
In particular, the last edge of the strip of stamps
is associated with the segment of river between its source and 
its first bridge, whereas the first edge of the fixed stamp
corresponds to the infinite segment of river
after the last bridge.
This last choice singles out the first edge of the first stamp
in such a way that no piece of the strip can wind around it. 
Finally (iv), the stamps can be drawn between the edges, 
the connection being indicated by the river.
Note that in the process one of the bridges has been erased by the initial cut, henceforth
we are left with $n-1$ bridges, hence $n-1$ stamps.
This gives a one to one mapping from semi--meanders of order $n$
to folded strips of $n-1$ stamps, thus proving \equimest.

An analogous construction allows one to relate the meander number $M_n$ to that of 
foldings of a {\it closed} strip of $2n$ stamps. This construction is dual to the 
direct equivalence mentioned in the previous section, 
in the sense that the road and the river are exchanged.

\subsec{Numerical data and a few conjectures}

In this section, we gather numerical data for multicomponent meanders and semi--meanders.

\noindent{\bf Meanders.}  A few preleminary remarks are in order. The number
of connected components for a meander of order $n$ may not exceed $n$,
as each connected component of the meander must cross the river an even number of times,
whereas the number of bridges is $2n$. 
We display the first values of the multi--component meander numbers $M_n^{(k)}$
in Table I below (some of these numbers can be found in \LZ\ and \SLO).

$$\vbox{\font\bidon=cmr8 \def\bidondon{\bidon} \bidondon \offinterlineskip
\halign{\tv \quad # \tv & 
\hfill \ # & \hfill # & \hfill # &  \hfill # 
& \hfill # &  \hfill # &  \hfill #
&  \hfill # &  \hfill # &  \hfill # 
&  \hfill # & \hfill # \tv \cr
\noalign{\hrule}
\tvi  $\scriptstyle k\backslash n$& 1 \hfill & 2 \hfill 
& 3 \hfill & 4 \hfill & 5 \hfill 
& 6 \hfill & 7 \hfill & 8 \hfill 
& 9 \hfill & 10 \hfill & 11 \hfill 
& 12 \hfill \cr
\noalign{\hrule}
\tvi 1 & 1 & 2 & 8 & 42 & 262 & 1828 & 13820 & 110954 & 
933458 & 8152860 & 73424650 & 678390116 \cr 
\tvi 2 &   & 2 & 12 & 84 & 640 & 5236 & 45164 & 406012 & 
3772008 & 35994184 & 351173328 & 3490681428 \cr
\tvi 3 &   &   & 5 & 56 & 580 & 5894 & 60312 & 624240 & 
6540510 & 69323910 & 742518832 & 8028001566 \cr
\tvi 4 &   &   &   & 14 & 240 & 3344 & 42840 & 529104 & 
6413784 & 76980880 & 919032664 & 10941339452 \cr
\tvi 5 &   &   &   &   & 42 & 990 & 17472 & 271240 & 
3935238 & 54787208 & 742366152 & 9871243896 \cr
\tvi 6 &   &   &   &   &   & 132 & 4004 & 85904 & 
1569984 & 26200468 & 412348728 & 6230748192 \cr
\tvi 7 &   &   &   &   &   &   & 429 & 16016 & 
405552 & 8536890 & 161172704 & 2830421952 \cr
\tvi 8 &   &   &   &   &   &   &   & 1430 & 
63648 & 1860480 & 44346456 & 934582000 \cr
\tvi 9 &   &   &   &   &   &   &   &   & 
4862 & 251940 & 8356656 & 222516030 \cr
\tvi 10 &   &   &   &   &   &   &   &   & 
& 16796 & 994840 & 36936988 \cr
\tvi 11 &   &   &   &   &   &   &   &   & 
&   & 58786 & 3922512 \cr
\tvi 12 &   &   &   &   &   &   &   &   & 
&   &   & 208012 \cr
\noalign{\hrule} 
}}$$
\noindent{\bf Table I:} The numbers $M_n^{(k)}$ of inequivalent
meanders of order $n$ with $k$ connected components, for $1 \leq k \leq n\leq 12$,
obtained by exact enumeration on the computer.
\vskip 1cm

The reader will check on Table I that 
\eqn\maxk{ M_n^{(n)}={1 \over 2n+1} {2n+1 \choose n} ={(2n)! \over n! (n+1)!}~,}
known as the Catalan number of order $n$, denoted by $c_n$.

Encouraged by the relative simplicity of $M_n^{(n)}$, we have tried various guesses
for the explicit form of the $M_n^{(n-j)}$, leading to the following 
conjecture\foot{We have actually proved the cases
$j=0,1,2,3$ recursively. 
Although some of them are quite tedious,
all the proofs are based on the same strategy, which we will explicitly show for 
the case $j=0$ in the next section. 
The same strategy should extend to higher values of $j$.}
values of $j=0,1,2,3,4,5$
\eqn\conjmean{\eqalign{
M_n^{(n)}&={( 2 n ) !\over n! ( n +1) !}
\cr
M_n^{(n-1)}&={2 ( 2 n ) !\over 
( n-2 ) ! (  n +2) !}
\cr
M_n^{(n-2)}&={2   ( 2 n ) !\over 
( n-3 ) ! ( n+4 ) !} (n^2+7n-2)
\cr
M_n^{(n-3)}&={4 (2n)!\over 
3 ( n-4) ! ( n+6 ) !} (n^4+ 20 {n^3}+ 107 {n^2}- 107 n+15)
\cr
M_n^{(n-4)}&={2 ( 2 n ) !\over 3 (n-5)! (n+8)!} 
({n^6}+39 {n^5}+ 547 {n^4}+ 2565 {n^3}- 5474 {n^2}
+ 2382 n-672)
\cr
M_n^{(n-5)}&={4  (2 n)!\over 15 (n-6)! (n+10)!} 
({n^8}+ 64 {n^7}+ 1646 {n^6} \cr
&\ \ \ \ \ \ \ \ \ + 20074 {n^5}
+83669 {n^4}- 323444 {n^3}+ 257134 {n^2}- 155604 n+45360)
\cr}}
This leads to the more general conjecture that the number $M_n^{(n-j)}$
of $(n-j)$--component meanders of order $n$ has the general form
\eqn\konjm{ M_n^{(n-j)}= { 2^j (2n)! \over
j! (n+2j)! (n-j-1)!} \times P_{2j-2}(n) ~,}
for $j \geq 1$, 
where $P_{2j-2}(x)$ is a
polynomial of degree $2j-2$, with integer coefficients. 
The first few coefficients of $P_{2j-2}$
read
\eqn\firstfewco{ P_{2j-2}(x)~=~x^{2j-2} +(j-1)(3j+1) x^{2j-3}+
{1 \over 6} (j-1)(27j^3 -23 j^2 -65 j -6) x^{2j-4}+...}
The expression \konjm\ is still valid for $j=0$, 
but the factor is not polynomial, it reads
\eqn\prefac{ P_{-2}(x)~=~{1 \over x(x+1)}~=~ x^{-2} -x^{-3}+x^{-4}-x^{-5}+..., }
and agrees with the large $x$ expansion \firstfewco\ for $j=0$.

This structure is strongly suggestive of algebraic recursion relations
between the $M$'s. We indeed found recursion relations between $M_n^{(n-j)}$
and the numbers of lower $j$, for the first few $j$'s. 
But the growing complexity of these relations when $j$ is increased is not
encouraging (typically, the structure of the recursion itself involves
the detailed structure of the $M_j$ meanders of order $j$). 

\noindent{\bf Semi--meanders.} Like in the meander case, the number $k$
of connected components
of order $n$ semi--meanders cannot exceed $n$. There is actually only one semi--meander
of maximal number of connected components: it corresponds to having $n$ concentric circles,
each crossing the half--river exactly once, hence
\eqn\firstmes{ {\bar M}_n^{(n)}~=~ 1~.}
The first values of ${\bar M}_n^{(k)}$ are displayed in Table II below, for
$1 \leq k \leq n \leq 14$ (some of these numbers can be found in \LUN\ and \SLO).

$$\vbox{\offinterlineskip
\halign{\tv\quad # \tv &
\hfill \ # & \hfill # & \hfill # & \hfill # 
& \hfill # & \hfill # & \hfill #
& \hfill # & \hfill # & \hfill # 
& \hfill # & \hfill #  & \hfill # 
& \hfill #  \tv \cr
\noalign{\hrule}
\tvi $k \backslash n$& $1$ \hfill & $2$ \hfill 
& $3$ \hfill & $4$ \hfill & $5$ \hfill 
& $6$ \hfill & $7$ \hfill & $8$ \hfill 
& $9$ \hfill & $10$ \hfill & $11$ \hfill 
& $12$ \hfill & $13$ \hfill & $14$ \hfill \cr
\noalign{\hrule}
\tvi $1$ & $1$ & $1$ & $2$ & $4$ & $10$ & $24$ & $66$ & $174$ & $504$ 
& $1406$ & $4210$ & $12198$ & $37378$ & $111278$ \cr
\tvi $2$ &   & $1$ & $2$ & $6$ & $16$ & $48$ & $140$ & $428$ & $1308$ 
& $4072$ & $12796$ & $40432$ & $129432$ & $413900$ \cr
\tvi $3$ &   &   & $1$ & $3$ & $11$ & $37$ & $126$ & $430$ & $1454$ 
& $4976$ & $16880$ & $57824$ & $197010$ & $675428$ \cr
\tvi $4$ &   &   &   & $1$ & $4$ & $17$ & $66$ & $254$ & $956$ 
& $3584$ & $13256$ & $49052$ & $179552$ & $658560$ \cr
\tvi $5$ &   &   &   &   & $1$ & $5$ & $24$ & $104$ & $438$ 
& $1796$ & $7238$ & $28848$ & $113518$ & $444278$ \cr
\tvi $6$ &   &   &   &   &   & $1$ & $6$ & $32$ & $152$ & $690$ 
& $3028$ & $12996$ & $54812$ & $228284$ \cr
\tvi $7$ &   &   &   &   &   &   & $1$ & $7$ & $41$ & $211$ &
$1023$ & $4759$ & $21533$ & $95419$ \cr
\tvi $8$ &   &   &   &   &   &   &   & $1$ & $8$ & $51$ 
& $282$ & $1451$ & $7112$ & $33721$ \cr
\tvi $9$ &   &   &   &   &   &   &   &   & $1$ & $9$ 
& $62$ & $366$ & $1989$ & $10227$ \cr
\tvi $10$ &   &   &   &   &   &   &   &   &   & $1$ 
& $10$ & $74$ & $464$ & $2653$ \cr
\tvi $11$ &   &   &   &   &   &   &   &   &   &  
& $1$ & $11$ & $87$ & $577$ \cr
\tvi $12$ &   &   &   &   &   &   &   &   &   &   
&   & $1$ & $12$ & $101$ \cr
\tvi $13$ &   &   &   &   &   &   &   &   &   &   
&   &   & $1$ & $13$ \cr
\tvi $14$ &   &   &   &   &   &   &   &   &   &   
&   &   &   & $1$ \cr
\noalign{\hrule} }}$$
\noindent{\bf Table II:} The numbers ${\bar M}_n^{(k)}$ of inequivalent
semi--meanders of order $n$ with $k$ connected components, 
for $1 \leq k \leq n \leq 14$, obtained by exact enumeration on the computer.
\vskip 1cm

Like in the meander case, it is straightforward (a little less tedious here)
to compute the first few ${\bar M}_n^{(n-j)}$, $j=0,1,2,...$ and 
to guess their general structure as functions of $n$ and $j$.
We find
\eqn\semipol{\eqalign{
{\bar M}_n^{(n)} &= 1  \cr
{\bar M}_n^{(n-1)} &= n-1 \cr
{\bar M}_n^{(n-2)} &= {1 \over 2!}(n^2 +    n   -  8 ) \cr
{\bar M}_n^{(n-3)} &= {1 \over 3!}(n^3 +  6 n^2 - 31 n   -    24 )\cr 
&+2 \delta_{n,4}\cr
{\bar M}_n^{(n-4)} &= {1 \over 4!}(n^4 + 14 n^3 - 49 n^2 -   254 n ) 
\cr
&+15 \delta_{n,5} + 5 \delta_{n,6}\cr
{\bar M}_n^{(n-5)} &= {1 \over 5!}(n^5 + 25 n^4 - 15 n^3 -  1105 n^2 
-1066 n +1680)\cr
&+87 \delta_{n,6}+42 \delta_{n,7}
+14 \delta_{n,8}\cr
{\bar M}_n^{(n-6)} &= {1 \over 6!}(n^6+39 n^5+145n^4-2895n^3-10226n^2 
+8616n+31680)\cr
&+ 456 \delta_{n,7}+292 \delta_{n,8}+126 \delta_{n,9} +42\delta_{n,10} \cr
{\bar M}_n^{(n-7)} &= {1 \over 7!}(n^7+56n^6+532n^5-5110 n^4-50141n^3 
-20146n^2+377208n \cr &+282240) 
+2234\delta_{n,8}+1720\delta_{n,9}+1008\delta_{n,10}
+396\delta_{n,11}+132\delta_{n,12} ~.\cr }}

Note that in addition to some polynomial structure as functions of $n$,
the numbers ${\bar M}_n^{(n-j)}$ also incorporate some constant corrections for the $j-1$
first values of $n=j+1,j+2,...,2j-1$.
The above suggests the following general form for the polynomial part of 
${\bar M}_n^{(n-j)}$
\eqn\semikon{\eqalign{ 
{\bar M}_n^{(n-j)}\bigg\vert_{\rm pol.} &= {1 \over j!}(n^j 
+{1 \over 2}j (3j-5) n^{j-1}
+{1 \over 24} j(j-1)(27 j^2-163 j+122) n^{j-2} \cr
&+{1 \over 48}j(j-1)(j-2)(27
j^3-354 j^2+1163 j -1224) n^{j-3} + \cdots ) ~. \cr}}
We also found the beginning of a pattern for the corrections to the polynomial part 
\semikon, of the form
\eqn\konsemi{ {\bar M}_n^{(n-j)}\bigg\vert_{\rm corr.}~=~ c_j \delta_{n,2j-2}+
3 c_j \delta_{n,2j-3}+ \cdots}
where $c_j={2j+1 \choose j}/(2j+1)$ is the Catalan number of
order $j$, and the sum goes down to a $\delta_{n,j+1}$ term.

\noindent{\bf Asymptotics.} The data of the Tables I and II (see also  \LZ\ for $M_n$)
enable one to evaluate 
numerically the asymptotic behaviour of $M_n$ and ${\bar M}_n$
which read respectively
\eqn\asympto{ \eqalign{
M_n~&\sim ~ {\rm const} \ {(12.25)^n \over n^{7/2}} \cr
{\bar M}_n ~&\sim ~ {\rm const} \ {(3.5)^n \over n^2 } \cr}}
The exponent
$7/2$ for meanders was conjectured to be exact in \LZ.

Remarkably, the entropy of meanders
\eqn\entrom{ s~=~ \lim_{n \to \infty} {1 \over n} {\rm Log} M_n~=~{\rm Log}~R~\sim 2.50 }
is likely to be exactly twice that of semi--meanders
\eqn\entrosem{ {\bar s}~=~ \lim_{n \to \infty} {1 \over n} {\rm Log} {\bar M}_n
~=~{\rm Log}~{\bar R}~\sim 1.25 ~.}
Considering $M_n$ (resp. ${\bar M}_n$) as
the number of foldings of a closed (resp. open with a fixed end) strip of $2n$
(resp. $n-1$) stamps, general statistical mechanical considerations suggest
that the leading behaviour of these numbers is exactly the same (i.e. $\propto R^N$),
when expressed in the total number of stamps $N=2n$ (resp. $N=n-1$) for
large $N$. This
leads to the relation ${\bar s}={s \over 2}\ (={\rm Log}~ {\bar R})$ between the corresponding 
thermodynamical entropies, i.e. $R={\bar R}^2$ between the leading terms.

\newsec{Arch statistics}

\fig{A generic meander is the superimposition of two arch 
configurations.}{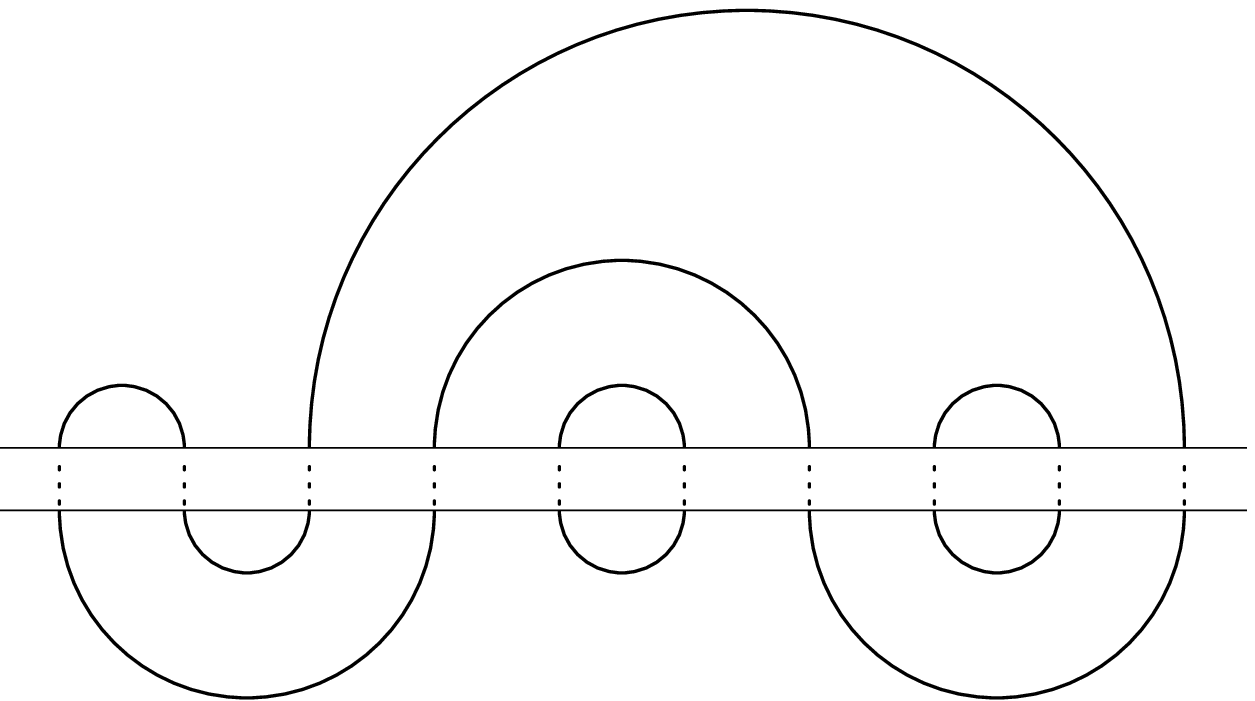}{7.truecm}
\figlabel\mearch

The most general meander of order $n$ with arbitrary number of connected components
is specified uniquely by its upper half (above the river) and lower half 
(below the river), as shown in Fig.\mearch.
Both halves form systems of $n$ non--intersecting arches connecting 
$2n$ bridges by pairs.  Any two arches are either disjoint or included,
one into the other. Any such system of arches will be referred to as an 
{\bf arch configuration} of order $n$.

\subsec{Catalan numbers}

\fig{Recursion principle for arch configurations. One sums over all 
positions $2j+2$ of the right
bridge of the leftmost arch, which separates the initial 
configuration into 
two configurations $X$ of order $j$  and $Y$ of order $n-j-1$, 
respectively below the leftmost arch 
and to its right.}{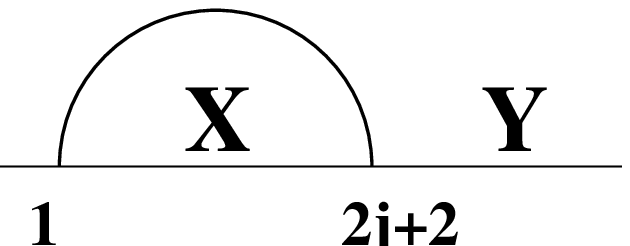}{5.truecm}
\figlabel\recursion

Let us first compute by recursion
the number $c_n$ of arch configurations of order $n$, linking $n$ pairs of bridges.
Starting from one such arch configuration, let us follow the arch linking
the leftmost bridge (position $1$) to another, say in position $2j+2$ 
(see Fig.\recursion: this position has to be even). 
This arch separates the configuration into two sub--configurations of arches, 
$X$ of order
$j$ (below the leftmost arch), and $Y$ of order $n-j-1$ 
(to the right of the leftmost arch).
Summing over the position of the right--hand bridge of the leftmost arch ($2j+2$),
we get the following simple recursion relation
\eqn\recucat{ c_n ~=~ \sum_{j=0}^{n-1} c_j \ c_{n-1-j}~, \quad \ n \geq 1~,}
where we have set $c_0=1$.
This rather simple example is nevertheless archetypical of a general type of
reasoning used to count any relevant number associated to arch  configurations. 
The scheme of Fig.\recursion\ is quite general.

The relation \recucat\ is the defining recursion relation of the 
celebrated Catalan numbers $c_n$, which count,
among other things, the numbers of parenthesings (with $n$ opening and 
$n$ closing parentheses) of words of $n+1$ letters. It follows immediately that
\eqn\catvat{ c_n~=~{1 \over 2n+1} {2n+1 \choose n}~=
~{(2n)! \over n! (n+1)!}~.}
In the following, we will need the generating function $C(x)$ of Catalan numbers,
\eqn\catagen{\eqalign{  C(x)~&=~ \sum_{n =0}^{\infty} ~ c_n ~ x^n \cr
&=~ 1+x + 2 x^2 + 5 x^3 + 14 x^4 + 42 x^5 +132 x^6+\cdots \cr}}
subject to  the algebraic relation
\eqn\algcat{ x C(x)^2 ~=~ C(x)-1~,}
due to \recucat. One gets
\eqn\gencat{ C(x)~=~{1- \sqrt{1-4x} \over 2x}~.}

\subsec{Arch configurations and meanders}

\fig{A particular $5$--component meander of order $5$, and the corresponding arch 
configuration.}{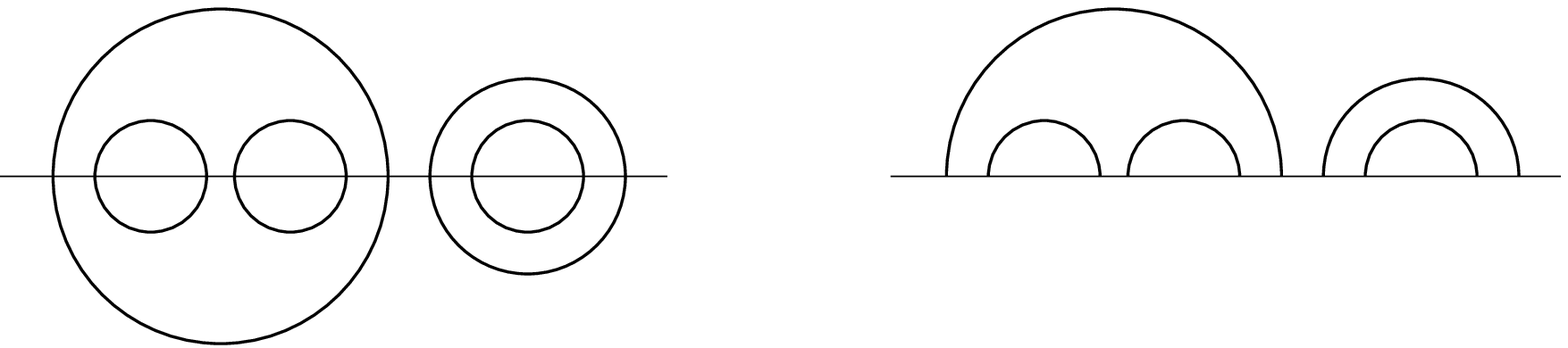}{8.truecm}
\figlabel\parenbridge

Our interest in arch configurations is motivated by the fact that both meanders 
and semi--meanders can be built out of them.  For instance, we get a direct 
one to one correspondence between $n$--component meanders of order $n$
and arch configurations of order $n$: any arch configuration is completed
by reflection wrt the river (see Fig.\parenbridge).
As a consequence, we have
\eqn\conseq{ M_n^{(n)}~=~c_n~,}
the total number of arch configurations of order $n$, in agreement with \maxk.

More generally, any 
multi--component meander of order $n$ is obtained by superimposing 
any two arch configurations of order $n$, one above the river, one below the river, 
and connecting them
through the $2n$ bridges.  
As a consequence, we find the sum rule
\eqn\sumrulm{ \sum_{k=1}^n ~ M_n^{(k)}~ = ~ (c_n)^2 ~,}
expressing the total number of multi--component meanders of order $n$ as the 
total number of couples of (top and bottom) arch configurations of order $n$.
This is readily checked on the data of Table I: the sums of numbers by columns
are equal to the square of the lowest number ($M_n^{(n)}=c_n$) in each column.

\fig{Any semi--meander (b) is obtained from the superimposition of an
arbitrary arch configuration (top of (a)) and a rainbow arch configuration
(bottom of (a)) connected through 
the $2n$ bridges ($n=5$ here), and by folding the river as indicated, thus identifying
the bridges by pairs. The process (a) $\leftrightarrow$ (b) is clearly
reversible.}{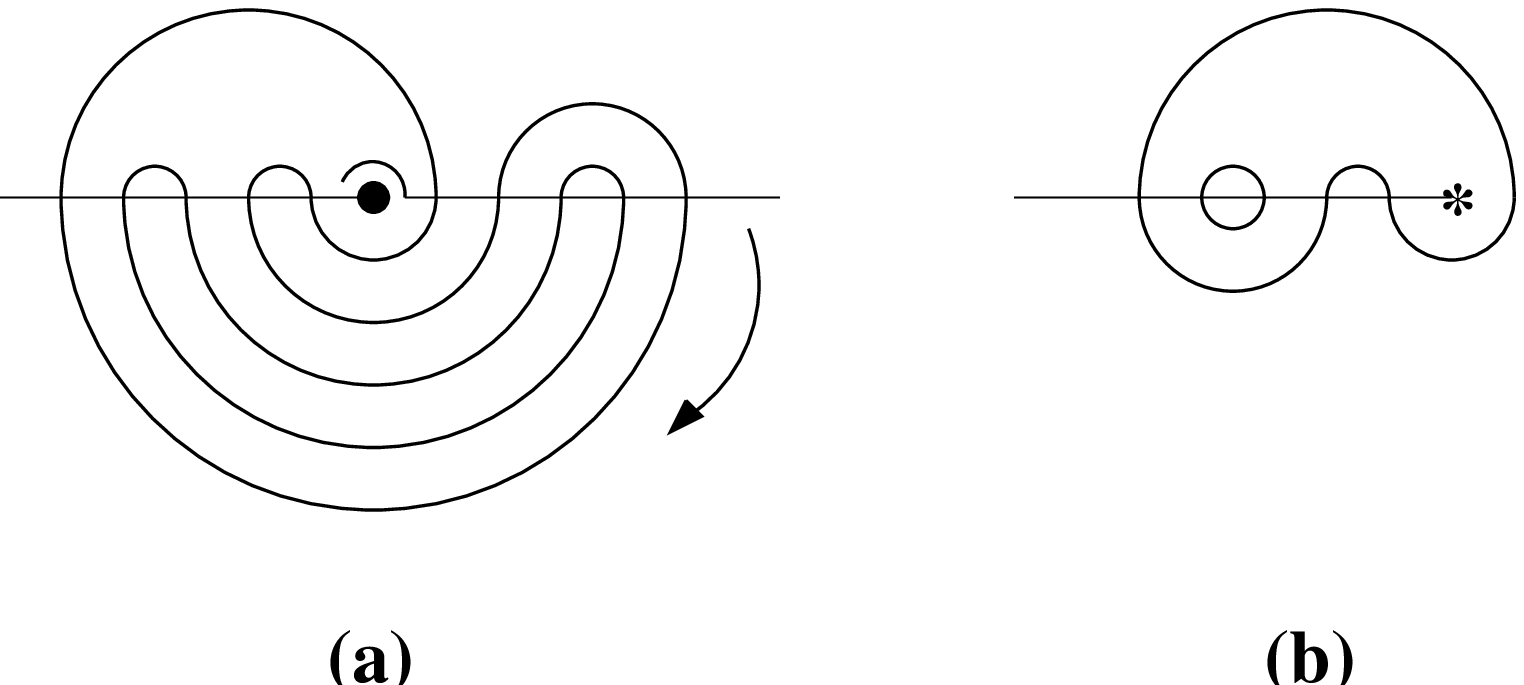}{8.truecm}
\figlabel\archsem

It is instructive to note that any multi--component
semi--meander of order $n$ may also be obtained as the superimposition of one arbitrary arch 
configuration of order $n$
above an infinite river, and a ``rainbow" arch configuration below it (see Fig.\archsem).
The rainbow arch configuration of order $n$, denoted by ${\cal R}_n$, consists
of $n$ arches linking the opposite pairs of bridges: $(1,2n)$, $(2,2n-1)$, ...,$(n,n+1)$.
Note that the number of bridges is doubled in this representation.
To recover the semi--meander, one simply has to fold the infinite river into 
a semi--infinite one, identifying the $2n$ bridges by pairs according
to the rainbow arches as indicated in Fig.\archsem.
As a consequence, we get the semi--meander version of the sum rule \sumrulm\ for
meanders
\eqn\sumsem{ \sum_{k=1}^n ~ {\bar M}_n^{(k)}~=~ c_n~,}
expressing the total number of multi--component semi--meanders of order $n$
as the total number $c_n$ of arch configurations of order $n$
(completed by the lower rainbow ${\cal R}_n$ to yield the semi--meanders). 
This sum rule is readily checked
on the data of Table II: the sums of numbers by columns are the Catalan numbers.

The main difficulty in the meander and semi--meander problems is to find a direct way,
for given arch configurations, to count the number of connected components
of the resulting meander or semi--meander. 
This however is far beyond reach. 
Nevertheless, it is instructive to gather more refined
statistical informations on the distribution of arches in random configurations,
in view of a tentative generalization to the arch statistics
of meanders with fixed number of connected components.

\subsec{Statistics of arches}

{\bf Exterior arches.} Let us first compute the distribution law of exterior arches. 
By exterior arches, we mean arches which have no other arch above them.
For instance, the arch configuration of Fig.\parenbridge\ has two
exterior arches, a rainbow ${\cal R}_n$ has only one exterior arch, etc...
Let $E(n,k)$ denote the number of arch configurations of order
$n$ with exactly $k$ exterior arches.  A simple recursion relation can be
obtained in the same spirit as for Catalan numbers, by following the 
general scheme of 
Fig.\recursion.
Starting from an order $n$ arch configuration with $k$ exterior arches,
let us consider the leftmost arch, starting at the leftmost bridge. It is clearly an
exterior arch.
Let $2j+2$ be the position of the right bridge of this arch.
Again, the arch separates the configuration into two arch configurations.
The one below the arch is an arbitrary configuration among the $c_j$ arch 
configurations of order $j$. 
The one to the right of the arch is an arch configuration of order 
$n-j-1$, with $k-1$ exterior arches. This leads to the recursion\foot{
One can also show that $E(n,k)=E(n,k+1)+E(n-1,k-1)$.}
\eqn\recexta{ E(n,k) ~=~ \sum_{j=0}^{n-1} c_j ~E(n-j-1,k-1)~,}
for $k \geq 1$ and $n \geq 1$, with the initial condition $E(n,0)=\delta_{n,0}$
and $E(n,k)=0$ for $k>n$.
This defines the numbers $E(n,k)$ uniquely, and it is easy to prove that
\eqn\resexta{ E(n,k)~=~ { k \over 2n-k} {2n-k \choose n}~=~ {k (2n-k-1)! \over
n! (n-k)!} ~.}

Another way of characterizing the distribution of arch configurations according to their
number of exterior arches is through its factorial moments, defined as
\eqn\facmoex{ \langle {k \choose l} \rangle_{\rm ext}~=~{ \sum_{k=1}^n {k \choose l}
E(n,k) \over \sum_{k=1}^n E(n,k)} ~,}
where ${k \choose l}\equiv k(k-1)...(k-l+1)/l!$.
Thanks to the identity
\eqn\idenpol{ \sum_{k=1}^n {k \choose l} E(n,k)~=~ {2l+1 \over 2n+1} {2n+1 \choose n-l}~
={2n \choose n-l} - {2n \choose n-l-1},}
which is proved in appendix A (note that for $l=0$ \idenpol\ gives the total
number $c_n$ of arch configurations of order $n$), we deduce that
\eqn\momfac{ \langle {k \choose l} \rangle_{\rm ext}~=~
(2l+1){n! (n+1)! \over (n-l)! (n+l+1)!}~.}
In particular, for $l=1$,
the average number of exterior arches for arch configurations
of order $n$ reads
\eqn\avext{ \langle k \rangle_{\rm ext}~=~ {3 n \over n+2}~.}

In the limit of infinite order, $n \to \infty$ (the thermodynamic limit), the $l$th
factorial moment of the distribution of exterior arches tends to $2l+1$. 
In particular, the average number of exterior arches tends to $3$.

{\bf Interior arches.} An interior arch of an arch configuration is an arch containing
no other arch below it. For instance the configuration of Fig.\parenbridge\ has
$3$ interior arches, a rainbow ${\cal R}_n$ has only one interior arch, etc...
Let $I(n,k)$ denote the
number of arch configurations of order $n$ with exactly $k$ interior arches.
As before (see Fig.\recursion), we can derive a recursion 
relation by following the leftmost exterior arch
of any order $n$ arch configuration with $k$ interior arches. Three situations
may occur: 

(i) the arch links the first bridge to the last one. 
Below this arch, we may have
any arch configuration of order $n-1$ with $k$ interior arches.

(ii) the arch links the first bridge to the second, hence it is an interior arch. 
To its right, we have any arch configuration of order $n-1$, with $k-1$ interior arches.

(iii) the arch links the first bridge to the one in position $2j+2$, $j=1,...,n-2$.
It separates the configuration into two configurations of respective orders $j$ 
(below the arch)
and $n-j-1$ (to the right of the arch), with respectively $q$ and $k-q$ interior
arches, $q=1,...,k-1$.

\noindent{}This leads to the recursion relation
\eqn\recint{ I(n,k)~=~ I(n-1,k)+I(n-1,k-1)+\sum_{j=1}^{n-2} \sum_{q=1}^{k-1}
I(j,q) I(n-j-1,k-q)~,}
for $n \geq k\geq 1$, and with the initial condition $I(1,k)=\delta_{k,1}$,
and $I(j,q)=0$ for $q>j$.
This allows one to solve for 
\eqn\solint{ I(n,k)~=~ {1 \over n} {n \choose k} {n \choose k-1}~=~{n! (n-1)! \over
k! (k-1)! (n-k)! (n-k+1)!}~.}
With these numbers, we compute
\eqn\compint{\sum_{k=1}^n {k \choose l} I(n,k)~=~ {1 \over n} {n \choose l} 
{2n-l \choose n-l+1} ~.}
A proof of this identity is given in appendix A. Hence the factorial moments of 
the distribution of interior arches read
\eqn\facint{\eqalign{
\langle {k \choose l} \rangle_{\rm int}~&=~ {\sum_{k=1}^n {k \choose l} I(n,k) \over
\sum_{k=1}^n I(n,k) } \cr
&=~ {{n \choose l} {2n-l \choose n-l+1} \over {2n \choose n-1}}~. \cr}}
For instance the average number of interior arches on an arch configuration 
of order $n$ is
\eqn\avint{ \langle k \rangle_{\rm int}~=~ {n+1 \over 2}~.}
{}From \facint, it is easy to see that the $l$th factorial moment behaves as
\eqn\compfacint{ \langle {k \choose l} \rangle_{\rm int} ~\sim ~ {n^l \over 2^l l!} }
for large $n$.
All these numbers tend to infinity when $n$ is large,
hence diverge in the thermodynamic limit. 

\fig{The $S$ duality between arch configurations is indicated by arrows,
for orders $n=1,2,3$.}{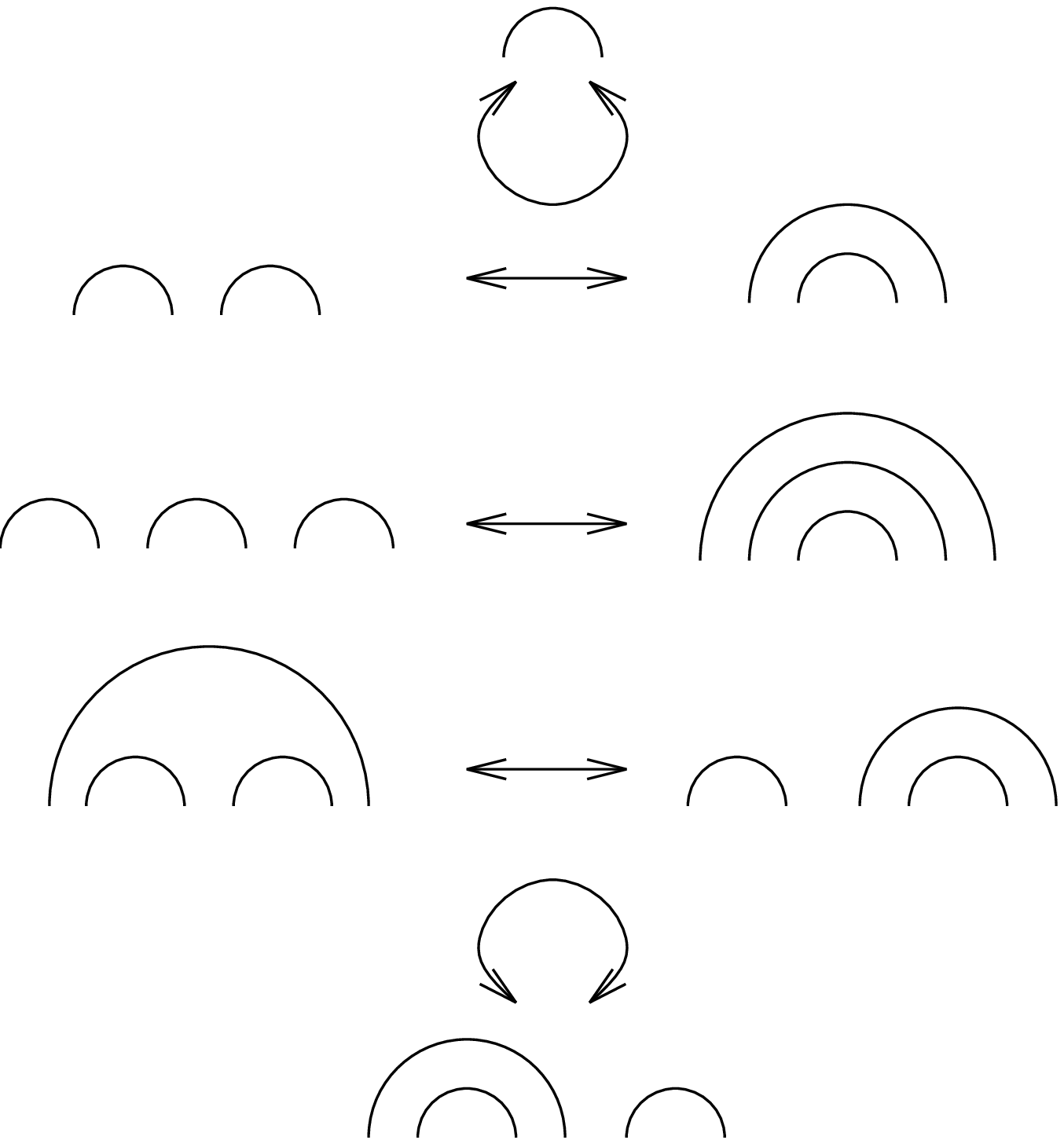}{8.truecm}
\figlabel\exdua

A last remark is in order. The numbers $I(n,k)$ \solint\ satisfy a symmetry relation
\eqn\symint{ I(n,k)~=~ I(n,n+1-k)~,}
which is in fact the expression of a duality between arch configurations.
Singling out the leftmost exterior arch, any arch configuration may be expressed as 
on Fig.\recursion.  The duality transformation is defined recursively as
\eqn\defdua{ \eqalign{
S(\emptyset)~&=~ \emptyset \cr
S({\epsfxsize=.9truecm \epsfbox{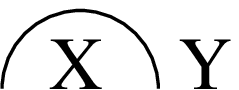}}\ \ \ )~&=~
{\epsfxsize=.9truecm \epsfbox{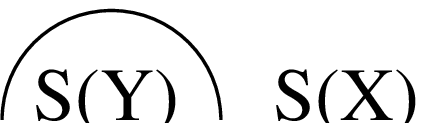}} \cr}}
This transformation is involutive ($S \circ S={\bf I}$) and exchanges
disjoint and nested arches as exemplified in Fig.\exdua.
The transformation $S$ changes the number of interior arches of a given 
configuration of order $n$ from $k$ to $n+1-k$, as is readily shown by recursion.
This explains the symmetry \symint.

{\bf Arches which are both interior and exterior.}  Let us compute the total number 
$EI(n)$ of arches which are both interior and exterior in all the arch configurations
of order $n$.  We have $EI(1)=1$.  By the recursion process of Fig.\recursion, 
we can relate the number $EI(n)$ to the numbers $EI(j)$, $j<n$ in the following manner.
Summing over the position $2j+2$ of the right bridge of  the leftmost exterior arch in
any order $n$ configuration, we find that only the configuration $Y$ of order $n-1-j$,
to the right of this arch, contributes to $EI(n)$ through
the total of $EI(n-1-j)$ arches which are both interior and exterior. This number 
comes $c_{j}$ times, corresponding to the arbitrary choice of the 
configuration $X$ of order $j$ below the leftmost exterior arch. 
When $j=0$, the leftmost arch is both interior and exterior, and there
are $c_{n-1}$ such configurations, hence in addition to the $EI(n-1)$ arches
which are both interior and exterior, we also have $c_{n-1}$ leftmost
arches, which are both interior and exterior. 
There is no contribution from $j=(n-1)$ (for $n \geq 2$) 
as the only exterior arch cannot be interior.
This yields
\eqn\recxtint{ EI(n)~=~ \sum_{j=1}^{n-2} c_j~EI(n-j-1)~+EI(n-1)+c_{n-1}~,}
valid for $n \geq 2$.
If we define $EI(0)=1$, this takes the form
\eqn\extintrec{ EI(n)~=~ \sum_{j=0}^{n-1} c_j ~ EI(n-1-j)~,}
hence 
\eqn\solexin{ EI(n)~=~c_n~.}
For all $n$, there is therefore exactly one arch in average 
which is both exterior and interior.

{\bf Arches of given depth.}  The depth of an arch in an arch configuration 
is defined as follows. The exterior arches are the arches of depth $1$.
The arches of depth $k+1$ are the exterior arches of the arch configuration
obtained by erasing the arches of depth $1$, $2$, ...,$k$.
For instance the rainbow configuration ${\cal R}_n$
of order $n$ has one arch of each depth 
between $1$ and $n$; the arch configuration of Fig.\parenbridge\ 
has $2$ arches of depth $1$,
and $3$ arches of depth $2$, which are also its interior arches.
Let us compute the {\it total} number $A(n,k)$ of arches 
of depth $k$ in {\it all} the arch
configurations of order $n$, by the usual reasoning of Fig.\recursion. 
Starting from any order $n$
arch configuration, consider as usual the leftmost arch.
It links the first bridge to the one in position $2j+2$, $j=0,...,n-1$.
The arch separates the configuration into one of order $j$ (below the arch, $X$
in Fig.\recursion),
whose depth $k-1$ arches give rise to depth $k$ arches of the initial configuration,
and one of order $n-j-1$
(to the right of the arch, $Y$ in Fig.\recursion), 
whose depth $k$ arches are depth $k$ arches 
of the initial configuration.

\noindent{}This translates into the recursion relation
\eqn\recdepth{ A(n,k)~=~ \sum_{j=0}^{n-1} \big( A(j,k-1) c_{n-j-1}+A(n-j-1,k)
c_j \big) ~,}
for $n\geq 0$ and $k \geq 2$, with $A(n,k)=0$ if $k >n$. The initial condition is 
the total number $A(n,1)$ of external arches of all order $n$ 
arch configurations\foot{Indeed the initial condition $A(1,1)=1$ 
would be sufficient, as the expression
\recdepth\ is recursive wrt $k+n$.},
namely
\eqn\initdepth{ A(n,1)~=~\sum_{j=1}^n j E(n,j)~=~ {3 \over 2n+1} {2n+1 \choose n-1}~,}
according to \idenpol\ for $l=1$. 
This fixes the $A(n,k)$ completely, and we find
\eqn\archnum{ A(n,k)~=~ {2k+1 \over 2n+1} {2n+1 \choose n-k}~
=~{2n \choose n-k}- {2n \choose n-k-1}.}
This is nothing but the rhs of the identity \idenpol, with the substitution $l \to k$
(indeed we have taken the $l=1$ value of this expression as initial condition).
We still lack of a good explanation for this phenomenon.
Note that the formula \archnum\ can be used as the definition of $A(n,0)=c_n$, the
total number of arch configurations, thus interpreting the depth $0$ arch number
to be $1$ for each configuration.  We can define the average number of depth $k$
arches as
\eqn\avdep{ \langle A(n,k) \rangle~=~ { A(n,k) \over A(n,0)}~=~
(2k+1){n! (n+1)! \over (n-k)! (n+k+1)!}~.}
These numbers coincide with the factorial moments \momfac\
of the distribution of external arches.
In particular, in the thermodynamic limit ($n \to \infty$), 
the average number of depth $k$ arches tends
to $2k+1$.
So the average arch configuration has $3$ exterior arches, $5$ arches of depth $2$, 
$7$ arches of depth $3$, etc ...

The total number of arches on a given configuration is $n$. Therefore, we have the following
sum rule
\eqn\sumrar{ \sum_{j=1}^\infty A(n,j) ~=~ n ~ A(n,0)~.}
The above result for the thermodynamic averages enables us to define an average
maximal depth in arch configurations of order $n$, denoted by $d(n)$, by
\eqn\maxdep{
\eqalign{
n~&=~ \sum_{j=1}^{d(n)} \lim_{n \to \infty} {A(n,j) \over A(n,0)} \cr
&=~ \sum_{j=1}^{d(n)} ~ (2j+1)~=~ d(n) ( d(n)+2)~.\cr}}
This gives 
\eqn\depmax{ d(n)~=~ \sqrt{n+1} -1~.}

\newsec{Recursion relations for semi--meanders}

In the following, we will mainly concentrate our efforts on semi--meanders. 
However, we will mention whenever possible the extensions of our results to meanders.

\subsec{The main recursion}

Let us present a simple algorithm for enumerating the semi--meanders of order $n$
with $k$ connected components. 
In this section, we will work in the infinite river/lower rainbow arch--framework for
semi--meanders (see Fig.\archsem\ (a)), namely consider a semi--meander 
as the superimposition
of the lower rainbow configuration ${\cal R}_n$, and some upper arch configuration.
 
\fig{The construction of all the semi--meanders of order $n+1$ with arbitrary number
of connected components from those
of order $n$. Process (I): (i) pick any exterior arch and cut it (ii)
draw its edges around the semi--meander and paste them below.
The lower part
becomes the rainbow configuration ${\cal R}_{n+1}$ of order $n+1$.
The process (I) preserves the number of connected components.
Process (II): draw a circle around the semi--meander of order 
$n$. The process (II) adds one connected component.}{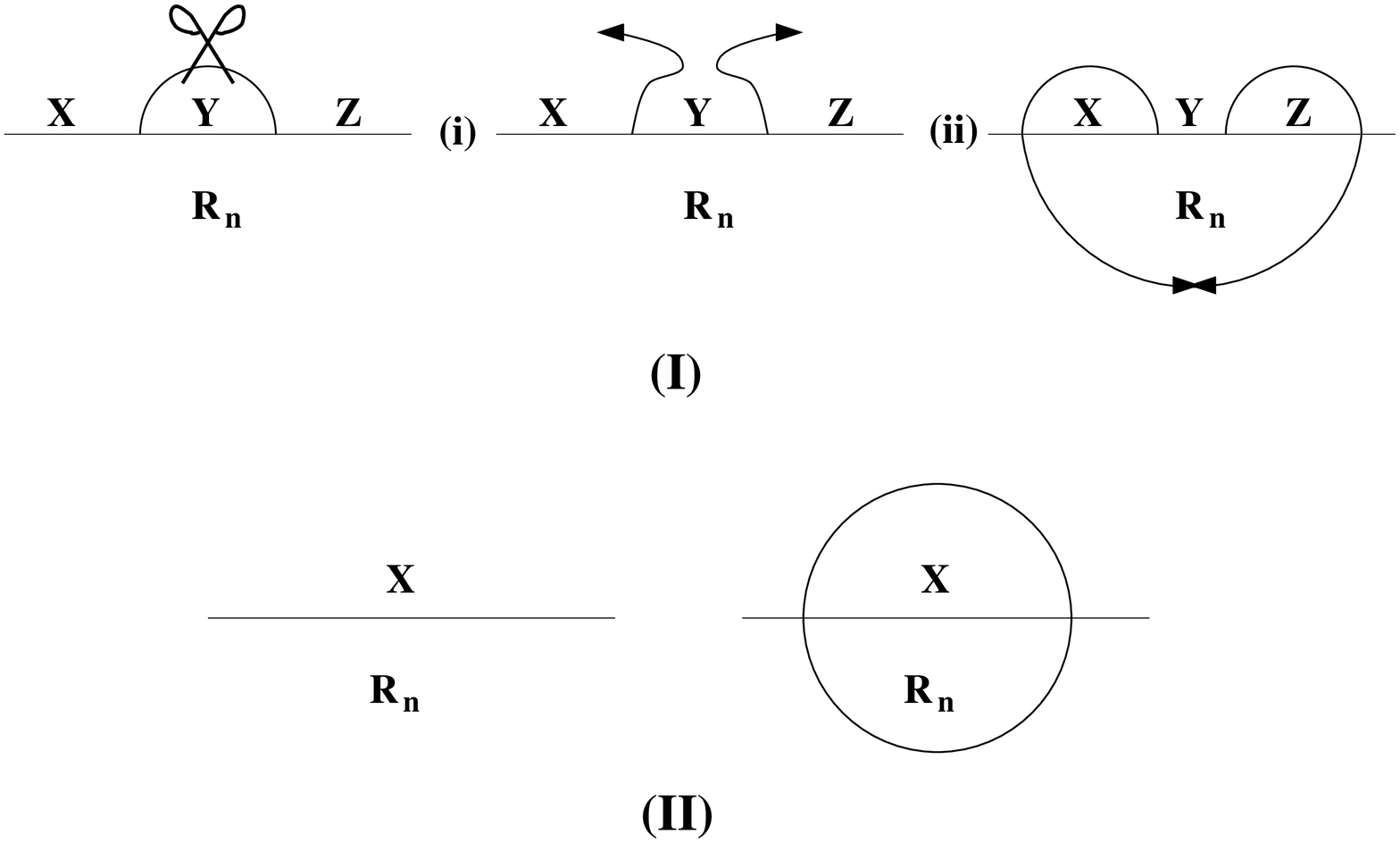}{12.truecm}
\figlabel\cutmeander

Starting from a semi--meander of order $n$
with $k$ connected components, let us construct several semi--meanders of order $n+1$
with $k$ connected components in the following way, as indicated in Fig.\cutmeander\
(I) (i)-(ii). 

(i) Pick any exterior arch of the upper arch
configuration of the semi--meander and cut it. 

(ii) Pull the two edges of the cut 
across the river (the left part of the exterior arch to the left,
the right one to the right), and paste them around the lower rainbow, 
thus increasing the rainbow configuration by one arch 
(${\cal R}_n \to {\cal R}_{n+1}$), 
and the number of bridges by $2$. 

The result is 
a certain semi--meander of order $n+1$ with the same number $k$ of connected components.
So for each semi--meander $\cal M$ of order $n$ with $k$ components, we can construct 
$E({\cal M})$ semi--meanders of order $n+1$ with $k$ components, where $E({\cal M})$
denotes the number of exterior arches of $\cal M$.  All these new
semi--meanders are clearly distinct.  
There is however another way of generating more semi--meanders of order $n+1$,
indicated in Fig.\cutmeander\ (II). 
Starting with any semi--meander of order $n$ with $k-1$ connected components, 
one just adds an extra circular loop around it, which transforms the lower rainbow
of order $n$ into ${\cal R}_{n+1}$, and adds $2$ bridges. The resulting semi--meander of
order $n+1$ has clearly $k$ connected components.
Such a semi--meander cannot be obtained from some order $n$ semi--meander by the 
procedure (I), because it has only one exterior upper arch, whereas the process
(I) produces at least two exterior arches.  

So all the order $n+1$ semi--meanders 
constructed by (I)-(II) are distinct. Conversely, given a semi--meander of order $n+1$
with $k$ components, two cases may occur: 

(a) it has only one exterior upper arch. In this case, it is surrounded by one circle, 
and therefore arises from the order $n$ semi--meander with $k-1$ components inside the
circle, through (II).

(b) it has at least two exterior upper arches. 
Cutting the lower exterior arch of the lower rainbow ${\cal R}_{n+1}$, pulling 
the edges of this arch above the upper configuration (the left edge by the
left side, the right one by the right side, both suppressing the left and 
right--most bridges), and finally pasting the two edges on the
upper side of the river (thus creating an exterior arch on the upper configuration), 
one gets an order $n$ semi--meander with $k$ components,
which leads to the initial semi--meander of order $n+1$ through (I)--(i)-(ii),
using the exterior arch constructed above.

\noindent{}This proves that the procedures (I)-(II) give a recursive
algorithm for constructing {\it all} the semi--meanders of order $n+1$ 
with $k$ connected components from the $k$ and $k-1$--component meanders of order $n$.
More precisely, these procedures can also be viewed as a recursive algorithm
for constructing all the arch configurations of order $n$ from those of order
$n-1$. When completed by a lower rainbow configuration ${\cal R}_n$, so as to give 
multi--component semi--meanders, we have the interesting property that (I) 
preserves
the number of connected components of the semi--meander, while (II) increases it 
by $1$, which allows us in principle to follow this number of components throughout
the construction.

The case $k=1$ of connected semi--meanders is special, in the sense that the 
corresponding arch configurations are obtained by successive actions of
the procedure (I) {\it only}. 
As a consequence, the number
of connected semi--meanders of order $n+1$ is equal to the total number of 
exterior upper arches of all the semi--meanders of order $n$.

\subsec{Expression in terms of arch numbers}

We now want to translate into some recursion relation the transformations (I)--(II).
It is best expressed in terms of the numbers of arches of given depth in the upper
configurations of the semi--meanders.
Recall that to build a semi--meander, one just has to close an upper arch configuration
using a lower rainbow arch configuration. Therefore, we can decompose the set of 
arch configurations of order $n$ into a partition, according to the number 
$k=1,2,...,n$ of connected components of the corresponding semi--meander.
Denoting respectively by ${\cal A}_n$ and $SM_{n,k}$ the sets of arch 
configurations of order $n$
and of semi--meanders of order $n$ with $k$ connected components, we have
therefore
\eqn\partiarch{ {\cal A}_n~=~\cup_{k=1}^n ~ SM_{n,k}~. }
The statistics of arches extensively studied in sect.3  only give some global
information on semi--meanders, irrespectively of their
number of connected components. Taking the latter into account is a subtle refining 
of the above study.  In particular we do not expect the statistics of arches to be 
the same when the number of connected components is fixed. 
We rather want to use sect.3  as a guideline for the study of semi--meander arch 
statistics. In particular, we will use analogous definitions for observable
quantities, and derive recursion relations between them.

For a given semi--meander ${\cal M}$
of order $n$ with $k$ components, let us denote by 
$A({\cal M},j)$ the number of arches of depth $j$ in its upper configuration,
and let
\eqn\soma{ A_k(n,j)~=~ \sum_{{\cal M} \in SM_{n,k}}  A({\cal M},j)~,}
denote the total number of arches of depth $j$ of the upper configurations
of the semi--meanders of order $n$ with $k$ connected components.
To make the contact with the numbers $A(n,j)$ of arches of depth
$j$ in the arch configurations of order $n$ (eq.\archnum), 
let us mention the obvious sum rule
\eqn\suarc{ \sum_{k=1}^n ~ A_k(n,j)~=~ A(n,j)~.}
This simply states that summing the numbers of upper arches
of depth $j$ over all (multi--component) semi--meanders of 
order $n$ amounts to counting the total number of arches of depth $j$ in all the
arch configurations of order $n$. This sum rule is parallel to the set decomposition
\partiarch, and simply evaluates the number of depth $j$ arches on both sides.

To write a recursion relation, we must follow the evolution of the depth of arches 
through the processes (I) and (II).  In the process (II), it is clear that the 
arches of depth $j-1$ in the order $n$ semi--meander become arches of depth $j$
in the order $n+1$ one, hence we get the
contribution
\eqn\contII{ A_{k-1}(n,j-1)=\sum_{{\cal M} \in SM_{n,k-1}}
A({\cal M},j-1) }
to the total number of arches of depth $j$.  
The process (I) of Fig.\cutmeander\ is more subtle. We start
from a meander of order $n$ with $k$ components, ${\cal M}_{n,k}$.
For each of its $E({\cal M}_{n,k})$ exterior arches, the process (I) affects 
the depth of its arches in the following way.  The arches of depth $j+1$ 
below the specified exterior arch (configuration denoted by $Y$ on Fig.\cutmeander)
become arches of depth $j$. The arches of depth
$j-1$ to the left ($X$) and to the right ($Z$) of this exterior arch become arches of depth 
$j$.  Summing over all the $A({\cal M}_{n,k},1)$
exterior arches of ${\cal M}_{n,k}$, we get 
the contribution
\eqn\sumamkn{A({\cal M}_{n,k},j+1)+
[A({\cal M}_{n,k},1)-1]~
A({\cal M}_{n,k},j-1) }
to the total number of arches of depth $j$.
Adding the contributions \contII\ and \sumamkn\ of the processes (II) and (I), we
finally get the main  recursion relation, for $k \geq 2$ and $j \geq 2$
\eqn\mainrecu{ A_k(n+1,j)~=~ A_{k-1}(n,j-1)+A_k(n,j+1)+\sum_{{\cal M}\in
SM_{n,k}} [A({\cal M},1)-1] ~A({\cal M},j-1)~.}
Like in the arch configuration case, we can introduce the number of arches of depth $0$
in a semi--meander, in such a way that
\eqn\defdepnul{ A({\cal M},0)~=~1 \qquad \forall {\cal M} \in SM_{n,k}~.}
In particular, 
\eqn\nulldef{ A_k(n,0)~=~ {\rm card}(SM_{n,k})~=~ {\bar M}_n^{(k)}~.}
For $j=1$, the recursion reads
\eqn\rekone{A_k(n+1,1)~=~ A_{k-1}(n,0)+ A_k(n,2) + 2 A_k(n,1)~.}
For $j=0$, the recursion relation becomes
\eqn\recujnull{ A_k(n+1,0)~=~ A_{k-1}(n,0)+A_k(n,1) ~.}
Eqs.\mainrecu\-\nulldef\-\rekone\ are all valid for $k \geq 2$. 

The case $k=1$ is different since it uses the procedure (I) only.
However, the above equations remain valid upon setting
$A_0(n,j)=0$ for any $n,j$. This leads to
\eqn\casekone{ A_1(n+1,j)~=~ A_1(n,j+1)+\sum_{{\cal M}\in
SM_{n,1}} [A({\cal M},1)-1] ~A({\cal M},j-1)~,}
valid for $j \geq 2$. For $j=1,0$, we have
\eqn\casenull{\eqalign{
A_1(n+1,1)~&=~ A_1(n,2)+2 A_1(n,1)~\cr 
A_1(n+1,0)~&=~A_1(n,1)~.\cr}}
The last equation expresses the fact that the total number ${\bar M}_{n+1}$ of connected
semi--meanders of order $n+1$ is equal to the total
number of exterior (depth $1$) arches on all the semi--meanders of order $n$.

The abovementioned leading behaviour \asympto\ of the connected semi--meander number
\eqn\leadsem{ {\bar M}_n ~\propto ~ {\bar R}^n ~,}
enables one to recover the value of $\bar R$ from the thermodynamic limit
of various averages over semi--meanders. 

We now denote by
\eqn\avseme{ \langle f({\cal M}) \rangle_{n,k}~=~
{1 \over A_k(n,0)} \sum_{{\cal M} \in SM_{n,k}}  f({\cal M}) }
the average of a function $f$ over the semi---meanders of order $n$ with 
$k$ connected components.
Dividing eq.\mainrecu\ by $A_k(n,0)$, we find a relation
between averages
\eqn\newmainrecu{\eqalign{
{A_k(n+1,0) \over A_k(n,0)} &\langle A({\cal M},j) 
\rangle_{n+1,k}~=~ 
{A_{k-1}(n,0) \over A_k(n,0)}\langle A({\cal M},j-1)\rangle_{n,k-1}\cr
&+\langle A({\cal M},j+1)+ 
[A({\cal M})-1]A({\cal M},j-1) \rangle_{n,k} ~,\cr}}
for $k\geq 2$. For $k=1$, we have
\eqn\newone{
{A_1(n+1,0) \over A_1(n,0)} \langle A({\cal M},j) 
\rangle_{n+1,1}~=~ \langle A({\cal M},j+1)+ 
[A({\cal M})-1]A({\cal M},j-1) \rangle_{n,1} ~,}
for $j\geq 2$ and for $j=1,0$
\eqn\newtwo{\eqalign{{A_1(n+1,0) \over A_1(n,0)}
\langle A({\cal M},1)\rangle_{n+1,1}~&=~
\langle A({\cal M},2)+ 2 A({\cal M},1) \rangle_{n,1}\cr
{A_1(n+1,0) \over A_1(n,0)}~&=~\langle A({\cal M},1)\rangle_{n,1}~. \cr}}
This last equation enables us to interpret $\bar R$ as the average
thermodynamic number of exterior arches
\eqn\avexta{ {\bar R}~=~ \lim_{n \to \infty} {A_1(n+1,0) \over A_1(n,0)}~=~
\lim_{n \to \infty} {{\bar M}_{n+1} \over {\bar M}_n}~=~
\langle A({\cal M},1)\rangle_{\infty,1}~.}

\subsec{Mean field approximation} 

In the thermodynamic limit, 
in view of the result for the statistics of arches,
we expect all these averages to tend to some finite values.
A standard
approximation when dealing with averages over some statistical 
distribution is the {\bf mean field} approximation. It expresses that the 
observable quantities (here the arch numbers) are fixed to their mean value.
In particular, in this approximation, the term correlating the arches
of depth $1$ and $j-1$ is replaced by
\eqn\approxmena{ \langle[A({\cal M})-1]A({\cal M},j-1) 
\rangle_{n,k} \longrightarrow
\langle[A({\cal M})-1]\rangle_{n,k} 
\langle A({\cal M},j-1) \rangle_{n,k}~.}
{}From now on, we focus our attention to the connected case $k=1$, where this
approximation gives the system
\eqn\meafeq{\eqalign{ 
{A_1(n+1,0) \over A_1(n,0)} \langle A({\cal M},j) 
\rangle_{n+1,1}~&=~ \langle A({\cal M},j+1)\rangle_{n,1} +\langle 
[A({\cal M})-1] \rangle_{n,1} \langle A({\cal M},j-1) \rangle_{n,1} \cr
{A_1(n+1,0) \over A_1(n,0)}
\langle A({\cal M},1)\rangle_{n+1,1}~&=~
\langle A({\cal M},2)+ 2 A({\cal M},1) \rangle_{n,1}\cr
{A_1(n+1,0) \over A_1(n,0)}~&=~\langle A({\cal M},1)\rangle_{n,1}~. \cr}}
Starting from the only semi--meander of order $n=1$,
(a circle crossing the river through two bridges), for which 
\eqn\inicir{\langle A({\cal M},j) \rangle_{1,1}~=~ \delta_{j,1}~,}
we use the system \meafeq\ recursively to get the quantities 
\eqn\obs{a_n(j)~=~\langle A({\cal M},j) \rangle_{n,1}~,}
for any $n \geq 1$. 
Let us prove that when $n \to \infty$ they converge to the values
\eqn\finicir{\langle A({\cal M},j) \rangle_{\infty,1}~=~3 \qquad \forall ~ j~.}

If the quantities \obs\ converge, their limits, denoted by
$a(j)= \langle A({\cal M},j) \rangle_{\infty,1}$, 
are subject to the recursion relation (as fixed points of the system \meafeq)
\eqn\meanrec{{\bar R}~ a(j)~=~a(j+1)+ [a(1)-1] a(j-1)~,}
valid for $j \geq 2$.
On the other hand, $a(1)={\bar R}$, from eq.\avexta.
We are thus left with the recursion relation
\eqn\recakj{ a(j+1)~=~ {\bar R} a(j)  -({\bar R}-1) a(j-1)~,}
for $j \geq 2$, whereas
\eqn\recfinmm{ a(2)~=~ a(1)~({\bar R}-2)~.}
Note that necessarily ${\bar R}>2$ for the average number $a(2)$ to be positive. 
With the initial conditions $a(0)=1$ and $a(1)={\bar R}$,
we find in the $1$--component case ($k=1$) that
\eqn\solone{a(j)~=~ {{\bar R} \over {\bar R}-2} ( 1+ ({\bar R}-3)({\bar R}-1)^{j-1})~.}
This is the mean field solution for the numbers of depth $j$ arches in connected
semi--meanders. These numbers are further constrained by the fact that the
average numbers of arches have to be non--negative, hence
\eqn\furcons{ {\bar R}~ \geq ~ 3~.}

Let us now prove that ${\bar R} \leq 3$. For this purpose, let us show by recursion
that the inequalities 
\eqn\inne{ 3 \geq \langle A({\cal M},1)\rangle_{n,1} \geq \langle
A({\cal M},2)\rangle_{n,1} \geq \cdots }
are satisfied for all $n$. The inequalities \inne\ are obvious for $n=1$.
Suppose \inne\ is true for some value of $n$, then using the recursion relations \meafeq,
and performing the difference of any two recursion relations for consecutive values of $j$,
we get
\eqn\varca{\eqalign{ a_n(1) a_{n+1}(1) &\leq 3 ~ a_n(1) \cr
a_n(1)(a_{n+1}(2)-a_{n+1}(1)) &\leq a_n(1)(a_n(1)-3) \cr
a_n(1)(a_{n+1}(j+1)-a_{n+1}(j))  &\leq (a_n(1)-1)(a_n(j)-a_n(j-1)) \cr}}
which imply \inne\ at order $n+1$.  In particular we always have $a_n(1) \leq 3$,
hence in the large $n$ limit, $a(1)={\bar R} \leq 3$.

This leaves us with the unique possibility
\eqn\meanapp{{\bar R}~=~3 \quad  \Rightarrow \quad a(j)~=~3 \quad\forall j \geq 0 ~,}
which is the actual limit of the mean field approximation to arch numbers
of given depth.
Note that in this case the recursion \solone\ is satisfied for $j=0,1$ as well.
Of course this is only a rough approximation of the exact $\bar R$
observed in \asympto, as is usually expected from a mean field approximation.

\subsec{Mean field is wrong for arch configurations}

The recursion relation \mainrecu\ can be used to recover some information on 
arch configurations, by summing it over all the numbers $k$ of connected components.
Recall that $\sum_k A_k(n,j)= A(n,j)$, the total number of arches of depth $j$ in
the arch configurations of order $n$. Hence, summing over $k=1,2,...,n+1$ in \mainrecu,
we get
\eqn\archrecu{A(n+1,j)~=~A(n,j-1)+A(n,j+1) + \sum_{{\cal M} \in {\cal A}_n}
[A({\cal M},1)-1] ~A({\cal M},j-1)~,}
where ${\cal A}_n$ denotes the set of arch configurations of order $n$. Using the explicit 
formula \archnum\ for the number of arches, we deduce that
\eqn\evalcorr{ \sum_{{\cal M} \in {\cal A}_n}[A({\cal M},1)-1] ~A({\cal M},j-1)~=~
2 A(n,j)~.}
This is obtained by noticing that the numbers $A(n,j)$ satisfy the linear
recursion relation
\eqn\linrec{ A(n+1,j)~=~A(n,j-1)+A(n,j+1) +2 A(n,j)~,}
for $n,j \geq 1$, as a consequence of the expression \archnum\ of $A(n,j)$ as
a difference between two binomial coefficients.  
This gives the value of the correlation function
between the numbers of arches of depth $1$ and those of depth $j-1$ in all 
arch configurations. Dividing \evalcorr\ by the number of arch configurations 
$A(n,0)=c_n$, we get a relation for averages. If we denote by
\eqn\notav{ \langle f({\cal M}) \rangle_n~=~{1 \over A(n,0)} 
\sum_{{\cal M} \in {\cal A}_n} f({\cal M})~,}
this relation reads
\eqn\avarch{ \langle [A({\cal M},1)-1] ~A({\cal M},j-1) \rangle_n ~=~
2 \langle A({\cal M},j) \rangle_n~.}
In the thermodynamic limit $n \to \infty$, as $\langle A({\cal M},j)\rangle_n
\to 2j+1$ (see eq.\avdep), we find that 
\eqn\diffav{\eqalign{
\langle [A({\cal M},1)-1] ~A({\cal M},j-1) &\rangle_n-  
\langle [A({\cal M},1)-1] \rangle_n \langle A({\cal M},j-1) \rangle_n \cr
&\buildrel{n \to \infty}\under\longrightarrow 2(2j+1) -(3-1)(2j-1) =4~.\cr}}
We see that the mean field approximation used in the 
previous section for semi--meanders
is clearly wrong in the case of 
arch configurations. Note also that the result \diffav\ is independent of $j$.

However, it is easy to solve the mean field equations for arch
configurations, which lead to a
growth $(2+\sqrt{2})^n$ for the total number of arch configurations, whereas the exact 
behaviour is $4^n$ (as readily seen from the large $n$
expression of the Catalan number $c_n$
by use of the Stirling  formula).

\subsec{Improved mean field approximation for semi--meanders}

This suggests to try a deformed thermodynamic
mean field--type ansatz for the numbers of arches in semi--meanders, 
of the form
\eqn\defmea{ \langle[A({\cal M})-1]A({\cal M},j-1) 
\rangle_{n,k} \buildrel{n \to \infty}\under\longrightarrow  K_k+
\langle[A({\cal M})-1]\rangle_{\infty,k} 
\langle A({\cal M},j-1) \rangle_{\infty,k}~,}
where $K_k$ are some constants to be determined.
For $k=1$, the main recursion relations \casekone\ and \casenull\ read, in the
thermodynamic limit
\eqn\maimpro{\eqalign{ {\bar R} ~a_1(j)~&=~ a_1(j+1) + K_1 + [a_1(1)-1]a_1(j-1)
\quad {\rm for} \ j \geq 2,\cr
a_1(2)~&=~ a_1(1)({\bar R}-2)~\cr
a_1(1)~&=~ {\bar R} ~a_1(0)\cr
a_1(0)~&=~1~.\cr}}
We determine the constant $K_1$ by imposing that the main recursion be 
satisfied for the value $j=1$ as well, in which case we get
\eqn\deterK{ {\bar R} ~a_1(1)={\bar R}^2= {\bar R}({\bar R}-2)+K_1+({\bar R}-1) \quad \Rightarrow \quad K_1={\bar R}+1~.}
Next the recursion is easily solved by noticing that
\eqn\solrecp{ a_1(j+1)-\alpha (j+1)~=~{\bar R}~ (a_1(j) -\alpha j)
-({\bar R}-1)~(a_1(j-1)-\alpha(j-1))~,}
where
\eqn\alsol{ \alpha={{\bar R}+1 \over {\bar R}-2}~.}
This is exactly the same recursion as in the mean field case \recakj, except that in 
the new variables $\alpha_1(j)=a_1(j)-\alpha j$, the initial conditions read 
\eqn\initcon{ \alpha_1(0)=1 \qquad \alpha_1(1)={{\bar R}^2-3 
{\bar R}-1 \over {\bar R}-2}~.}
The solution reads
\eqn\solimproved{ a_1(j)~=~{{\bar R}+1 \over {\bar R}-2}j+ 
{1 \over ({\bar R}-2)^2}(3+({\bar R}^2-4{\bar R} +1) ({\bar R}-1)^j)~.}
Like in the mean field case, the solution should not grow exponentially 
with the depth $j$.
This imposes that the
prefactor of $({\bar R}-1)^j$ in \solimproved\ vanishes, i.e.
\eqn\impr{ {\bar R}^2-4{\bar R}+1~=~0 \quad \Rightarrow \quad {\bar R}=2+\sqrt{3}=3.732...}
hence 
\eqn\solajimpro{ a_1(j)~=~ (1+\sqrt{3})j + 1~.}
The value of $\bar R$ found in \impr\ is now above the observed value \asympto.

When applied to the arch configurations, this improved mean field ansatz turns 
out to yield the exact thermodynamic solution 
\eqn\solter{  \langle A({\cal M},j) \rangle_{\infty}~=~ 2j+1 ~,}
which lead 
to the correct growth $4^n$ for the number of arch configurations of order $n$.

It is possible that even more refined
improvements of the mean field ansatz give better approximations for semi--meanders. 
One can for instance think
of replacing the constant $K_1$ above \defmea\ by some specific function of 
$j$, but
such a function is no longer fixed by initial conditions.

\subsec{Other approximations using the main recursion}

Starting from a particular arch configuration,
let us generate its ``descendents"
by repeatedly using the processes (I) and (II) of Fig.\cutmeander.
After a given sequence of $n$ ((I) or (II)) steps, let us denote by 
$(m,e)$ respectively the {\it total} number
of arch configurations generated and their average number of exterior arches.
Then let us adopt the following ``mean field type" recursive algorithm 
for an extra action by (I) or (II)
\eqn\recalgom{\eqalign{
(m,e) ~ & \buildrel{(I)} \under\longrightarrow ~ (me,e) \cr
(m,e) ~ & \buildrel{(II)} \under\longrightarrow ~  (m,1) ~.\cr}}
The second line of \recalgom\ is clear, as (II) builds one arch configuration
out of each inital 
arch configuration,
with only one exterior arch. 
The first line incorporates two suppositions. 
First, the number of arch configurations
generated is approximated by a mean field value, 
namely the product of the initial number $m$ of arch configurations by the
{\it average} number $e$ of exterior arches.  
Second, the average number
of exterior arches is supposed to be unchanged.  

Let us now choose as starting point $(m=1,e)$ the upper
arch configuration of a semi--meander. 
Repeated actions of (I) only will generate its descendents which are 
themselves semi--meanders. Implicitly, \recalgom\ supposes a certain number
of properties on the starting point
of the recursion, like for instance the fact that repeated actions of (I)
keep the average number of exterior arches $e$ unchanged, while the number of 
semi--meanders obtained is multiplied by $e$ each time.  
Thus in this scheme, $e$ will be identified with $\bar R$, governing the large
$n$ behaviour of the number of semi--meanders ${\bar M}_n \sim {\bar R}^n$.
The scheme \recalgom\ is
only intended as an approximation and for that purpose the initial
semi--meander is supposed to be very large, with a number $e$ of exterior arches
equal to the average $\bar R$. 
Let us try to evaluate $e$ by identifying
the total number $p_n(e)$ of configurations generated after all possible
sequences of $n$ actions of (I) or (II)
with the total number $c_n \sim 4^n$ (this assumes that $c_n$ gives
the correct large $n$ behaviour
of the total number of descendents of a given configuration, independently of this
configuration).
We start with only one semi--meander, hence $p_0(e)=1$.
After one step, according to eq.\recalgom, we have generated
$p_1(e)=e+1$ arch configurations. 
Let us decompose the number $p_n(e)$ into
\eqn\decompn{ p_n(e)~=~p^{(I)}_n(e)~+~p^{(II)}_n(e)~,}
where we make the distinction between the total number of arch 
configurations obtained by
the process (I) (resp. (II)) in the last step \recalgom\ from the $p_{n-1}(e)$ 
previous ones. The algorithm \recalgom\ leads to the recursion relations
\eqn\recupnsm{\eqalign{
p_n^{(I)}(e)~&=~ e ~p_{n-1}^{(I)}(e)~+~p_{n-1}^{(II)}(e)\cr
p_n^{(II)}(e)~&=~ p_{n-1}^{(I)}(e)~+~p_{n-1}^{(II)}(e).\cr}}
In terms of the vectors $P_n(e)= (p_n^{(I)}(e),p_n^{(II)}(e))^t$, this takes the
matrix form $P_n(e) = M(e) P_{n-1}(e)$, with
\eqn\matrim{ M(e)~=~ \pmatrix{ e & 1 \cr 1 & 1 \cr}~. }
For the total number of arch configurations generated to behave
like $4^n$, we simply have to write that the
largest eigenvalue of this matrix is $4$, namely that the determinant of 
$M(e)- 4 {\bf I}$ vanishes 
(this also fixes the other eigenvalue of $M(e)$ to be $2/3$). 
This gives
\eqn\estimean{ e~=~ {11 \over 3}~=~ 3.666...}
This estimate of $\bar R$ is the closest we can get to the 
numerical estimate \asympto. But the sequence of approximations used to get 
\estimean\ should certainly be refined.

A last remark is in order. One could wonder how much the above estimate depends
on the starting point of the algorithm.  In particular, starting with any semi--meander
with a finite number $k$ of connected components and a large order, we end up with the 
same estimate \estimean\ for the average number of exterior arches. This in turn 
infers an estimate for ${\bar R}_k={\bar R}_1={\bar R}=e=11/3$ for the numbers 
${\bar R}_k$ governing the large order behaviour of ${\bar M}_n^{(k)}\sim ({\bar R}_k)^n$.
Moreover the average number of exterior arches for all the arch configurations
(generated in our scheme by both (I) and (II)) reads, after $n$ steps,
\eqn\avarnst{ \langle {\rm ext} \rangle_n ~=~ { e. p_n^{(I)} + 1.p_n^{(II)}
\over p_n^{(I)}+p_n^{(II)} }~.}
For large $n$, the vector $( p_n^{(I)} ,p_n^{(II)})$ tends, up to
a global normalization, to the eigenvector $(3,1)$ associated to the largest eigenvalue 
$\lambda_{\rm max}=4$ of the matrix $M(11/3)$ \matrim. 
Therefore the average number of exterior arches for all arch configurations is 
estimated as
\eqn\estiavex{  \langle {\rm ext} \rangle~=~ {3e+1 \over 4}=3~,}
which coincides with the exact value, as given by \avext.

\newsec{Matrix model for meanders}

Field theory, as a computational method, involves expansions over graphs
weighted by combinatorial factors.  In this section, we present
a particular field theory which precisely generates planar
graphs with a direct meander interpretation.  The planarity of theses
graphs is an important requirement, which ensures that the arches of the
meander do not intersect each other, when drawn on a planar surface.
The topology of the graphs in field theoretical expansions is best 
taken into account in matrix models, where the size $N$ of the matrices
governs a topological expansion in which the term of order $N^{2-2h}$
corresponds to graphs with genus $h$. 
The planar graphs (with $h=0$) are therefore obtained by taking
the large $N$ limit of matrix models (see for instance \DGZ\ for a review
on random matrices).

\subsec{The matrix model as combinatorial tool}

Random matrix models are useful combinatorial tools for the enumeration of (connected)
graphs \DGZ. Typically, one considers the following integral over Hermitian matrices
of size $N \times N$
\eqn\matint{ Z(g,N)~=~{\int dM~e^{-N {\rm Tr}( {M^2 \over 2} - g{M^4 \over 4}) }
\over \int dM~e^{-N {\rm Tr}( {M^2 \over 2})} } }
where the integration measure is
\eqn\haar{ dM~=~\prod_{i=1}^N dM_{ii} \prod_{i<j} d {\rm Re}M_{ij}
d {\rm Im}M_{ij} ~.}
The rules of Gaussian matrix integration are simple enough to provide us with a
trick for computing the expression \matint\ as a formal series expansion in powers of $g$.
For instance,
\eqn\gaumatrule{\langle M_{ij}M_{kl} \rangle_{\rm Gauss}~=~
{\int dM~e^{-N {\rm Tr}( {M^2 \over 2})} M_{ij} M_{kl} \over
\int dM~e^{-N {\rm Tr}( {M^2 \over 2})} }~=~ {\delta_{il}\delta_{jk} \over N}~.}
Expanding $Z(g,N)$ in powers of $g$, we are left with the computation of 
\eqn\computre{ Z(g,N)~=~\sum_{V=0}^\infty {(Ng)^V \over V!} \langle 
({\rm Tr}{ M^4 \over 4} )^V \rangle_{\rm Gauss}~.}

\fig{A ribbon graph with $V=2$ vertices, $E=4$ edges, and
$L=4$ oriented loops.}{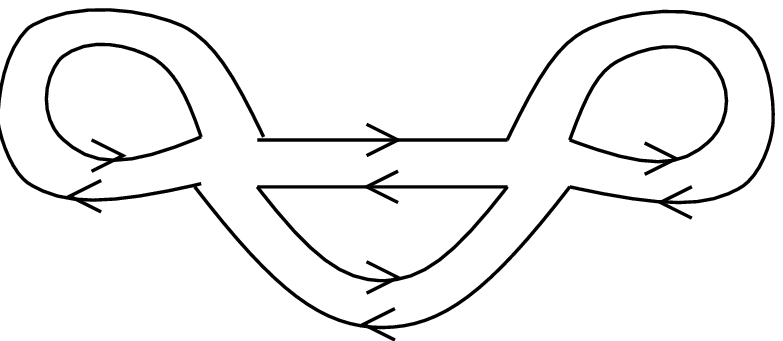}{8.truecm}
\figlabel\exagraph

The result \gaumatrule\ suggests to represent any term in the small 
$g$ expansion as a sum over graphs
constructed as follows. 
\fig{The Feynman rules for the one matrix model: the matrix 
indices are
conserved along the oriented lines, 
which form closed loops. Edges receive a factor $1/N$, vertices $(Ng)$, 
and all the matrix indices have to be summed over.}{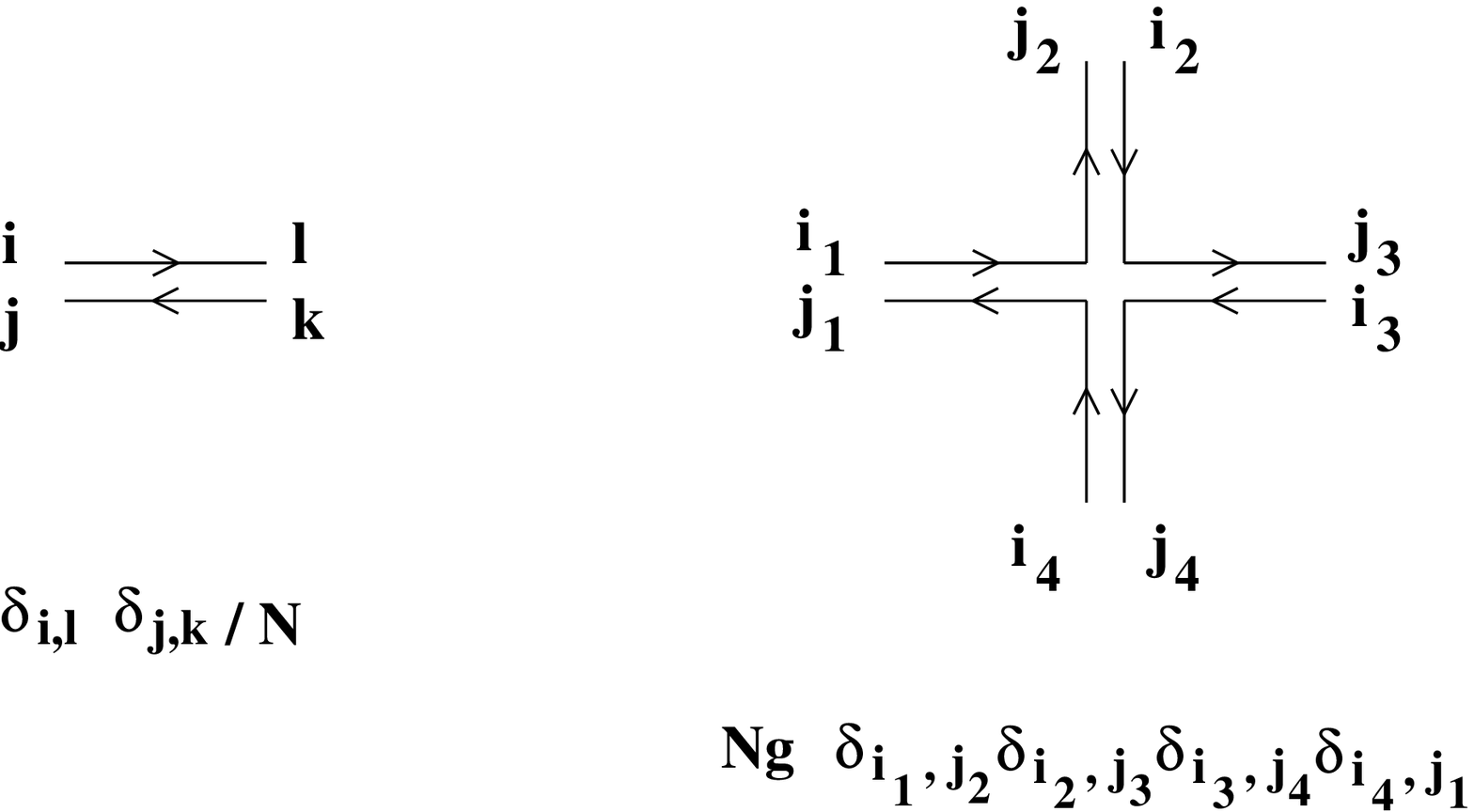}{9.truecm}
\figlabel\onematrule
\noindent{}The graphs considered are ribbon graphs, i.e. with edges made of double--lines, 
oriented with opposite orientations (see Fig.\exagraph\ for an example). 
These two oriented lines represent the circulation of matrix indices 
$i,j=1,2,...,N$. Namely each end of an edge is associated with a matrix element
$M_{ij}$, the index $i$ being carried by the line pointing from
(resp. $j$ by the line pointing to) this end of the edge,
as shown in Fig.\onematrule. An edge is therefore
interpreted as the propagator between the states sitting at its extremities, 
according to \gaumatrule, hence
each edge will be weighed by a factor $1/N$ (hence an overall factor $N^{-E}$,
where $E$ is the total number of edges of the graph). 
To compute the term of order $V$ in the small $g$ expansion \computre, one simply 
has to sum
over all the ribbon graphs connecting the $V$ four--valent vertices corresponding 
to the $V$ terms ${\rm Tr}M^4$ (an obvious $4$--fold cyclic symmetry 
absorbs the factors $1/4$). Each vertex has to be weighed by a factor $Ng$.
These Feynman rules are summarized in Fig.\onematrule.
An overall weight also comes from the summation over all the matrix indices
$i=1,2,...,N$ running on the oriented loops of the graph. This gives a global
factor $N^{L}$ for each graph, where $L$ is the number of loops of the graph.
For instance, the ribbon graph of Fig.\exagraph\ receives a total weight
$(Ng)^2 \times N^{-4} \times N^4=(Ng)^2$.

The second trick is the fact that this sum can restricted to {\it connected} graphs
only, by taking the logarithm of the function \matint\
\eqn\matfin{F(g,N)~=~ {\rm Log}~ Z(g,N)~=~ \sum_{{\rm conn.}\ {\rm graphs}\ \Gamma}
g^V N^{V-E+L} \times {1 \over |{\rm Aut}(\Gamma)|}~,}
where $|$Aut($\Gamma$)$|$ denotes the order of the automorphism group of $\Gamma$,
i.e. the number of permutations of its (supposedly labelled)
vertices leaving the graph invariant. 
This symmetry factor results from the incomplete compensation
of the factor $1/V!$ by the number of equivalent graphs (with different
labelling of the vertices).
Finally, we identify the power of $N$
as the Euler--Poincar\'e characteristic of the graph $\Gamma$
\eqn\eupoin{ \chi(\Gamma)~=~ V-E+L~=~ 2 - 2 h~,}
which can be taken as the definition of the genus $h$ of the graph.
The number of oriented loops 
is indeed equal to that of faces $F$ of the cellular complex induced by the graph,
hence we can use the more standard definition of the 
Euler--Poincar\'e characteristic $\chi=V-E+F=V-E+L$. 

Various techniques for direct computation of the integral \matint\ have made it possible
to enumerate connected graphs with arbitrary genus, and derive many of their properties.
In  this section, we consider a matrix model adapted to the meander enumeration problem.

\subsec{The model}

\fig{A sample black and white graph. The white loop is represented in thin
dashed line. There are $10$ intersections.}{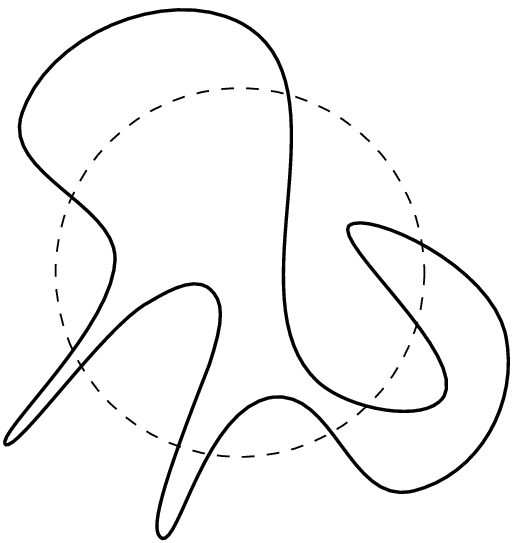}{7.truecm}
\figlabel\bandw

The enumeration of (planar) meanders is very close to that of $4$--valent (genus $0$)
graphs made of two self--avoiding loops (say one black and one white), intersecting
each other at simple nodes \LZ.  
The white loop stands for the river, closed at infinity.
The black loop is the road.  Such a graph will be called a black and white graph.
An example is given in Fig.\bandw.
The fact that the river becomes a loop replaces the order of the bridges by a 
cyclic order, and identifies the
regions above the river and below it. 
Hence the number of meanders $M_n$ is $2\times 2n$ ($2$ for the up/down symmetry and
$2n$ for the cyclic symmetry) times that of inequivalent black and white graphs with $2n$
intersections, weighed by the symmetry factor $1/|{\rm Aut}(\Gamma)|$.
The same connection holds between $M_n^{(k)}$ and the black and white
graphs where the black loop has $k$ connected
components.
\fig{A particular black and white graph with $6$ intersections, and its two 
associated meanders. The automorphism group of the black and white graph is
$\IZ_6$.}{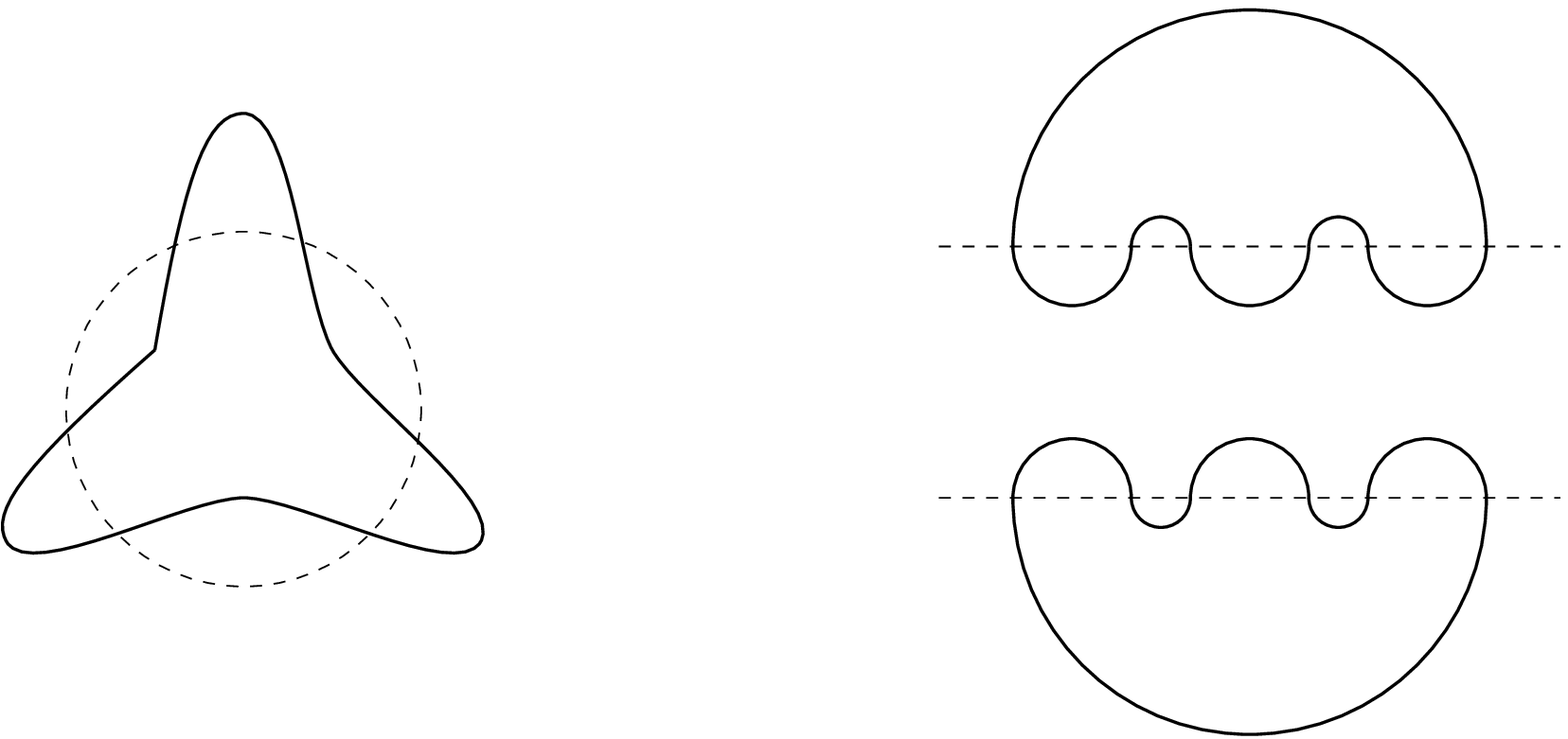}{8.truecm}
\figlabel\exbw
\noindent{}For illustration, we display a particular black and white 
graph $\Gamma$ in Fig.\exbw, together with
its two corresponding meanders of order $3$. 
The automorphism group of this black and white graph is $\IZ_6$,
with order ${\rm Aut}(\Gamma)=|\IZ_6|=6$. The two meanders come with an overall factor
$1/(2\times 6)$, hence contribute a total $2 \times 1/12=1/6$, 
which is precisely the desired symmetry factor.

In analogy with the ordinary matrix model \matint, a simple way of 
generating black and white graphs
is the use of the multi--matrix integral (with $m+n$ hermitian
matrices of size $N$ denoted by $B$ and $W$)
\eqn\multimat{ Z(m,n,c,N)~=~ {1 \over \kappa_N}\int \prod_{\alpha=1}^m dB^{(\alpha)}
\prod_{\beta=1}^n dW^{(\beta)} e^{-N{\rm Tr}~P(B^{(\alpha)},W^{(\beta)})} ~,}
where the matrix potential reads
\eqn\matpot{P(B^{(\alpha)},W^{(\beta)} )~=~
\sum_\alpha {(B^{(\alpha)})^2
\over 2} +\sum_\beta {(W^{(\beta)})^2 \over 2} -{c \over 2}\sum_{\alpha,\beta}
B^{(\alpha)} W^{(\beta)} B^{(\alpha)} W^{(\beta)} ~,}
and the normalization constant $\kappa_N$ is such that $Z(m,n,c=0,N)=1$.
In the following, the $\alpha$ and $\beta$ indices will be referred to as
color indices.

\fig{The Feynman rules for the black and white matrix model.
Solid (resp. dashed) double--lines correspond to black (resp. white)
matrix elements, whose indices run along
the two oriented lines. 
An extra color index $\alpha$ (resp. $\beta$) indicates the number
of the matrix in its class, $B^{(\alpha)}$, $\alpha=1,2,...,m$ (resp.
$W^{(\beta)}$, $\beta=1,2,...,n$). The only allowed vertices
are $4$--valent, and have alternating black  and white edges: they describe
simple intersections of the black and white loops.}{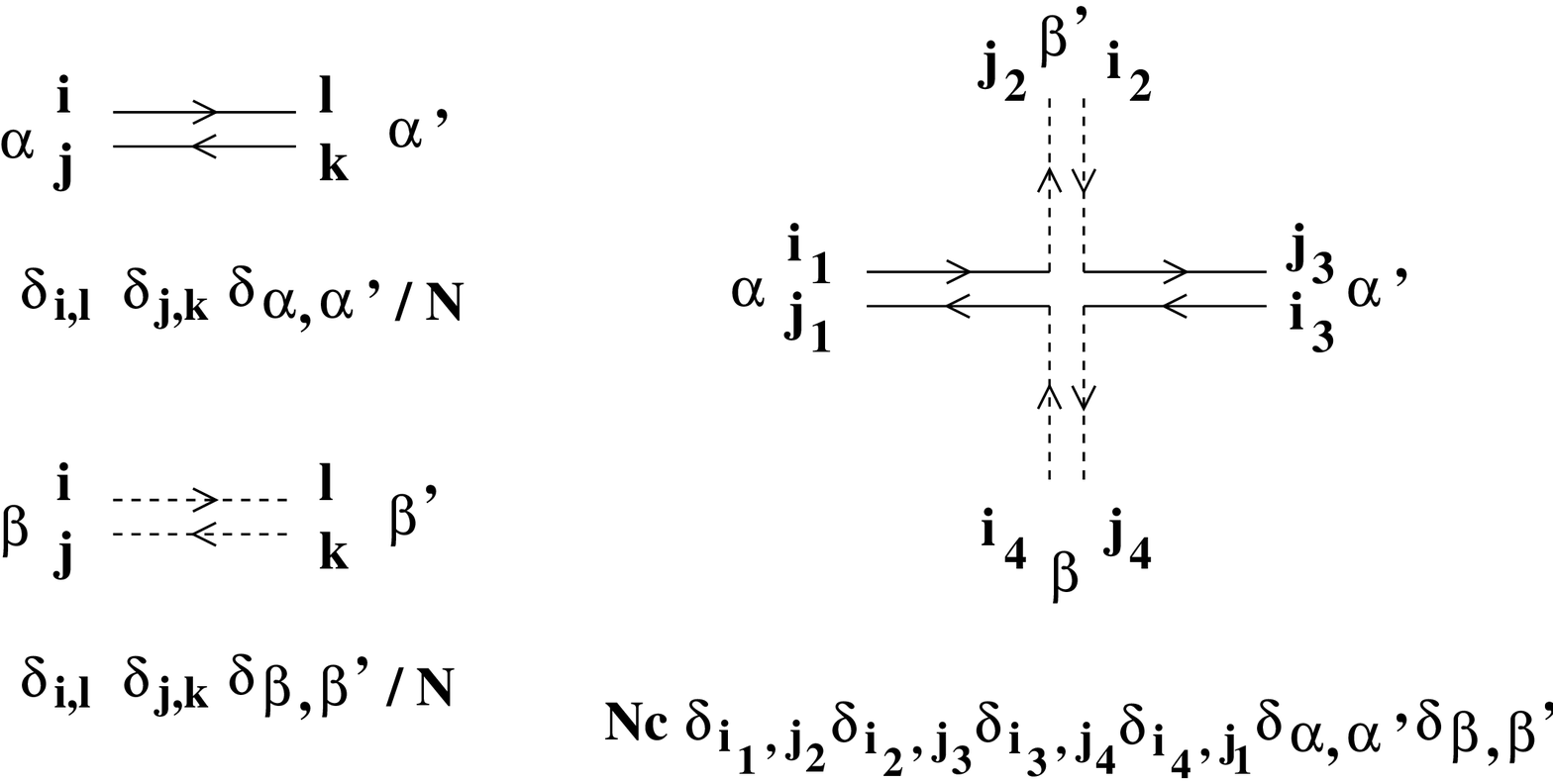}{9.truecm}
\figlabel\feymean

Like in the case \matint, the logarithm
of the function \multimat\ can be evaluated pertubatively as a
series in powers of $c$. A term of order $V$ in this expansion is readily evaluated
as a Gaussian multi--matrix integral. It can be obtained as a sum over
$4$--valent connected graphs, whose $V$ vertices have to be connected by
means of the two types
of edges
\eqn\edgesty{ \eqalign{
&{\rm black}\ {\rm edges} \ \ \  \langle [B^{(\alpha)}]_{ij} 
[B^{(\alpha')}]_{kl}\rangle ~=~
{\delta_{il} \delta_{jk} \over N} \delta_{\alpha \alpha'}\cr
&{\rm white}\ {\rm edges} \ \ \  \langle [W^{(\beta)}]_{ij} 
[W^{(\beta')}]_{kl}\rangle ~=~
{\delta_{il} \delta_{jk} \over N} \delta_{\beta \beta'}~,\cr }}
which have to alternate around each vertex.  The corresponding Feynman
rules are summarized in Fig.\feymean.
This is an exact realization of 
the desired connected black and white graphs, 
except that any number of loops\foot{The reader must distinguish between these
loops, made of double--lines of a definite color, from the oriented loops along which
the matrix indices run.}  of each color is 
allowed.  
In fact, each graph receives a weight
\eqn\weighgra{ N^{2-2h} ~ c^{V} ~ m^b ~n^w~,}
where $b$ (resp. $w$) denote the total numbers of black (resp. white) loops.

A simple trick to reduce the number of say white loops $w$ to 
one is to send the number 
$n$ of white matrices $W$ to $0$, and to retain only the 
contributions of order $1$ in $n$.
Hence
\eqn\matmean{ f(m,c,N)~=~\lim_{n \to 0} {1 \over n} {\rm Log}~Z(m,n,c,N)~=~
\sum_{b.\& w. \  {\rm conn.}\ {\rm graphs} \ \Gamma\atop 
{\rm with} \ {\rm one}\ w \ {\rm loop}}  N^{2-2h} ~ c^{V} ~ m^b ~ 
{1 \over |{\rm Aut}(\Gamma)|}~.}

If we restrict this sum to the leading order $N^2$, namely
the genus $0$ contribution ($h=0$), we finally get a relation
to the meander numbers in the form
\eqn\meanmat{\eqalign{ 
f_0(m,c)~&=~\lim_{N\to \infty} {1 \over N^2} f(m,c,N) \cr
&=~ \sum_{p=1}^\infty {c^{2p} \over 4p} \sum_{k=1}^p M_p^{(k)} m^k \cr}}
where the abovementioned relation between the numbers of black and white graphs and
multi--component meanders has been used to rewrite the expansion \matmean.

\subsec{Meander numbers as Gaussian averages of words}

The particular form of the matrix potential \matpot\ allows one
to perform the exact integration
over say all the $W$ matrices (the dependence of $P$ on $W$ is Gaussian),
with the result
\eqn\partint{ Z(m,n,c,N)~=~ {1 \over \theta_N} \int \prod_{\alpha=1}^m dB^{(\alpha)} 
\det \big[ {\bf I}\otimes {\bf I} - 
c \sum_\alpha B^{(\alpha)\ t} \otimes B^{(\alpha)} \big]^{-n/2}
e^{-N {\rm Tr} \sum_\alpha {(B^{(\alpha)})^2 \over 2}} ~,}
where ${\bf I}$ stands for the $N \times N$ identity matrix, $\otimes$ denotes the
usual tensor product of matrices, and the superscript $t$ stands for the
usual matrix transposition. 
The prefactor $\theta_N$ is fixed by the
condition $Z(m,n,c=0,N)=1$. With this form, it is easy to take the logarithm and to
let $n$ tend to $0$, with the result
\eqn\matsuite{\eqalign{ 
f(m,c,N)~&=~-{1 \over 2\theta_N} \int \prod_{\alpha=1}^m dB^{(\alpha)}
{\rm Tr}({\rm Log} \big[ {\bf I}\otimes {\bf I} - c \sum_\alpha B^{(\alpha)\ t} 
\otimes B^{(\alpha)} \big] ) 
e^{-N {\rm Tr} \sum_\alpha {(B^{(\alpha)})^2 \over 2} }
\cr
&=~ \sum_{p=1}^\infty  {c^p \over 2p} \langle
{\rm Tr}(\sum_{\alpha=1}^m B^{(\alpha)\ t} \otimes B^{(\alpha)})^p 
\rangle_{\rm Gauss} \cr
&=~ \sum_{p=1}^\infty  {c^p \over 2p}
\sum_{1 \leq \alpha_1,...,\alpha_p \leq m} \langle \big\vert {\rm Tr} 
(B^{(\alpha_1)} ... B^{(\alpha_p)}) \big\vert^2 \rangle_{\rm Gauss}~,\cr}}
where we still use the notation $\langle ...\rangle_{\rm Gauss}$ 
for the multi--Gaussian average
over the matrices $B^{(\alpha)}$, $\alpha=1,2,...,m$. 
The modulus square simply comes from
the hermiticity of the matrices $B^{(\alpha)}$, namely
\eqn\proher{ {\rm Tr}(\prod B^{(\alpha_i)\ t})~=~{\rm Tr}(\prod B^{(\alpha_i)\ *})~=~
{\rm Tr}(\prod B^{(\alpha_i)})^*~.}
Taking the large $N$ limit \meanmat, it is a known fact \DGZ\
that correlations should
factorize, namely
\eqn\factocor{ \langle  \big\vert {\rm Tr}(\prod_{i=1}^p B^{(\alpha_i)}) \big\vert^2 
\rangle_{\rm Gauss} 
\buildrel{N \to \infty}\under\longrightarrow \big\vert \langle{\rm Tr}(\prod_{i=1}^p
B^{(\alpha_i)}) 
\rangle_{\rm Gauss} \big\vert^2 ~.}
By parity, we see that only even $p$'s give non--vanishing contributions, and comparing
with \meanmat\ we find a closed expression for the meander numbers of order $n$
with $k$ connected components
\eqn\luttefin{ \sum_{k=1}^n M_n^{(k)} m^k ~=~ 
\sum_{1 \leq \alpha_1,...,\alpha_{2n} \leq m}
\big\vert  \lim_{N \to \infty} \langle
{1 \over N}{\rm Tr}(\prod_{i=1}^{2n} B^{(\alpha_i)}) \rangle_{\rm Gauss} \big\vert^2~.}
This expression is only valid for integer values of $m$, but as it is
a polynomial of degree $n$ in $m$ (with vanishing constant coefficient), 
the $n$ first values $m=1,2,...,n$ of $m$ determine it completely.  So we only have to
evaluate the rhs of \luttefin\ for these values of $m$ to  determine all the coefficients
$M_n^{(k)}$.

\fig{The connected toric meander of order $1$: 
it has only $1$ bridge.}{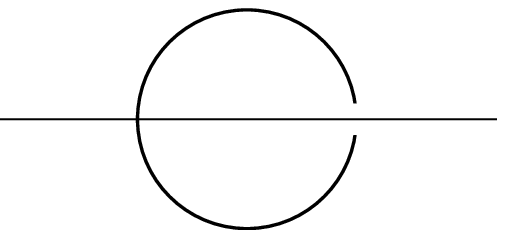}{5.truecm}
\figlabel\tormean
The relation \luttefin\ suggests to introduce higher genus meander numbers,
denoted by $M_p^{(k)}(h)$, with $M_{2n}^{(k)}(0)=M_n^{(k)}$ (note that the indexation
is now by the number of intersections, or bridges), through the generating function
\eqn\higenme{
\sum_{h=0}^ \infty \sum_{k=1}^{\infty} 
M_p^{(k)}(h) m^k N^{2-2h} ~=~ \sum_{1 \leq \alpha_1,...,\alpha_{p} \leq m}
\langle \big\vert {\rm Tr}(\prod_{i=1}^{p} B^{(\alpha_i)})\big\vert^2 
\rangle_{\rm Gauss} ~,}
which incorporates the contribution of all genera in the Gaussian averages.
Note that the genus $h$ is that of the corresponding black and white graph
and not that of the river or the road alone. 
In particular, the river (resp. the road) may be contractible
or not in meanders of genus $h>0$.
As an example the $M_1^{(1)}=1$ toric meander is represented in Fig.\tormean.

\fig{A typical graph in the computation of the rhs of \higenme.
The two $p$--valent vertices corresponding to the two traces of
words are represented as racks of $p$ double legs ($p=10$ here). 
The connected components of the resulting
meander (of genus $h=0$ on the example displayed here) 
correspond to loops of matrices $B^{(\alpha)}$. 
This is indicated by a different coloring of the various connected components.
Summing over all values
of $\alpha_i$ yields a factor $m$ per connected component, 
hence $m^3$ here.}{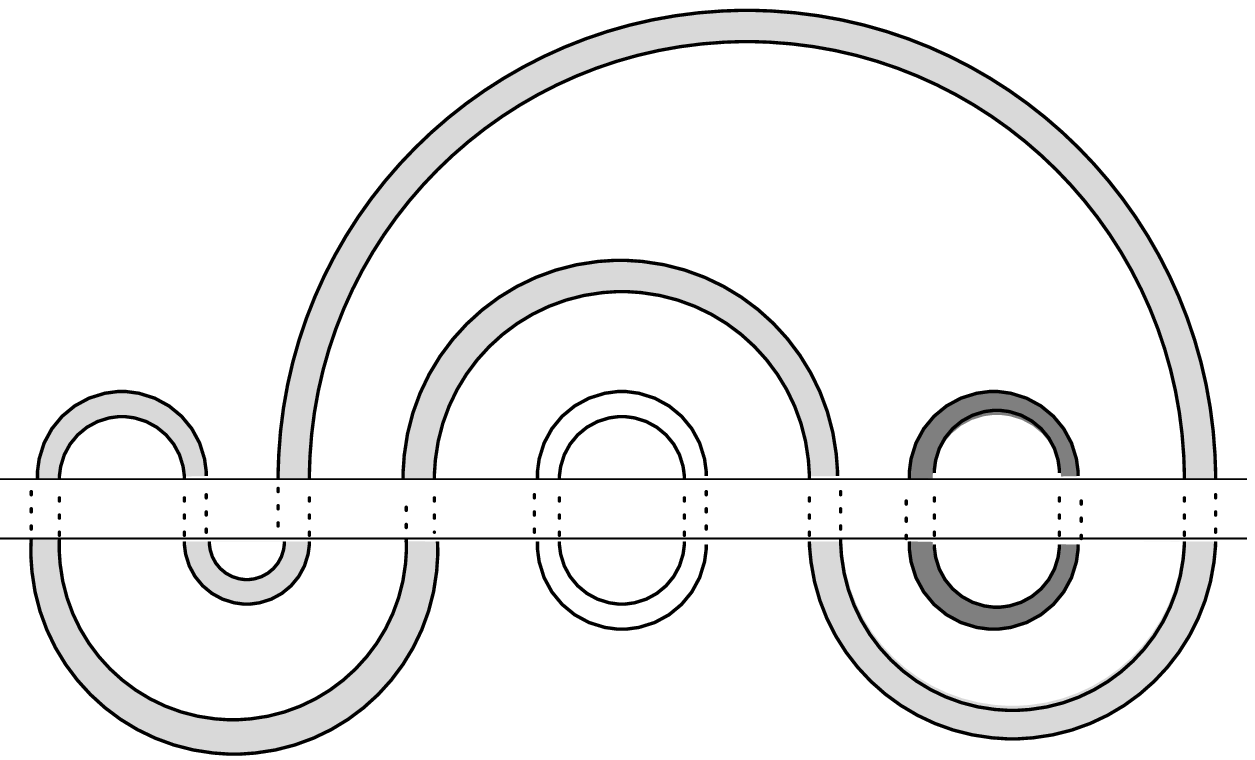}{8.truecm}
\figlabel\reinter

The relation \higenme\ can also be proved directly as follows. 
Its rhs is a sum over correlation functions of the traces of certain
words (products of matrices) with themselves. 
More precisely, using the hermiticity of the matrices $B^{(\alpha)}$, the complex
conjugate of the trace
${\rm Tr} (\prod_{1 \leq i\leq 2n} B^{(\alpha_i)})$ can be rewritten as 
\eqn\retrac{{\rm Tr} (\prod_{1 \leq i\leq 2n} B^{(\alpha_i)})^*~=~
{\rm Tr} (\prod_{1 \leq i\leq 2n} B^{(\alpha_{2n+1-i})})~,}
i.e. in the form of an analogous trace, with the order of the $B$'s reversed.
According to the Feynman rules of the previous
section in the case of only black matrices, 
such a correlation can be computed graphically as follows. The two traces
correspond to two $p$--valent vertices, and the Gaussian average is computed
by summing over all the graphs obtained by connecting pairs of legs (themselves
made of pairs of oriented double--lines) by means of edges.  
Re--drawing these vertices as small racks of $p$ legs as in Fig.\reinter, we get
a sum over all multi--component, multi--genera meanders
(compare Fig.\reinter\ with Fig.\mearch).  
More precisely,
the edges can only
connect two legs with the {\it same} matrix label $\alpha$, which can be interpreted as
a color: indeed, we have to sum over all colorings of the graph by means of $m$ colors.
But this coloring is constrained by the fact that the colors of the legs of the 
two racks have to be identified two by two (the color of both first legs is
$\alpha_1$,...,of both $p$-th legs is $\alpha_{p}$). 
This means that each connected component of the resulting meander 
is painted with a 
color $\alpha \in\{1,2,...,m\}$.
A graph of genus $h$
comes with the usual weight $N^{2-2h}$.  
Summing over all
the indices $\alpha_1,...,\alpha_{p}=1,2,...,m$, we get an extra factor of
$m$ for each connected component of the corresponding meander, which proves the
relation \higenme.

In the genus $0$ case, we must only consider planar graphs, which correspond 
to genus $0$ meanders by the above interpretation. 
Due to the planarity of the graph, the two racks of $p=2n$ legs each
are connected to themselves
through $n$ edges each, and are no longer connected to each other: they form
two disjoint arch configurations of order $n$. 
This explains the factorization mentioned in eq.\factocor, 
and shows that the
genus $0$ meanders are obtained by the superimposition of two arch configurations.
The beauty of eq.\luttefin\ is precisely to keep track of the number
of connected components $k$ in  this picture, by the $m$--coloring of the connected
components.

This last interpretation leads to a straightforward generalization of \higenme\
to semi--meanders and many meander--related numbers.

\subsec{Matrix expressions for semi--meanders and more}

In view of the above interpretation, we immediately get the generalization of 
eq.\luttefin\ to semi--meanders as
\eqn\semimat{\eqalign{ 
&\sum_{k=1}^n {\bar M}_n^{(k)} m^k ~=~ 
\sum_{1 \leq \alpha_1,...,\alpha_{n} \leq m} \cr
\lim_{N \to \infty} 
{1 \over N}
&\langle {\rm Tr}(B^{(\alpha_1)}B^{(\alpha_2)}...B^{(\alpha_n)}B^{(\alpha_n)}
B^{(\alpha_{n-1})}...
B^{(\alpha_1)}) \rangle_{\rm Gauss}~.\cr}}
To get this expression, we have used the $m$--coloring of the matrices to produce 
the correct rainbow-type connections between the loops of matrices. 

More generally, this provides a number of matrix integral identities for 
the following generalized semi--meanders.
Let us label the arch configurations of order $n$ by a permutation $\mu \in S_{2n}$,
the symmetric group over $2n$ objects, in such a way that if we label the bridges
of the arch configuration $1,2,...,2n$, the permutation $\mu$ indicates the 
pairs of bridges linked by arches, namely, for any $i=1,2,...,2n$, 
$\mu(i)$ is the bridge linked to $i$ by an arch.  By definition, $\mu$ is made of
$n$ cycles of length $2$, it is therefore an element of the class $[2^n]$ of $S_{2n}$.
Note that an element of this class generally does not lead to an arch configuration,
because the most general pairing of bridges has intersecting arches. 
A permutation $\mu \in [2^n]$ will be called {\bf admissible} if it leads
to an arch configuration.
Let ${\cal A}_\mu$ be the arch configuration associated to some admissible
$\mu \in [2^n]$.
We can define some generalized semi--meander number 
${\bar M}_n^{(k)}({\cal A}_\mu)$ associated to ${\cal A}_\mu$
as the number of meanders of order $n$ with $k$ connected components whose 
lower arch configuration is ${\cal A}_\mu$.
With this definition, 
\eqn\idenme{ {\bar M}_n^{(k)}({\cal R}_n)~=~{\bar M}_n^{(k)}~,}
where ${\cal R}_n$ is the rainbow configuration of order $n$, associated to the 
permutation $\mu(i)=2n+1-i$, for $i=1,2,...,2n$.
In other words, ${\bar M}_n^{(k)}({\cal A}_\mu)$ is the number of 
closures by some arch configurations
of order $n$ of the lower arch configuration ${\cal A}_\mu$ which have $k$
connected components.  Eq.\semimat\ extends immediately to
\eqn\semigen{\eqalign{ 
\sum_{k=1}^n {\bar M}_n^{(k)}({\cal A}_\mu)~ m^k ~&=~ 
\sum_{1 \leq \alpha_1,...,\alpha_{2n} \leq m  \atop
\alpha_i=\alpha_{\mu(i)},\ i=1,...,2n} \cr
\lim_{N \to \infty} 
{1 \over N}
&\langle {\rm Tr}(B^{(\alpha_1)}
B^{(\alpha_2)}...B^{(\alpha_{2n})}) \rangle_{\rm Gauss}~,\cr}}
where the structure of the lower arch configuration ${\cal A}_\mu$ is 
encoded in the conditions $\alpha_i=\alpha_{\mu(i)}$, $i=1,...,2n$, which
identifies the colors of the arches according to ${\cal A}_\mu$.
Some of these numbers will be computed in sect.6.3 below.

Higher genus generalizations are straightforward, by simply removing the large
$N$ limit in the above expressions, namely
\eqn\semineg{\eqalign{ 
\sum_{k\geq 1,h\geq 0} {\bar M}_{2n}^{(k)}({\cal A}_\mu,h)~ m^k ~ N^{1-2h}~&=~ 
\sum_{1 \leq \alpha_1,...,\alpha_{2n} \leq m  \atop
\alpha_i=\alpha_{\mu(i)},\ i=1,...,2n} \cr
&\langle {\rm Tr}(B^{(\alpha_1)}
B^{(\alpha_2)}...B^{(\alpha_{2n})}) \rangle_{\rm Gauss}~.\cr}}
Note that the number $2n$ of bridges is even here, as we are considering arbitrary genus
closures of a given arch configuration of order $n$. In genus $h=0$, we recover the
numbers
${\bar M}_{2n}^{(k)}({\cal A}_\mu,0)={\bar M}_{n}^{(k)}({\cal A}_\mu)$ defined
by \semigen.
In the particular case ${\cal A}_\mu={\cal R}_n$, this defines higher genus
semi--meander numbers ${\bar M}_{2n}^{(k)}(h)={\bar M}_{2n}^{(k)}({\cal R}_n,h)$,
with the correspondence ${\bar M}_{2n}^{(k)}(0)={\bar M}_n^{(k)}$.
The resulting higher genus semi--meanders are obtained generically by allowing the $4$ 
ends of two given arches to alternate along the river. As the arches cannot intersect
each other, this requires increasing the genus of the graph. 
On the other hand, the lower rainbow configuration is contractible, hence the
genus is also that of the contracted graph obtained by the folding process of 
Fig.\archsem.

It is instructive to calculate the sum over all genera of these numbers, while
keeping track of the numbers of connected components. We simply take $N=1$, in
which case the Gaussian average becomes an ordinary Gaussian average over
real scalars $(b^{(1)},...,b^{(m)})\in \IR^m$
\eqn\gaurn{
\eqalign{
\sum_{k\geq 1,h\geq 0} {\bar M}_{2n}^{(k)}({\cal A}_\mu,h)~ m^k ~&=~
\langle \big(\sum_{i=1}^m (b^{(i)})^2 \big)^{n} \rangle_{\rm Gauss}\cr
&=~ \lambda_m \int_0^\infty r^{m-1} ~r^{2n}~e^{-{r^2 \over 2}} dr\cr
&=~ m(m+2)(m+4)...(m+2n-2)~,\cr}}
where the normalization constant has been fixed by the $n=1$ case (the result is $m$).
For $m=1$, the above
simply counts the total number of pairings between $2n$ legs, namely
$(2n-1)!!=1.3.5...(2n-1)$. Note that the result \gaurn\ is independent of the lower 
arch configuration, it holds in particular for semi--meanders. The result \gaurn\
is a polynomial of degree $n$, with leading coefficient $1$ corresponding to the only
(genus $0$) meander with $n$ connected components, obtained by reflecting the lower
arch configuration ${\cal A}_\mu$ wrt the river.
For meanders, we simply get
\eqn\gaurm{
\eqalign{
\sum_{k\geq 1,h\geq 0} {M}_p^{(k)}(h)~ m^k ~&=~
\langle \big(\sum_{i=1}^m (b^{(i)})^2 \big)^{p} \rangle_{\rm Gauss}\cr
&=~ m(m+2)(m+4)...(m+2p-2)~.\cr}}
Note that in this case the polynomial is of degree $p$, with leading coefficient $1$,
corresponding to the (genus $1$) meander made of a collection of $p$ loops 
intersecting the river only once.

\subsec{Computing averages of traces of words in matrix models}

As a warming up, let us first compute the rhs of eq.\semimat\ in the case
of one matrix $m=1$, namely
\eqn\onecat{ \gamma_n~=~ \lim_{N \to \infty} {1 \over N} \langle 
{\rm Tr}(B^{n}) \rangle_{\rm Gauss}~.}
By parity,  we see that $\gamma_{2s+1}=0$ for all integer $s$.
A simple method usually applied for computing Gaussian averages uses
the so--called {\bf loop equations} of the matrix model.
In the case of one matrix, these are obtained as follows. 
We write that the matrix integral of a total derivative vanishes, namely
\eqn\lopone{\eqalign{
0~&=~ \int dB~{\partial \over \partial B_{ji}} \bigg[ (B^{s+1})_{kl}
e^{-N{\rm Tr}{B^2 \over 2}} \bigg] \cr
\Rightarrow \quad 0&=~ \langle -N (B^{s+1})_{kl}B_{ij}+
\sum_{r=0}^{s} (B^r)_{kj} (B^{s-r})_{il} \rangle_{\rm Gauss}~.\cr}}
Taking $i=l$ and $j=k$, and summing over $i,j=1,...,N$, we finally get
\eqn\looponemat{ \langle ~{\rm Tr}(B^{s+2}) ~\rangle_{\rm Gauss}~=~
{1 \over N}\sum_{r=0}^s \langle ~{\rm Tr}(B^r) ~{\rm Tr}(B^{s-r}) 
~\rangle_{\rm Gauss}~.}
In the large $N$ limit, due to the abovementioned factorization property, 
only the even powers of $B$ contribute by parity, 
and setting $s+2=2n$, this becomes
\eqn\larn{ \gamma_{2n}~=~\sum_{r=0}^{n-1} \gamma_{2r} \gamma_{2n-2r-2}~,}
valid for $n \geq 1$, and $\gamma_0=1$. This is exactly the defining recursion 
\recucat\ for
the Catalan numbers, hence $\gamma_{2n}=c_n$, whereas $\gamma_{2n+1}=0$. 
In this case, the equation \semimat\ reduces therefore to the sum rule \sumsem\
for semi--meanders.
Similarly, when $m=1$,
the equation \luttefin\ reduces to the sum rule \sumrulm\ for meanders.
An important remark is in order. It sould not be surprising that the general
recursion principle for arches of Fig.\recursion\ resembles the large $N$ limit
of loop equations in matrix models. 
The role of the leftmost arch is played in the latter
case by the differentiation wrt the matrix element $B_{ji}$: 
it can act at all the matrix positions in the word, which cut it into two even words,
and can be graphically interpreted as just 
the right bridge position of the leftmost exterior arch in the pairings of matrices
necessary to compute the Gaussian average of the trace of the word.

More generally, the loop equations for the Gaussian $m$--matrix model enable us to derive a
general recursion relation for traces of words. The most general average
of trace of word in $m$ matrices in the large $N$ limit is denoted by
\eqn\defwo{\eqalign{\gamma_{p_1,p_2,...,p_{mk}}^{(m)} &~=\cr
\lim_{N \to \infty} {1 \over N}
\langle {\rm Tr}\big( &(B^{(1)})^{p_1}  (B^{(2)})^{p_2} ... (B^{(m)})^{p_m}
(B^{(1)})^{p_{m+1}}...(B^{(m)})^{p_{mk}} \big) \rangle_{\rm Gauss}~.\cr}}
In the above, some powers $p_j$ may be zero, but no $m$ consecutive of them vanish 
(otherwise the word could be reduced by erasing the $m$ corresponding pieces).
Of course $2p=\sum_i p_i$ has to be an even number for \defwo\ to be non--zero,
by the usual parity argument.
For $m=1$, we recover $\gamma_p^{(1)}=\gamma_p$.
If $\omega=\exp(2i \pi/m)$ denotes the primitive $m$--th root of unity, 
then we have the following recursion relation between large $N$ averages 
of traces of words, for $m \geq 2$
\eqn\limavtr{ \gamma_{p_1,p_2,...,p_{mk}}^{(m)}~=~-\sum_{j=1}^{mk-1} ~\omega^j
\gamma_{p_1,...,p_j}^{(m)}~\gamma_{p_{j+1},...,p_{mk}}^{(m)}~.}
When $j$ is not a multiple of $m$, it is understood in the above that the multiplets
$(p_1,...,p_j)$ and $(p_{j+1},...,p_{km})$ have to be completed by zeros so a to form
sequences of $m$--uplets. For instance, we write $\gamma_3^{(3)}=\gamma_{3,0,0}^{(3)}=
\gamma_{0,0,3}^{(3)}$. Note also that if only $q<m$ matrices are actually used
to write a word, the corresponding $\gamma^{(m)}$ can be reduced to a $\gamma^{(q)}$
by erasing the spurious zeros (for instance, $\gamma_{3,0,0}^{(3)}=\gamma_3^{(1)}$).
Together with the initial condition 
$\gamma_{0,...,0,2n+1,0,...,0}^{(m)}=\gamma_{2n+1}=0$ and
$\gamma_{0,...,0,2n,0,...,0}^{(m)}=\gamma_{2n}=c_n$, this gives a compact recursive 
algorithm to compute all the large $N$ averages of traces of words in any multi--Gaussian
matrix model.

A direct proof of eq.\limavtr\ goes as follows. Throughout this argument, we refer to the 
labels $\alpha=1,2,...,m$ of the matrices as colors. 
The quantity $\gamma_{p_1,..,p_{mk}}^{(m)}$
can be expressed as the sum over all possible planar pairings 
of matrices of the same color 
in the corresponding word, i.e. the sum over all arch configurations of order $p$
(encoded in {\it admissible} permutations $\mu$ of $1,2,...,2p$),
preserving the color of the matrices 
(the matrices sitting
at positions $i$ and $\mu(i)$ have the same color $\alpha$). 
Such a color--preserving arch configuration appears exactly once
in the lhs of \limavtr. 
We will show the relation \limavtr\ by proving that each such term also
comes with a coefficient $1$ in the rhs of \limavtr.
Let us evaluate the rhs of \limavtr\ in this language.
\fig{A sample color--preserving pairing of matrices for $m=3$ matrices.
Each block of matrices is denoted by a single letter $A$, $B$, $C$, according
to its color $1$, $2$, $3$. The precise pairing of matrices within blocks is
not indicated for simplicity. The available separator positions are indicated
by arrows. One checks that (i) the positions
are consecutive modulo $3$, and (ii) the number of available positions is
$5=2 \times 3 -1$.}{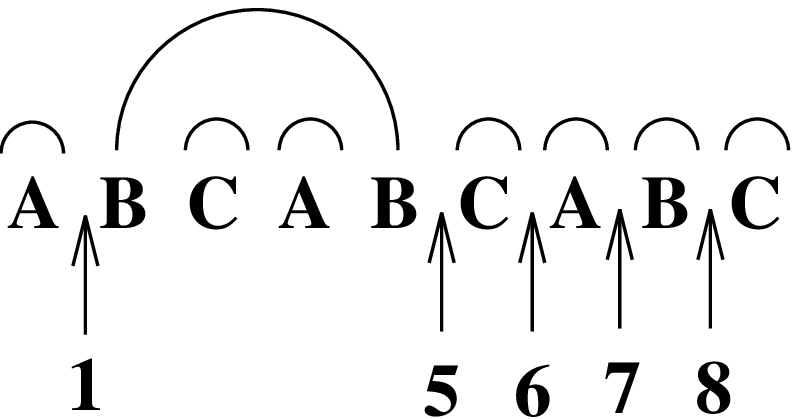}{8.truecm}
\figlabel\propcol
\noindent{}The index $j$ may be viewed as the position of a ``separator", which cuts the color--preserving
arch configurations into two disconnected pieces. The separator positions
are labeled by the index $j=1,2,...,mk-1$.
%Let us suppose that $p_1\geq 1$, which is not restrictive, up to a redefinition of colors.
%In general the block $(B^{(1)})^{p_1}$ of matrices of color $1$ is linked to some other block
%$(B^{(1)})^{p_{ml+1}}$ by means of an arch between one matrix of the first block and
%one of the second.  Such a term only appears in the rhs of \limavtr\ for values of the
%separator index $j \geq m l+1$ (otherwise the two blocks could not be connected to each other).
For a given color--preserving
arch configuration in the lhs of \limavtr,
the only terms of the rhs of \limavtr\ contributing to it are those where the 
separator index $j$ takes its values at positions inbetween the exterior arches linking various 
blocks of the same color.  These positions will be referred to as available positions.  

Available positions satisfy the two following properties, 
illustrated in Fig.\propcol.

(i) any two successive available positions
are labelled by {\it successive} integers modulo $m$: if an available separator position
sits between two blocks of colors $\alpha$ and $\alpha+1$, hence at a position 
of the form $j=s m+\alpha$, then the block of color 
$\alpha+1$ is linked to other blocks of the same color, and the next available 
separator position sits between a block of color $\alpha+1$ and a block of color $\alpha+2$,
hence at a position of the form $j=t m+ \alpha+1$. These two positions are consecutive modulo $m$.

(ii) the first available separator position is $j=ml+1$: it sits
to the right of the first set of related blocks 
of color $1$. The last available separator position is of the form $j=q m+(m-1)$, as it
sits to the left of the rightmost set of related blocks of color $m$.  Hence, thanks to 
property (i), the
total number of available separator positions is of the form $rm-1$.

Consequently, the total contribution of a given color--preserving
admissible pairing in the rhs of \limavtr\ is of the form
\eqn\contrili{ (r-1)~ \sum_{j=1}^m (- \omega^s) + \sum_{j=1}^{m-1} (-\omega^s)~=~1~.}
So we have proved that the total contribution of each admissible pairing in the
rhs of \limavtr\ is $1$. 
This completes the proof of \limavtr.

Let us show explicitly how to use the recursion \limavtr\ to compute the thermodynamic
average of the trace of a particular word in $m=3$ matrices.
We wish to compute 
\eqn\wordthree{ \gamma_{2,1,2,0,1,0}^{(3)}~=~\lim_{N \to \infty} {1 \over N}
\langle {\rm Tr}\big( A^2 B C^2 B \big) \rangle_{\rm Gauss}~,}
where we denote by $A$, $B$, $C$ the matrices with respective colors $1$, $2$, $3$.
Applying \limavtr\ we get
\eqn\applili{\eqalign{
\gamma_{2,1,2,0,1,0}^{(3)}~&=~ -j \gamma_2^{(1)} \gamma_{2,2}^{(2)}
-j^2 \gamma_{2,1}^{(2)} \gamma_{2,1}^{(2)} \cr
&~- \gamma_{2,1,2}^{(3)} \gamma_1^{(1)}
-j \gamma_{2,1,2}^{(3)}\gamma_1^{(1)} -j^2 \gamma_{2,1,2,0,1,0}^{(3)}~, \cr}}
where $j=\exp(2i \pi/3)$, and the various
numbers $m$ of matrices have been reduced to their minimal value.
Regrouping the $\gamma_{2,1,2,0,1,0}^{(3)}$'s, and using the fact that 
$\gamma_{2s+1}^{(1)}=\gamma_{2s+1}=0$
for integer $s$, we get
\eqn\morappli{ (1+j^2)\gamma_{2,1,2,0,1,0}^{(3)}~=~ -j \gamma_2^{(1)} \gamma_{2,2}^{(2)} ~.}
Using the recursion \limavtr\ for $m=2$, we get
\eqn\aplimor{ \gamma_{2,2}^{(2)}~=~ \gamma_2^{(1)} \gamma_2^{(1)} ~=~ (c_1)^2~.}
Finally, we get 
\eqn\aplifin{ \gamma_{2,1,2,0,1,0}^{(3)}~=~(c_1)^3~=~1~.}
This corresponds to the only way of pairing matrices of the same color in the sequence
$$ \epsfxsize=5.truecm \epsfbox{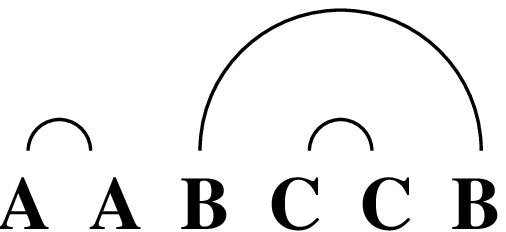} $$

\subsec{Equivalent coloring problems}

\fig{The $5$ semi--meanders of order $3$.}{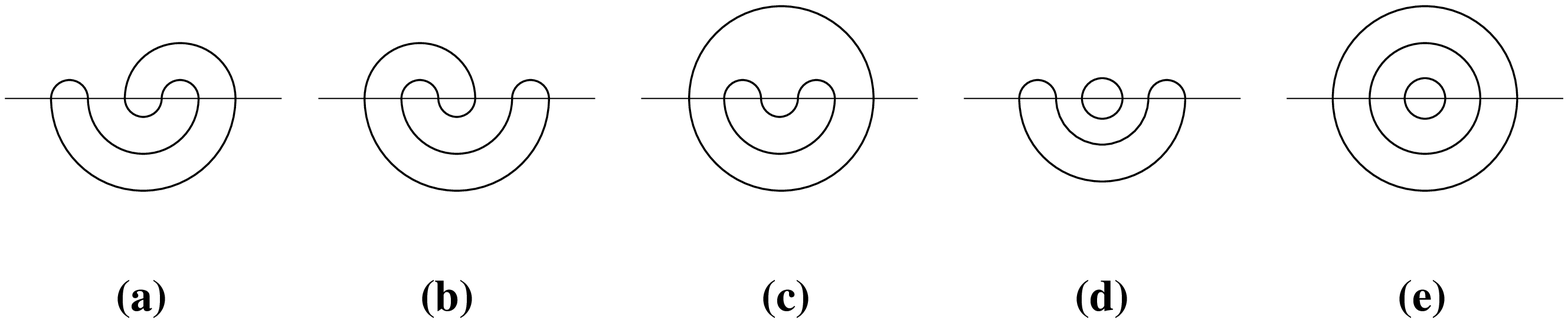}{11.truecm}
\figlabel\fivem

The language of sect.5.5 above leads naturally to the introduction of the number 
${\bar \mu}_n^{(k)}$ of $k$--colored multi--component semi--meanders of order 
$n$. By $k$--colored, we mean that exactly $k$ distinct colors are used, but
we do not distinguish between colorings differing only by a permutation of the colors. 
For illustration, let us take $n=3$. 
The $5$ semi--meanders of order $3$ are displayed in Fig.\fivem, and denoted $(a)$,
$(b)$, $(c)$, $(d)$, $(e)$.  Only the last one $(e)$ can be colored with $k=3$
distinct colors, hence ${\bar \mu}_3^{(3)}=1$. With $k=2$ colors, there are $3$
ways of coloring $(e)$, and $1$ way of coloring $(c)$ and $(d)$, hence a total
of ${\bar \mu}_3^{(2)}=5$. Finally, each connected semi--meander $(a)$ and $(b)$
can be colored in a unique way
with $k=1$ color, hence ${\bar \mu}_3^{(1)}=5$.

In terms of these numbers, the rhs of eq.\semimat\ 
is expressed as a sum over the number $k$ of colors used, with a total choice of
$m(m-1)(m-2)...(m-k+1)$ colors among $m$, weighed by the numbers ${\bar \mu}_n^{(k)}$.
This implies a relation between coloring numbers $\bar \mu$ and meander numbers
$\bar M$
\eqn\colintsem{ {\bar m}_n(m)~=~\sum_{k=1}^n {\bar M}_n^{(k)}~m^k~
=~\sum_{k=1}^n {\bar \mu}_n^{(k)}~
m(m-1)...(m-k+1)~.}
Remarkably, the numbers $\bar \mu$ do not carry direct information about the numbers 
of connected components of the semi--meanders, but their knowledge is sufficient to 
know the $\bar M$ exactly, by use of the Stirling numbers. Note also 
that eq.\colintsem\ expresses the polynomial
${\bar m}_n(x)$ in two different bases, $\{ x^k,k=1,...,n\}$ and
$\{ x(x-1)...(x-k+1),k=1,...,n\}$, of the space of degree $n$
polynomials with vanishing 
constant coefficient. For $n=3$, we simply have
\eqn\exfiv{ x^3+ 2 x^2+ 2 x~=~ x(x-1)(x-2)+ 5 x(x-1) + 5x~.}

The same type of expression holds for meanders, namely
\eqn\fivemean{ \sum_{k=1}^n M_n^{(k)}m^k~=~\sum_{k=1}^n \mu_n^{(k)}~ m(m-1)...(m-k+1)~,}
where $\mu_n^{(k)}$ denotes the number of $k$--colored multi--component meanders
of order $n$.

\subsec{Combinatorial expression using symmetric groups}

\fig{An arch configuration of order $3$ and the corresponding interpretation as a
ribbon graph, with $V=1$ six--valent vertex and $E=3$ edges. On the intermediate
diagram, the arches have been doubled and oriented. These oriented arches 
indicate the pairing of bridges, i.e. represent the action of $\mu$. Similarly,
the oriented horizontal segments indicate the action of the shift permutation $\sigma$.
Each oriented loop corresponds to a cycle
of the permutation $\sigma \mu$.}{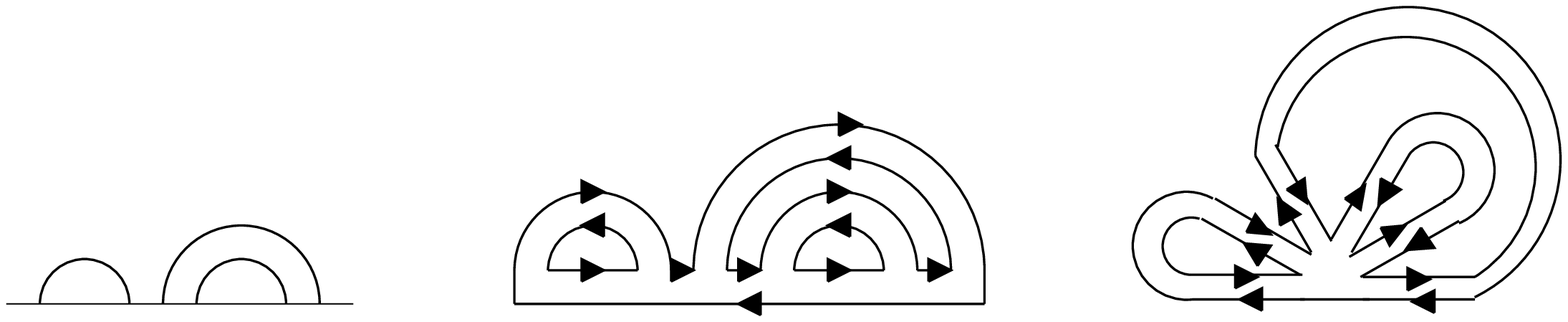}{9.truecm}
\figlabel\rib

{\bf Admissibility condition.} In sect.5.4 above, we have seen how an arch configuration
of order $n$ could be encoded in an admissible permutation $\mu$ belonging to the class
$[2^n]$ of $S_{2n}$. Let us write the admissibility condition explicitly. 
This condition states that arches do not intersect each other, namely that the ribbon
graph (see Fig.\rib) with only one $2n$--valent vertex (the $2n$ bridges), whose legs
are connected according to the arch configuration, is {\it planar}, i.e. of genus $h=0$.
This graph has $V=1$ vertex, and $E=n$ edges (arches). Let us compute its number $L$
of oriented loops in terms of the permutation $\mu$.
Let $\sigma$ denote the ``shift" permutation, namely $\sigma(i)=i+1$, $i=1,2,...,2n-1$ and
$\sigma(2n)=1$.  Then an oriented loop in the ribbon graph is readily seen to correspond to
a {\it cycle} of the permutation $\sigma \mu$. Indeed, the total number of loops is
$L={\rm cycles}(\sigma \mu)$, the number of cycles of the permutation $\sigma \mu$.
The admissibility condition reads
\eqn\admi{\eqalign{
\chi~=~2=L-E+V~&=~1-n+{\rm cycles}(\sigma \mu)~\cr
\Leftrightarrow \quad {\rm cycles}(\sigma \mu)~&=~n+1~.\cr}}
Note that if we demand that the ribbon graph be of genus $h$, the above condition becomes 
\eqn\higconad{ {\rm cycles}(\sigma \mu)~=~n+1-2h~.}

{\bf Connected components.} Given an admissible permutation $\mu \in [2^n]$, let us
now count the number of connected components of the corresponding semi--meander of
order $n$. Let $\tau$ be the ``rainbow" permutation $\tau(i)=2n+1-i$. Note that $\tau$
changes the parity of the bridge label.  On the other hand, the admissible permutation
$\mu$ is readily seen to also change the parity of the bridge labels. 
As a consequence,
the permutation $\tau \mu$ preserves the parity of bridge labels. 
In other words, even bridges are never mixed with odd ones.
The successive iterations of the permutation $\tau \mu$ describe its cycles. 
The corresponding meander will be connected iff these cycles are maximal, namely $\mu$
has two cycles of length $n$ (one for even bridges, one for odd bridges), i.e.
$\tau \mu \in [n^2]$.  
We get a purely combinatorial expression
for connected semi--meander numbers
\eqn\purecom{
{\bar M}_n~=~ {\rm card}\{ \mu \in [2^n]~|~ {\rm cycles}(\sigma \mu)=n+1,\ 
{\rm and}\ \tau \mu \in [n^2] \}~.}

More generally, the semi--meander corresponding to $\mu$ 
will have $k$ connected components iff $\tau \mu$ has exactly $k$ pairs of cycles 
of equal length (one over even bridges, one over odd ones).

{\bf Character expressions.} The above conditions on various permutations are best expressed
in terms of the characters of the symmetric group. Denoting by $[i^{\nu_i}]$
the class of permutations with $\nu_i$ cycles of length $i$, and labelling the 
representations of $S_{2n}$ by Young tableaux $Y$ with $2n$ boxes
as customary, the characters
can be expressed as
\eqn\defchar{ \chi_Y([i^{\nu_i}])~=~ \det \big( p_{i+\ell_i -j}(\theta_.) \big)
\bigg\vert_{t_\nu}~,}
where the Young tableau has $\ell_i$ boxes in its $i$-th line, counted from the 
top, $t_\nu=\prod_i {\theta_i^{\nu_i} \over \nu_i!}$,
$p_m(\theta_.)$ is the $m$-th Schur polynomial of the
variables $\theta_1$, $\theta_2$,...
\eqn\schupol{ p_m(\theta_.)~=~ \sum_{k_i \geq 0, i=1,2,...\atop
\Sigma i k_i =m} \prod_i {\theta_i^{k_i} \over k_i!}~,}
and we used the symbol $f(\theta_.)\vert_{t_\nu}$ for
the coefficient of the monomial $\prod_i{\theta_i^{\nu_i} \over \nu_i!}$ in the polynomial
$f(\theta_.)$.
As group characters, the $\chi_Y$'s satisfy the orthogonality relation
\eqn\ortochar{ \sum_Y \chi_Y([\lambda]) \chi_Y([\mu])~=~ {(2n)! \over |[\lambda]|}~
\delta_{[\lambda],[\mu]}~,}
where the sum extends over all Young tableaux with $2n$ boxes,
$[\lambda]$ denotes the class of a permutation $\lambda\in S_{2n}$, and
$|[\lambda]|$ the order of the class. 
The order of the class $[i^{\nu_i}]$ is simply
\eqn\orcla{ |[i^{\nu_i}]|~=~ {(2n)! \over \prod_i i^{\nu_i} \nu_i!}~.}
The orthogonality relation \ortochar\ provides us with a means of expressing any condition on 
classes of permutations in terms of characters.  
It leads to the following compact expression for
the connected semi--meander numbers
\eqn\compsem{\eqalign{
{\bar M}_n~&=~ \sum_{[i^{\lambda_i}]\in S_{2n} \atop \Sigma \lambda_i=n+1}
\sum_{\mu \in [2^n]} ~\delta_{[\sigma \mu],[i^{\lambda_i}]}~\delta_{[\tau \mu],[n^2]} \cr
&=~ \sum_{[i^{\lambda_i}]\in S_{2n} \atop \Sigma \lambda_i=n+1}
\sum_{\mu \in S_{2n}} \sum_{Y,Y',Y''} 
{|[2^n]|~|[i^{\lambda_i}]|~|[n^2]| \over ( (2n)!)^3} \cr
&\times \chi_Y([\mu]) \chi_Y([2^n]) \chi_{Y'}([\sigma \mu])\chi_{Y'}([i^{\lambda_i}])
\chi_{Y''}([\tau \mu])\chi_{Y''}([n^2])~.\cr}}

Analogous expressions hold for (higher genus)
semi--meanders with $k$ connected components and 
for meanders as well.  These make completely explicit the calculation of the various
meander--related numbers.  Unfortunately, the characters of the symmetric group are
not so easy to deal with, and we were not able to
use them in an efficient way to enhance our numerical data.

\newsec{Properties}

In this section, we derive a number of identities and inequalities involving
various meander and semi--meander numbers.  We also compute exactly the number
of closings of some particular lower arch configurations, made of repetitions
of simple motives. 

\subsec{Sum rules}

In addition to the sum rules \sumrulm\ \sumsem\ and \gaurn\ \gaurm\
for (higher genus) meander and semi--meander numbers,
let us derive the following
\eqn\sumore{\eqalign{
\sum_{k=1}^n M_n^{(k)}~ (-1)^{k-1}~&=~ \left\{\matrix{ 0 & \ \hbox{if $n$ is even} \cr
(c_p)^2 & \ \hbox{if $n=2p+1$} \cr}\right. \cr
\sum_{k=1}^n {\bar M}_n^{(k)}~(-1)^{k-1}~&=~  \left\{
\matrix{ 0 & \ \hbox{if $n$ is even} \cr
c_p & \ \hbox{if $n=2p+1$} \cr}\right. ~. \cr}}
The reader can easily check these sum rules with the data of tables I and II above.

To prove both sum rules, we introduce the notion of {\bf signature} of an arch configuration.
Let us denote by $|{\cal A}|$ the order of an arch configuration $\cal A$. The signature
of $\cal A$ is defined recursively by the initial value on the vacuous configuration
and its behaviour under the processes (I) and (II) of sect.4.1,   
which enable us to construct any arch configuration.  We set
\eqn\defrecsig{ \eqalign{
{\rm sig}(\emptyset)~&=~1\cr
{\rm sig}((I).{\cal A})~&=~ (-1)^{|{\cal A}|}~{\rm sig}({\cal A})\cr
{\rm sig}((II).{\cal A})~&=~ (-1)^{|{\cal A}|+1}~{\rm sig}({\cal A})~.\cr }}
It is important to notice that the signature behaves differently under (I) and (II),
hence depends on the number of connected components in semi--meanders for instance.
More precisely, if ${\cal M}$ is a semi--meander of order $n$
with $k$ connected components, ${\cal M}\in SM_{n,k}$, and ${\cal P}\in SM_{n,1}$,
then they both originate from the single arch of order $1$ (with signature $-1$)
by a total of $(n-1)$ actions of (I) and (II), 
each adding one arch,
except that $\cal P$ is obtained by acting $(n-1)$ times with (I), whereas $\cal M$
is obtained by acting $(k-1)$ times with (II), and $(n-k)$ times with (I).
Due to \defrecsig, this implies a difference in the signatures
\eqn\relasemcom{ {\rm sig}({\cal M})~=~ (-1)^{k-1} ~{\rm sig}({\cal P})~,}
whereas after $(n-1)$ actions of (I), adding one arch each time, the signature of 
$\cal P$ reads
\eqn\sigp{ {\rm sig}({\cal P})~=~ (-1)^{1+{n(n-1) \over 2}}~.}
Thus, the signature of a semi--meander only depends on its order $n$ and its 
number of connected components $k$, namely
\eqn\sigcal{ {\rm sig}({\cal M})~=~ (-1)^{k+{n(n-1) \over 2}}, 
\quad \forall \ {\cal M}\in
SM_{n,k}~.}

In fact this signature can also be easily related to the numbers of arches of 
given depth in $\cal A$
as follows
\eqn\siganddep{ {\rm sig}({\cal A})~=~(-1)^{\sum_{j \geq 1} j A({\cal A},j) }~,}
where $A({\cal A},j)$ (see sect.3) denotes the number of arches of depth $j$ in $\cal A$.
One readily checks that eq.\siganddep\ solves the recursion relations \defrecsig.
For instance, the signature of the rainbow configuration of order $n$ is
\eqn\rainsig{ {\rm sig}({\cal R}_n)~=~ (-1)^{n(n+1) \over 2} ~,}
in agreement with \relasemcom\ for $k=n$.
To conclude, the signature of a semi--meander only depends on its order $n$ and its 
number of connected components $k$, namely
\eqn\sigcal{ {\rm sig}({\cal M})~=~ (-1)^{k+{n(n-1) \over 2}}, \quad \forall \ {\cal M}\in
SM_{n,k}~.}

The sum rule \sumore\ for semi--meanders is obtained by computing the sum 
of all the signatures of multi--component semi--meanders of order $n$. 
Let us introduce
\eqn\moredefs{\eqalign{ 
s(n)~&=~ \sum_{{\cal A} \in A_n} {\rm sig}({\cal A})~\cr
&=~ (-1)^{1+{n(n-1) \over 2}}~\sum_{k=1}^n {\bar M}_n^{(k)}~(-1)^{k-1}~, \cr}}
where the first sum extends over all the arch configurations of order $n$. 
Using the usual recursive reasoning of Fig.\recursion, the leftmost arch of a configuration
separates the configuration into two configurations of smaller orders, say $\cal X$ of
order $j$
and $\cal Y$ of order $n-1-j$ (the only configuration of order $0$ is just $\emptyset$). 
As the $j$ arches of $\cal X$ are shifted downwards in the global configuration, we obtain 
the recursion relation
\eqn\recusofn{\eqalign{ 
s(n)~&=~ \sum_{j=0}^{n-1} \sum_{{\cal X}\in A_j}\sum_{{\cal Y}\in A_{n-1-j}}
{\rm sig}({\cal X}) {\rm sig}({\cal Y}) (-1)^{j+1}\cr
&=~ \sum_{j=0}^{n-1} (-1)^{j+1} s(j)~s(n-1-j)~.\cr}}
With $s(0)=1$, we immediately find that $s(2p)=0$ for $p=1,2,...$, and that
\eqn\catsn{ s(2p+1)~=~\sum_{j=0}^{p-1} s(2j+1) ~s(2p-2j-1)~,}
with $s(1)=-1$, hence for $n=2p+1$
\eqn\solsnca{ s(2p+1)~=~ (-1)^{p+1} ~ c_p~.}
Comparing with \moredefs, and noticing that 
$1+n(n-1)/2=1+(2p+1)(2p)/2=p+1$ modulo $2$, this
completes the proof of \sumore\ for semi--meanders.

In the case of meanders, let us define the signature of a given meander ${\cal M}$ with 
upper and lower arch configurations ${\cal A}_{\rm up}$ and ${\cal A}_{\rm down}$ respectively,
as the product\foot{Note that in a previous definition, the signature of a
semi--meander was referring to its upper arch configuration only. 
Considered as a particular meander, its signature is simply multiplied by that 
of the raibow of order $n$ \rainsig.}
\eqn\defsigmeand{ {\rm sig}( {\cal M})~=~ {\rm sig}({\cal A}_{\rm up})~
{\rm sig}({\cal A}_{\rm down}) ~.}
With this definition, we find that the signature of a meander only depends 
on its order and on its
number of connected components through
\eqn\calsigmd{ {\rm sig}( {\cal M})~=~(-1)^{k+n}~.}
Therefore, 
\eqn\calfinlu{\eqalign{
\sum_{k=1}^n M_n^{(k)}~ (-1)^{k-1}~
&=~ (-1)^{n-1}\sum_{{\cal A}_{\rm up}\in A_n}
\sum_{{\cal A}_{\rm down} \in A_n} {\rm sig}({\cal A}_{\rm up})~
{\rm sig}({\cal A}_{\rm down}) ~\cr
&=~(-1)^{n-1} s(n)^2 \cr
&=~\left\{\matrix{ 0 &\  \hbox{if $n=2p$} \cr
(c_p)^2 & \ \hbox{if $n=2p+1$} \cr}\right. ~\cr}}
This completes the proof of \sumore\ for meanders. 

Combining \sumrulm\ and \sumsem\ with \sumore, we get the equivalent set of rules
\eqn\eqserul{\eqalign{
\sum_{k=1}^n M_{2n}^{(2k-1)}~&=~{c_{2n}^2\over 2}~=~\sum_{k=1}^n M_{2n}^{(2k)} \cr
\sum_{k=1}^{n+1} M_{2n+1}^{(2k-1)}~&={c_{2n+1}^2+c_n^2 \over 2} \cr
\sum_{k=1}^n M_{2n+1}^{(2k)}~&=~{c_{2n+1}^2-c_n^2 \over 2} \cr
\sum_{k=1}^n {\bar M}_{2n}^{(2k-1)}~&=~{c_{2n}\over 2}~=~\sum_{k=1}^n {\bar M}_{2n}^{(2k)} \cr
\sum_{k=1}^{n+1} {\bar M}_{2n+1}^{(2k-1)}~&={c_{2n+1}+c_n \over 2} \cr
\sum_{k=1}^n {\bar M}_{2n+1}^{(2k)}~&=~{c_{2n+1}-c_n \over 2} ~.\cr}}

\subsec{Inequalities}

The above sum rules \eqserul\ lead to some immediate inequalities, by simply using
the positivity of all the meandric numbers involved, namely
\eqn\ineqm{\eqalign{
M_{2n}^{(k)}~&\leq {c_{2n}^2 \over 2} \cr
M_{2n+1}^{(2k-1)}~&\leq {c_{2n+1}^2+c_n^2 \over 2}\cr
M_{2n+1}^{(2k)}~&\leq~{c_{2n+1}^2-c_n^2 \over 2} \cr
{\bar M}_{2n}^{(k)}~&\leq~{c_{2n}\over 2}\cr
{\bar M}_{2n+1}^{(2k-1)}~&\leq{c_{2n+1}+c_n \over 2} \cr
{\bar M}_{2n+1}^{(2k)}~&\leq~{c_{2n+1}-c_n \over 2} ~,\cr}}
for all allowed $k$'s.

In particular, according to the leading behaviour $c_n \sim 4^n$ when $n \to \infty$, 
this gives an upper bound on the numbers $R$ and $\bar R$ \entrom\ \entrosem\
\eqn\maxR{ R~\leq~16 \qquad {\bar R} ~\leq~4~.}
Defining as before
$R_k$ and ${\bar R}_k$ as the leading terms in respectively meander
and semi--meander numbers with $k$ connected components for large order $n$
(with $R_1=R$ and ${\bar R}_1={\bar R}$)
\eqn\behak{ M_n^{(k)} \sim (R_k)^n \qquad {\bar M}_n^{(k)} \sim ({\bar R}_k)^n~,}
this also gives the bounds
\eqn\maxkr{ R_k~\leq~16 \qquad {\bar R}_k ~\leq~4~.}

The matrix model formulation of sect.5 gives rise to more interesting inequalities,
relating meander and semi--meander numbers.
They are based on the Cauchy--Schwarz inequality for the hermitian product $(A,B)$ over
complex matrices defined by
\eqn\herpro{ (A,B)~=~{\rm Tr}(A B^{\dagger})~,}
where the superscript $\dagger$ stands for the hermitian conjugate $B^\dagger=B^{t\ *}$,
namely
\eqn\causchw{ |(A,B)|^2 \leq (A,A)~(B,B)~.}
Starting from the expression \luttefin, let us write
\eqn\usefin{ A=B^{(\alpha_1)}B^{(\alpha_2)}...B^{(\alpha_l)}\quad {\rm and} \quad
B^\dagger=B^{(\alpha_{l+1})}...B^{(\alpha_{2n})}~,}
then \causchw\ becomes
\eqn\incau{ \big\vert{\rm Tr}(\prod_{i=1}^{2n} B^{(\alpha_i)})\big\vert^2 \leq
{\rm Tr}\big(\prod_{i=1}^l B^{(\alpha_i)} \prod_{j=1}^l B^{(\alpha_{l+1-j})}\big)
{\rm Tr}\big(\prod_{i=l+1}^{2n} B^{(\alpha_i)} \prod_{j=1}^{2n-l} 
B^{(\alpha_{2n+1-j})}\big)~.}
Taking the thermodynamic Gaussian averages of both sides and summing over the 
colors $\alpha_i=1,2,...,m$, we finally get
\eqn\fineg{ \sum_{k=1}^n M_n^{(k)} ~m^k ~\leq~ \big( \sum_{i=1}^l {\bar M}_l^{(i)}~m^i \big)
\big(\sum_{j=1}^{2n-l} {\bar M}_{2n-l}^{(j)}~m^j \big)~,}
where the rhs has been identified by use of \semimat.
This inequality between polynomials results in many inequalities for 
the meander and semi--meander numbers themselves.

In the trivial case $l=0$, i.e. for the choice $A={\bf I}$, we have
\eqn\partin{ \sum_{k=1}^n M_n^{(k)} ~m^k ~\leq~ \sum_{k=1}^{2n}
{\bar M}_{2n}^{(k)}~m^k~.}
This inequality is in fact obviously satisfied coefficient by 
coefficient for $k=1,2,...,n$, as a meander
of order $n$ is a particular case of semi--meander of order $2n$ 
(which does not wind around the source of the river), namely
\eqn\coin{ M_n^{(k)}~\leq~ {\bar M}_{2n}^{(k)} \quad \forall \ k=1,2,...,n.}
This implies the inequalities between the leading terms \maxkr\
\eqn\leadine{ R_k~\leq {\bar R}_k^2~ \forall \ k=1,2,...}

Before taking the large $N$ limit, \incau\ also leads to an inequality 
involving higher genus meander and semi--meander numbers
\eqn\higeineq{\eqalign{ 
\sum_{h \geq 0, k\geq 1} M_p^{(k)}(h)~m^k~N^{2-2h}~\leq~
&\big( \sum_{h \geq 0, k\geq 1} {\bar M}_{2l}^{(k)}(h)~m^k~N^{1-2h} \big) \cr
&\times \big( \sum_{h \geq 0, k\geq 1} {\bar M}_{2p-2l}^{(k)}(h)~m^k~N^{1-2h} \big)~.\cr}}

\fig{The algorithm for closing a given upper arch configuration into a connected
meander. The upper arch configuration $\cal A$
is of order $6$. The path starts on the
leftmost exterior arch of $\cal A$: the first lower arch 
connects it to the second exterior arch, and is numbered $1$. 
Then the path goes to the left, and connects the depth $2$ arches of $\cal A$
through lower arches 
number $2$, $3$ and $4$. The leftmost depth $2$ arch of $\cal A$ is connected 
to the depth $3$ one through the lower arch number $5$, and finally the depth $3$ arch of 
$\cal A$ is connected to the leftmost exterior arch of $\cal A$ 
through the lower arch number $6$.}{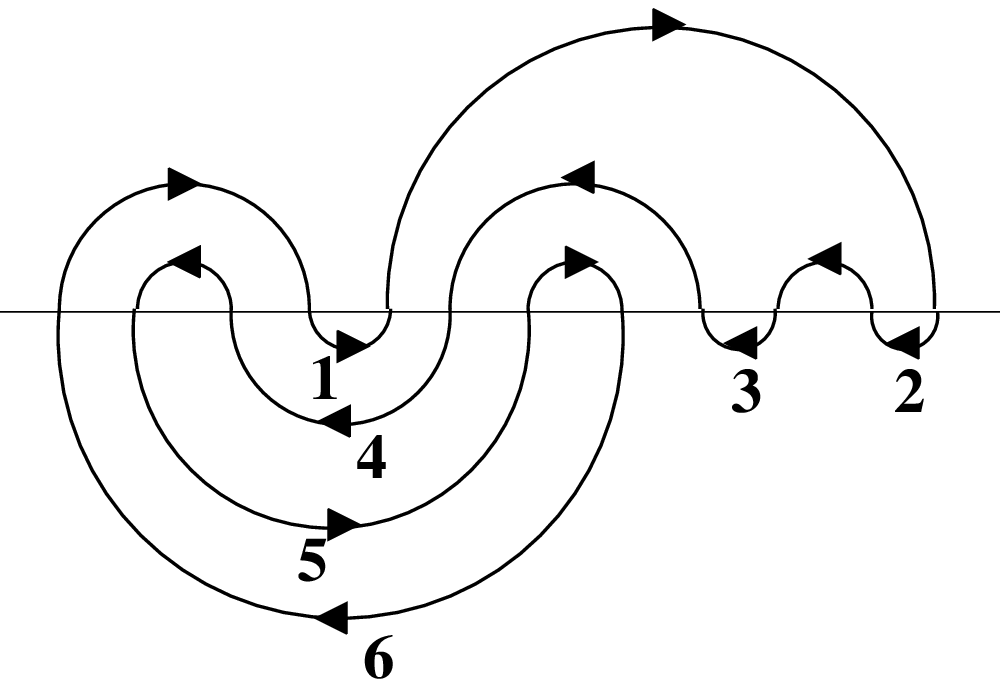}{8.truecm}
\figlabel\algoclo

A minoration of the connected meander numbers $M_n$ is obtained as follows.
For any given upper arch configuration $\cal A$, we have a simple algorithm for
constructing its closing by a lower arch configuration, such that the resulting meander is
connected.  
Consider the exterior arches of $\cal A$, and draw lower arches connecting
any two consecutive exterior arches, as shown in Fig.\algoclo: this creates a path 
(oriented, say from the left to the right)
connecting the exterior arches of 
$\cal A$. In a second step, the path is continued by connecting from the right to 
the left the arches of {\it depth 2} in $\cal A$, through lower arches which 
are either disjoint from the previous ones, or contain some of them.
The lower arch configuration is further continued by connecting the arches of increasing
depth $j$ in $\cal A$, through lower arches. The resulting path zig--zags through
the arches of $\cal A$, describing those of even depth from right to left
and those of odd depth from left to right. Once the last arch of $\cal A$ is described, 
a last lower arch is constructed, connecting this last arch to the leftmost exterior 
arch we started with. The total being closed and going through all the arches of $\cal A$
as well as the lower ones, we have constructed a connected meander. Therefore the total number
of connected meanders of order $n$, $M_n$, is larger than that of arch configurations 
of order $n$, $c_n$
\eqn\inegfinal{ M_n ~\geq ~ c_n~.}
This implies that 
\eqn\ineqfin{ R ~ \geq ~ 4~ \qquad {\bar R} ~ \geq ~2~,}
where we have used the inequality \leadine\ for $k=1$.

\subsec{Special cases}

In this section, we compute exactly some of the generalized semi--meander
numbers of sect.5.4. 
We will consider the closings of some particular lower arch configurations ${\cal A}_\mu$.
It turns out that the numbers can always be found by relatively simple
recursions provided the arch configuration ${\cal A}_\mu$
is the {\it repetition} (say $p$ times) of some small order arch configuration. 

\noindent{\bf Single arch.} The simplest example
is ${\cal A}_1$, obtained as the repetition ($n$ times) of single 
arches of order $1$, and corresponding to the 
permutation $\mu$: $\mu(2i-1)=2i$, $i=1,2,...,n$.
Let us compute the corresponding numbers ${\bar M}_n^{(k)}({\cal A}_1)$ by the 
method of loop equations described in sect.5.5. 
The generating function $\alpha_n(m)$
for these numbers satisfies \semigen, namely
\eqn\arcrep{\alpha_n(m)~=~\sum_{k=1}^n {\bar M}_n^{(k)}({\cal A}_1)~m^k~=~
\lim_{N \to \infty}{1 \over N} \langle {\rm Tr}\big( \sum_{j=1}^m (B^{(j)})^2 \big)^n
\rangle_{\rm Gauss}~.}
Let us use the loop equation method of sect.5.4 to derive a recursion relation 
for $\alpha_n(m)$, by writing that the multi--matrix
integral of a total derivative vanishes
\eqn\intotder{
0~=~\int \prod_{\alpha=1}^m dB^{(\alpha)} {\partial \over \partial B_{ji}^{(\beta)} }
\big[ [\sum_{r=1}^m (B^{(r)})^2]^{n-1} B^{(\beta)} \big]_{kl} 
e^{-N{\rm Tr}\sum_{s=1}^m {B^{(s)} \over 2} }~,}
hence, taking $i=l$ and $j=k$, and summing over $i,j=1,2,...,N$
\eqn\avdepo{\eqalign{ 
N\langle {\rm Tr}\big( [\sum_{r=1}^m (B^{(r)})^2]^{n-1} &(B^{(\beta)})^2
\big) \rangle_{\rm Gauss}=\langle N{\rm Tr}\big( [\sum_{r=1}^m (B^{(r)})^2]^{n-1}\big)\cr
&+\sum_{j=1}^{n-1} {\rm Tr}\big([\sum (B^{(r)})^2]^{j-1}B^{(\beta)}\big)
{\rm Tr}\big([\sum (B^{(r)})^2]^{n-1-j} B^{(\beta)}\big) \cr
&+ {\rm Tr}\big([\sum (B^{(r)})^2]^{j-1}\big){\rm Tr}\big([\sum (B^{(r)})^2]^{n-1-j} 
(B^{(\beta)})^2 \big) \rangle_{\rm Gauss}~.\cr}}
By the factorization property \factocor, the term on the second line vanishes by imparity.
Summing then over $\beta=1,2,...,m$ and letting $N \to \infty$,
we finally get a recursion relation for $\alpha_n(m)$ \arcrep\
\eqn\trader{ \alpha_n(m)~=~(m-1)\alpha_{n-1}(m)+ 
\sum_{j=0}^{n-1} \alpha_j(m) ~\alpha_{n-1-j}(m)~,}
whith the initial condition $\alpha_0(m)=1$. This determines uniquely
the polynomials $\alpha_n(m)$ of degree $n$ in $m$. Remarkably, we find
\eqn\resal{ \alpha_n(m)~=~ \sum_{k=1}^n I(n,k)~m^k~,}
where the numbers $I(n,k)$ are given by \solint\ ! 
In particular, we find the average number of connected components in closings
of ${\cal A}_1$
\eqn\avconone{ \langle {\rm comp}_1 \rangle_{n}~=~ {\sum_{k=1}^n k I(n,k) \over
c_n} ~=~ {n+1 \over 2}~.}
The result \resal\ is readily checked by multiplying both sides of the recursion
relation \recint\ by $m^k$ and summing over $k$, which yields \trader.
So the number of arch configurations of order $n$ with $k$ interior 
arches is also the number of closings of ${\cal A}_1$ with $k$ connected components.
Note that there is $I(n,1)=1$ connected closing of ${\cal A}_1$
only, which may be obtained by 
the closing algorithm presented at the end of the previous section.

This remarkable coincidence can be explained as follows.
Let us constuct a one--to--one
mapping $U$
between arch configurations, which sends arch configurations which 
close ${\cal A}_1$ with $k$ connected components onto arch configurations with $k$ 
interior arches. The map $U$ is defined recursively as follows
\eqn\defumap{\eqalign{
U({\epsfxsize=.6truecm \epsfbox{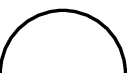}})~&=~
{\epsfxsize=.6truecm \epsfbox{archone.eps}} \cr
U(X~Y)~&=U(X)~U(Y)\cr
U({\epsfxsize=1.9truecm \epsfbox{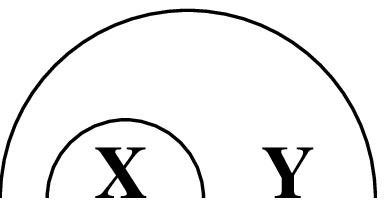}})~&=~
{\epsfxsize=2.6truecm \epsfbox{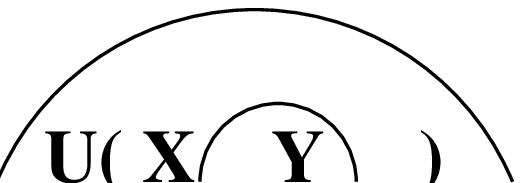}}~.\cr}}
\fig{The mapping $U$ between arch configurations is
indicated by arrows, for the cases of order $1$, $2$, $3$. In these cases $U$
appears to be involutive, but this is only an accident, 
as one readily checks for $n \geq 4$.}{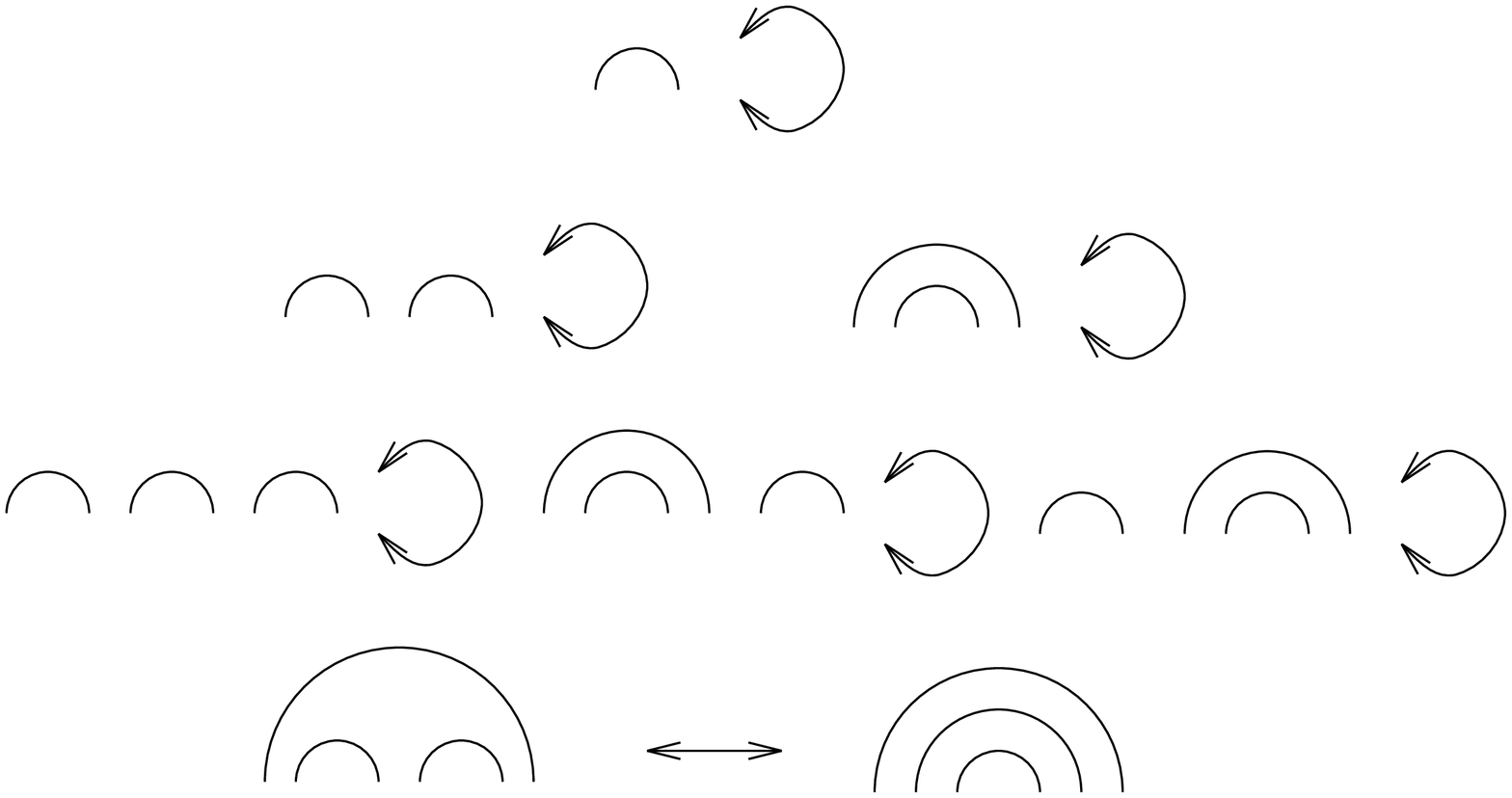}{8.truecm}
\figlabel\exu
\noindent{}A few examples of the action of $U$ are given in Fig.\exu.
The coincidence between the number of connected components after closing a
given configuration
by ${\cal A}_1$, and the number of interior arches in its image by $U$ needs
only to be checked on a configuration with one exterior arch, because
the second relation of \defumap\ can be used to reduce any arch configuration
into a succession of arch configurations with one exterior arch only, whereas
the numbers of connected components in the closure by ${\cal A}_1$ and of
interior arches both add up in a succession of arch configurations. 
Counting the number $k$ of connected components in the lhs of the last line 
of \defumap, we find
\eqn\counco{ k~=~\#{\rm c.c.}({\epsfxsize=1.9truecm \epsfbox{uone.eps}})~=~
\# {\rm c.c.}(X)+\# {\rm c.c.}({\epsfxsize=.9truecm \epsfbox{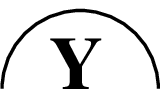}})~}
On the other hand, the number of interior arches in the rhs is
\eqn\inrhs{ i~=~\#{\rm int.}U(X)+\#{\rm int.}
U({\epsfxsize=.9truecm \epsfbox{archy.eps}})~.}
Therefore, comparing \counco\ and \inrhs,
we get the desired result by recursion on the order of the arch configuration:
the number of connected components in the closing by ${\cal A}_1$ is equal to
the number of interior arches after the action of $U$.

\noindent{\bf Double arch.} The repetition (say $n$ times) of a double arch 
(rainbow of order $2$) gives an arch configuration ${\cal A}_2$ of order $2n$.
Let us compute the number ${\bar M}_{2n}^{(k)}({\cal A}_2)$ of closings of
this configuration with $k$ connected components.  As pointed out in the previous section,
the loop equation method is parallel to the general recursion principle of Fig.\recursion\
for arch configurations. To further illustrate this similarity, let us compute
${\bar M}_{2n}^{(k)}({\cal A}_2)$ in the language of arches, by deriving some
recursion relations. When trying to apply the general process of Fig.\recursion,
we immediately see that the recursion will not involve closings of ${\cal A}_2$
only, but requires the introduction of some new arch configuration closings.

Let us introduce the notations 
\eqn\notaclo{\eqalign{\lambda_{2n+1}^{(k)}~&=~{\bar M}_{2n+1}^{(k)}
({\cal A}_2\  {\epsfxsize=.4truecm \epsfbox{archone.eps}}\ \ )\cr 
\mu_{2n}^{(k)}~&=~{\bar M}_{2n}^{(k)}({\cal A}_2 )\cr
\nu_{2n+3}^{(k)}~&=~{\bar M}_{2n+3}^{(k)}(
{\epsfxsize=.4truecm \epsfbox{archone.eps}} \ 
{\epsfxsize=.9truecm \epsfbox{archone.eps}}\!\!\!\!\!\!\!\!\!{\cal A}_2 \ \ \,
{\epsfxsize=.4truecm \epsfbox{archone.eps}})~,\cr}}
for $n \geq 0$. The configuration ${\cal A}_2$ and its slight modifications 
listed in \notaclo\ form a system which is closed under the recursion principle
of Fig.\recursion. 
\fig{The three recursion relations involving the numbers $\lambda_{2n+1}^{(k)}$,
$\mu_{2n+2}^{(k)}$ and $\nu_{2n+3}^{(k)}$ respectively. 
The leftmost arch is indicated in thick solid line. 
The symbol $\Sigma$ indicates a sum
over the position of the right bridge of the leftmost arch.
}{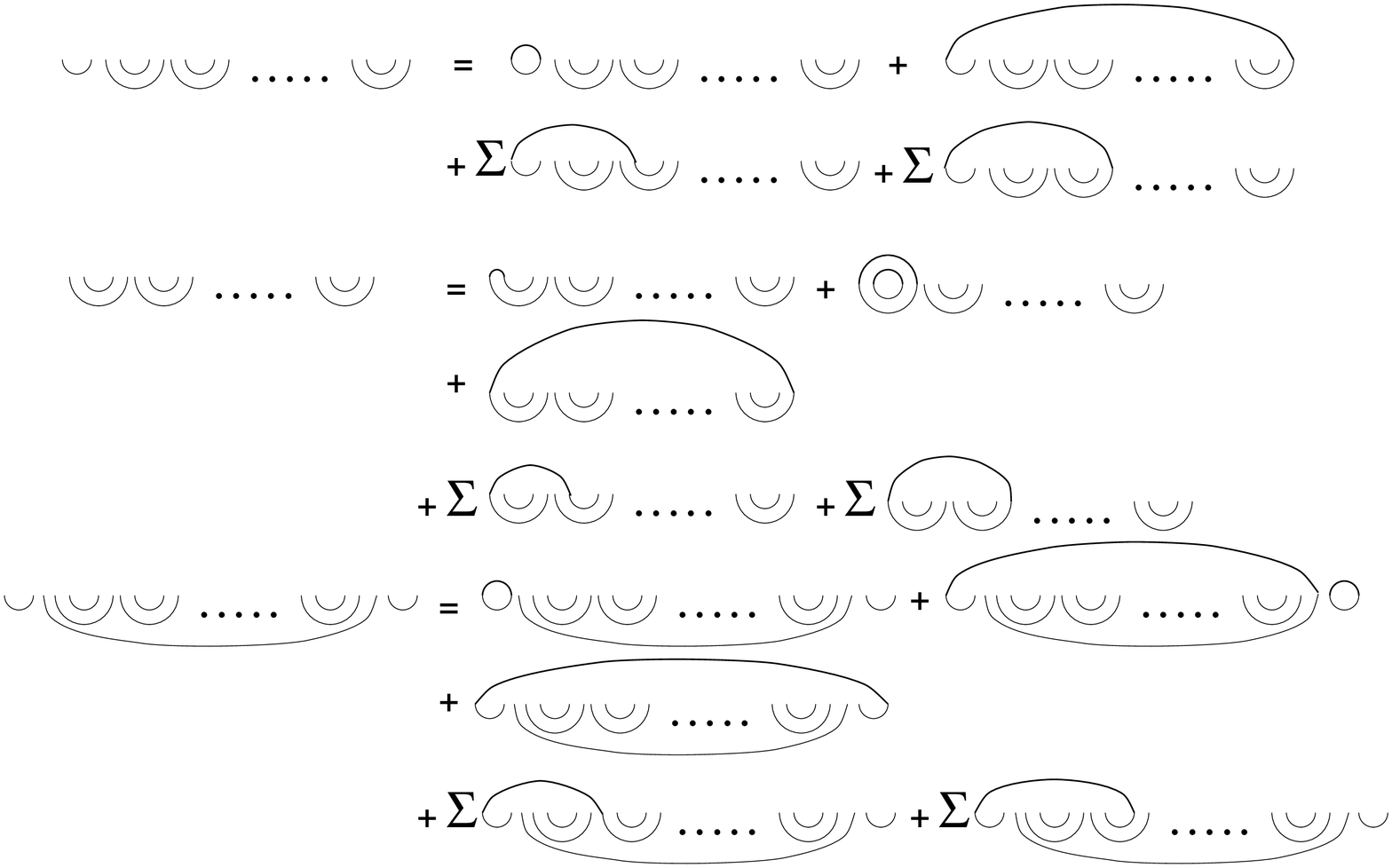}{10.truecm}
\figlabel\rectwoar
\noindent{}More precisely, we find the three recursion relations,
graphically depicted in Fig.\rectwoar\
\eqn\recthreetwo{\eqalign{
\lambda_{2n+1}^{(k)}~&=~\mu_{2n}^{(k-1)}+\mu_{2n}^{(k)}+\sum_{j=0}^{n-1}\sum_{i=1}^k
\lambda_{2j+1}^{(i)} \lambda_{2n-2j-1}^{(k+1-i)}\cr
&+\sum_{j=1}^{n-1} \sum_{i=1}^{k-1} \mu_{2j}^{(i)} \mu_{2n-2j}^{(k-i)}\cr
\mu_{2n+2}^{(k)}~&=~ \lambda_{2n+1}^{(k)}+\mu_{2n}^{(k-2)}+\nu_{2n+1}^{(k)}\cr
&+\sum_{j=0}^{n-1} \sum_{i=1}^k \lambda_{2j+1}^{(i)} \mu_{2n-2j}^{(k+1-i)}+
\sum_{j=1}^n \sum_{i=1}^{k-1} \mu_{2j}^{(i)} \nu_{2n+1-2j}^{(k-i)}\cr
\nu_{2n+3}^{(k)}~&=~ \mu_{2n+2}^{(k-1)}+\lambda_{2n+1}^{(k-1)}+\mu_{2n+2}^{(k)}\cr
&+\sum_{j=0}^{n} \sum_{i=1}^k \nu_{2j+1}^{(i)} \lambda_{2n+1-2j}^{(k+1-i)}+
\sum_{j=1}^n \sum_{i=1}^k \mu_{2j}^{(i)} \mu_{2n+2-2j}^{(k+1-i)}~, \cr}}
valid for $n \geq 0$ and $k \geq 1$.
In terms of the generating functions
\eqn\generectre{\eqalign{
\lambda(z,x)~&=~ \sum_{n \geq 0} \sum_{k=1}^{2n+1} \lambda_{2n+1}^{(k)} x^k z^{2n+1}
=zx+...\cr
\mu(z,x)~&=~ \sum_{n \geq 0} \sum_{k=1}^{2n+2} \mu_{2n+2}^{(k)} x^k z^{2n+2} 
=z^2(x+x^2)+...\cr
\nu(z,x)~&=~ \sum_{n \geq 0} \sum_{k=1}^{2n+3} \nu_{2n+3}^{(k)} x^k z^{2n+3}
=z^3(x+3 x^2+x^3)+...\cr}}
we find the following system of algebraic equations
\eqn\alsyeq{\eqalign{
\lambda(z,x) ({1 \over z} -{\lambda(z,x) \over x})~&=~ (1+\mu(z,x))(x + \mu(z,x))\cr
(\mu(z,x)+x)({1 \over z} -{\lambda(z,x) \over x})~&=~ (1+\mu(z,x))
(\nu(z,x)+ zx^2) + {x\over z}\cr
(\nu(z,x)+z x^2)({1 \over z} -{\lambda(z,x) \over x})~&=~ (x+\mu(z,x))
(x+{ \mu(z,x) \over x})~.\cr}}
Combining these three equations, we easily eliminate the functions $\lambda$ and $\nu$
and get a polynomial equation of order $4$ for $\mu(z,x)$
\eqn\polfour{z^2(1+\mu)(x+\mu)(x+x^2+2 \mu)^2-x \mu (x+x^2+\mu)=0~,}
which, together with the initial condition $\mu(z,x)=z^2(x+x^2)+...$, determines the numbers
$\mu_{2n}^{(k)}$ completely. The functions $\lambda$ and $\nu$ are simply expressed in terms
of $\mu$ thanks to \alsyeq. For instance, 
\eqn\lammu{\lambda(z,x)~=~ { x \mu(z,x) \over z(x+x^2+2 \mu(z,x))}~,}
in the sense of formal series expansion in powers of $z$.

It is interesting to check the validity of the result \alsyeq\ by taking $x=1$,
in which case the numbers $\sum_{1 \leq k \leq 2n} \mu_{2n}^{(k)}=c_{2n}$
simply count the total number of arch configurations of order $2n$, namely
\eqn\redone{ \mu(z,1)~=~ \sum_{n \geq 1} c_{2n}~z^{2n} ~=~ C_+(z)-1~,}
where $C_+$ is the even part of the Catalan generating function \catagen\
\eqn\cplus{ C_+(z)~=~{ \sqrt{1 + 4 z} -\sqrt{1-4z} \over 4z}~.}
The relation \alsyeq\ reduces for $x=1$ to
\eqn\cpleq{ 4 z^2 C_+(z)^4~=~ C_+(z)^2 -1~,}
in agreement with \cplus.
For $x=-1$, we find that $\mu(z,-1)=0$, which translates into the sum rule
\eqn\toarsr{ \sum_{k=1}^{2n} ~(-1)^k~ \mu_{2n}^{(k)}~=~0~.}

Let us now use the equation \alsyeq\ to derive the leading behavior of the average
number of connected components over closings of ${\cal A}_2$. This number reads
\eqn\avcompto{ \langle {\rm comp}_2 \rangle_{2n}~=~{{\dot \mu}_{2n}\over c_{2n}}~=~ 
{ \sum_{k=1}^{2n} k ~ \mu_{2n}^{(k)} \over
c_{2n} }~.}
We extract from \alsyeq\ the generating function 
${\dot \mu}(z)=\partial_x \mu(z,x)\vert_{x=1}$ of 
${\dot \mu}_{2n}$ by differentiating it wrt $x$ and then setting $x=1$
\eqn\difalsy{ {\dot \mu}(z)~=~ { (C_+(z)+4)(C_+(z)-1)-16 z^2 C_+(z)^3 \over
C_+(z) ( 16 z^2 C_+(z)^2 -2)}~.}
{}From the study of the limit $z \to 1/4^-$ in this last expression, we find the asymptotics
of ${\dot \mu}_{2n}$ namely
\eqn\asymtoy{\eqalign{{\dot \mu}(z)~&\sim {\sqrt{2} -1\over 2\sqrt{1-(4z)^2}}\cr
\Rightarrow~ {\dot \mu}_{2n} ~&\sim {\sqrt{2} -1\over \sqrt{2}}
{ 4^{2n} \over \sqrt{\pi} (2n)^{1 \over 2}}~.\cr}}
Dividing this by the large $n$ asymptotics of $c_{2n}~\sim 4^{2n}/(\sqrt{\pi} (2n)^{3/2})$,
we finally get the leading behavior of the average number of connected components in closings
of ${\cal A}_2$
\eqn\cocompy{  \langle {\rm comp}_2 \rangle_{2n}~=~ 2n (1-{1 \over \sqrt{2}})~.}
This is to be compared with the analogous expression \avconone\
for ${\cal A}_1$ closings, which gives asymptotically
$\langle {\rm comp}_1 \rangle_{n}\sim n/2$. There are much less connected components 
per bridge in average closings of ${\cal A}_2$ than ${\cal A}_1$.

The connected numbers $\mu_{2n}^{(1)}={\bar M}_{2n}({\cal A}_2)$ are generated by the 
coefficient of $x$ in the series expansion of $\mu(z,x)$, namely
\eqn\asymu{ \mu(z,x)~=~x \mu(z)+O(x^2)
~=~ x \sum_{n=1}^\infty {\bar M}_{2n}({\cal A}_2)~z^{2n} + O(x^2)~,}
hence as a consequence of \polfour,
the series $\mu(z)$ satisfies an algebraic equation of order $2$
\eqn\algtwop{ z^2 (1+2\mu(z))^2 -\mu(z)~=~0~,}
which we can rewrite as 
\eqn\rewtp{ 2 z^2 (2 \mu(z)+1)^2 =(2 \mu(z)+1) -1~.}
We recognize the second degree equation for the generating function 
for Catalan numbers
\algcat, upon the substitution $2 \mu(z)+1 = C(2z^2)$. 
We deduce that
\eqn\resatwo{ \mu_{2n}^{(1)}~=~{\bar M}_{2n}({\cal A}_2)~=~ 
2^{n-1} c_n \qquad \forall \ n \geq 1~.}
Analogously, we find that
\eqn\othresp{\eqalign{ \lambda_{2n+1}^{(1)}~&=~ 2^n c_n \cr
\nu_{2n+1}^{(1)}~&=~ 2^{n-1} (c_{n+1} -c_n)~.\cr}}
The $\lambda_{2n+1}^{(1)}$ connected
closings of ${\cal A}_2$ completed by a single arch are
obtained, starting from the ${\cal A}_2$ connected closings of order $(2n)$,
by the process (I)
of Fig.\cutmeander.  
Indeed, the resulting meander is of order $(2n+1)$ and has a single
lower exterior arch surrounding ${\cal A}_2$. By cyclicity along the infinite river,
this arch can be sent to the right of ${\cal A}_2$, thus forming one of 
the $\lambda_{2n+1}^{(1)}$ meanders.
Therefore, $\lambda_{2n+1}^{(1)}$ is equal to the total number of exterior arches
in connected closings of ${\cal A}_2$ 
(each exterior arch gives rise to one action by (I)). 
The average number of exterior arches 
in connected closings of ${\cal A}_2$ reads therefore
\eqn\extome{ \langle {\rm ext}_2 \rangle_{2n}~=~{ \lambda_{2n+1}^{(1)}
\over \mu_{2n}^{(1)} }~=~2,}
independent of $n$.
On the other hand, 
due to the asymptotic behaviour of the Catalan numbers ($c_n \sim 4^n$, $n \to \infty$),
we find that for large $n$
\eqn\asymatwo{ M_{2n}({\cal A}_2)~ \sim (2 \sqrt{2})^{2n}~.}

\noindent{\bf More cases.} The triple arch case ${\cal A}_3$
(repetition $n$ times of the rainbow
${\cal R}_3$ of order $3$) involves the introduction of $9$ slightly different arch
configurations.
We will not give the details of the solution here, and simply give the result:
the generating function for the number of connected closings of ${\cal A}_3$,
\eqn\genthr{ \theta(z)~=~ \sum_{n \geq 1} {\bar M}_{3n}({\cal A}_3)~z^{3n}~,}
satisfies the polynomial equation of order $4$
\eqn\solthreep{ 2 z^3 (1+ \theta(z))(1+3 \theta(z))^3 - \theta(z)(1+2 \theta(z))^2=0~.}
The asymptotic behaviour of the numbers ${\bar M}_{3n}({\cal A}_3)$ is found by solving the 
coupled system of equations
\eqn\cousys{\eqalign{ 2 z^3 (1+\theta)(1+3 \theta)^3 -\theta(1+2 \theta)^2~&=~0\cr
{\partial \over \partial \theta}\big( 2 z^3 (1+\theta)(1+3 \theta)^3 -\theta(1+2 \theta)^2
\big) ~&=~0~.\cr}}
The solution $(\theta^*,z^*)$ is such that 
\eqn\asymtrhe{ {\bar M}_{3n}({\cal A}_3)~\sim (z^*)^{-3n}~=~ (2.842923..)^{3n}~.}
The above behaviour is to be compared with that of connected
closings of single arch repetitions $\sim 1^{n}$
and double arch repetitions $\sim (2\sqrt{2})^{2n}=(2.828..)^{2n}$.
We see that very little entropy has been gained when replacing double 
arches by triple arches.  
Intuitively, these numbers should form an increasing sequence (with increasing order
of the elementary raibow which is repeated), which may very well
converge to the number $\bar R$, governing the large $n$ asymptotics of semi--meander
numbers. 

Another case of interest is the repetition ${\cal B}_p$ 
($n$ times) of the arch configuration of order $p$
obtained as a repetition of $(p-1)$ single arches of order $1$ below a single arch, i.e.
corresponding to the permutation $\mu(1)=2p$, $\mu(2i)=2i+1$, $i=1,2,...,p-1$.
The generating function
\eqn\genebp{ \beta(z)~=~ \sum_{n=1}^\infty {\bar M}_{np}({\cal B}_p)~z^{np}~}
for the numbers of connected closings of ${\cal B}_p$ can be shown to satisfy 
a polynomial equation of order $p$
\eqn\polorn{ z^p (1+p \beta(z))^p- \beta(z)~=~0~.}
This leads to the large $n$ behaviour
\eqn\larnbehs{ {\bar M}_{np}({\cal B}_p)~\sim~ \big( p^{p+1 \over p} (p-1)^{1-p \over p}
\big)^{np}~.}
For $p=2$, we recover the result \asymatwo\ for double arch repetitions.

Note that, in view of the above cases, an empiric rule seems to emerge, indicating
that, for fixed order, the number of connected closings of a motive is increased by 
nesting the arches rather than juxtaposing them. 
If this is true, it means that the number of closings of a given motive
is maximal with the rainbow (semi--meander). 
We have checked this conjecture numerically up to order $12$.

\newsec{Irreducible meanders, exact results}

A multi--component
meander is said to be {\bf $k$--reducible} if one (proper)
subset of its connected components
can be detached from it by cutting the river $k$ times between the bridges.
A multi--component meander is said to be {\bf $k$--irreducible} if it is {\it not}
$k$--reducible, i.e. if no (proper)
subset of its connected components can be detached from it by cutting the
river $k$ times between the bridges.
\fig{Reducibility and irreducibility for meanders. The cuts reducing the meanders
are indicated by arrows. The meander $(a)$ is $1$--reducible, 
i.e. the succession of two meanders along the river. 
The meander $(b)$ is $1$--irreducible and $2$--reducible. 
The meander $(c)$ is $1$--and $2$--irreducible and 
$3$--reducible.}{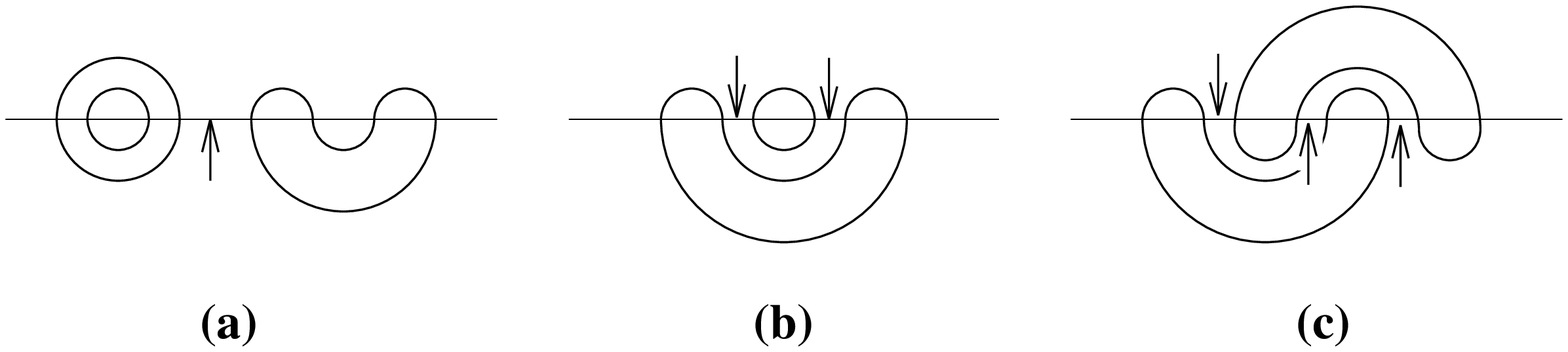}{10.truecm}
\figlabel\exired
\noindent{}In Fig.\exired, we give a few examples to illustrate the notion 
of reducibility and irreducibility of meanders. 
The same definition applies to the semi--meanders in the formulation
with a semi--infinite river.

\subsec{$1$--irreducible meanders and semi--meanders}

A meander is $1$--irreducible if it is not the succession along
the river of at least two meanders. 
We can enumerate all the meanders by their growing number of 
$1$--irreducible components.
Denoting by $P_n^{(k)}$ the total number of $1$--irreducible meanders
of order $n$ with $k$ connected components, we compute
\eqn\numerate{ M_n^{(k)} ~=~ \sum_{n_1+n_2+...=n\atop
k_1+k_2+...=k} P_{n_1}^{(k_1)} P_{n_2}^{(k_2)} ... ~,}
hence the generating functions
\eqn\genefuncme{\eqalign{
M(z,x)~&=~ \sum_{n \geq k \geq 1} M_n^{(k)}~z^{2n}~x^k \cr
P(z,x)~&=~ \sum_{n \geq k \geq 1} P_n^{(k)}~z^{2n}~x^k \cr}}
satisfy the equation
\eqn\eqgemp{ M(z,x)~=~ P(z,x)+P(z,x)^2+P(z,x)^3+...~=~{P(z,x) \over 1 - P(z,x)}~.}

An analogous reasoning for semi--meanders leads to the relation
\eqn\semivermp{ {{\bar P}(z,x) \over 1- P(z,x)}~=~ { {\bar M}(z,x)}~,}
between the generating functions
\eqn\genefuncsme{\eqalign{
{\bar M}(z,x)~&=~ \sum_{n \geq k \geq 1} {\bar M}_n^{(k)}~z^{2n}~x^k \cr
{\bar P}(z,x)~&=~ \sum_{n \geq k \geq 1} {\bar P}_n^{(k)}~z^{2n}~x^k \cr}}
of respectively semi--meander and $1$--irreducible semi--meander numbers 
of order $n$ with $k$ connected components, and $P(z,x)$ defined in \genefuncme.

\subsec{$2$--irreducible meanders and semi--meanders}

The $2$--irreducible meanders have been studied in \LZ\ extensively (under the name
of {\it irreducible meanders}). They use the following  equivalent characterization 
of $2$--irreducible meanders: 
for any subset of connected components of a $2$--irreducible meander, 
its set of bridges is not consecutive. Indeed, if it were not the case, 
one could cut the river before the first bridge and after the last one and
detach the corresponding piece. One easily checks on Fig.\exired\ $(c)$ that 
the two sets of bridges of the two connected components of the 
meander are intertwined.

For completeness, we reproduce here their computation of the numbers $q_n$
of $2$--irreducible meanders of order $n$, by slightly generalizing
their argument to include the numbers $Q_n^{(k)}$ of $2$--irreducible meanders
of order $n$ with $k$ connected components. The idea is to enumerate the
$M_n^{(k)}$ meanders by focussing on their leftmost $2$--irreducible component, 
namely the largest $2$--irreducible subset of its connected components, 
containing the leftmost bridge.
\fig{A general meander. The leftmost $2$--irreducible component is represented
in thick solid lines. The positions marked by $\#$ can be decorated with
any meanders to get the most general meander with this leftmost $2$--irreducible
component.}{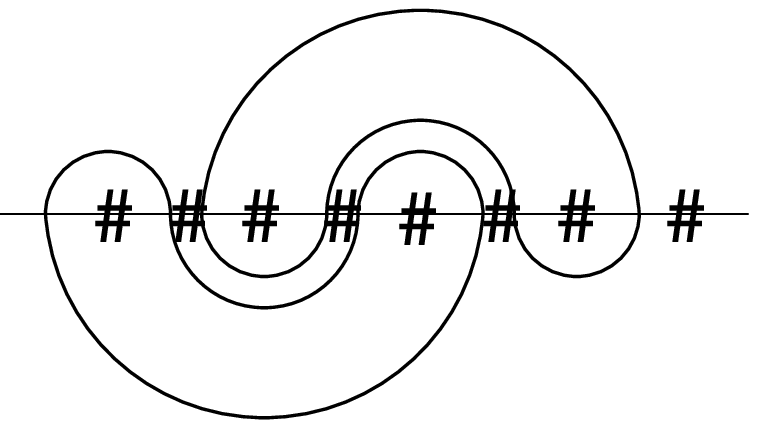}{8.truecm}
\figlabel\enumea
\noindent{}In Fig.\enumea, the leftmost $2$--irreducible piece of a meander is depicted in 
thick solid line. Suppose this piece is of order $p$ and has $l$ connected components.
The most general meander having this leftmost $2$--irreducible piece can be obtained
by decorating any segment of river between two consecutive bridges
with arbitrary meanders of respective orders 
$n_1$, $n_2$, ... $n_{2p}$, $p+n_1+...+n_{2p}=n$ (there
are $2p$ positions, indicated by the symbols $\#$ in Fig.\enumea, for these 
possible decorations), and with respective numbers of connected components $k_1$, $k_2$,
... $k_{2p}$, with $l+\sum k_i = k$. We get the relation
\eqn\relameir{ M_n^{(k)} ~=~ \sum_{p+n_1+...+n_{2p}=n \atop
l+k_1+...+k_{2p}=k} Q_p^{(l)}~M_{n_1}^{(k_1)}~M_{n_2}^{(k_2)}~...~M_{n_{2p}}^{(k_{2p})}~,}
with the convention that $M_0^{(0)}=1$.
In terms of the generating functions $M(z,x)$ of eq.\genefuncme\ and
\eqn\geneirto{ Q(z,x)~=~\sum_{n \geq k \geq 1} Q_n^{(k)} ~z^{2n}~x^k~,}
the relation \relameir\ reads
\eqn\relaingen{ M(z,x)~=~ Q(z (1+ M(z,x)),x)~.}
In the special case $x=1$, if we denote by 
\eqn\defbz{\eqalign{ 
B(z)~&=~1+ M(z,1)~=~ \sum_{n=0}^\infty (c_n)^2 z^{2n}~\cr
q(z)~&=~1+ Q(z,1)~=~ \sum_{n=0}^\infty  q_n ~ z^{2n}~,\cr}}
then \relaingen\ reduces to 
\eqn\finlutms{ B(z)~=~ q(z B(z))~.}
The radius of convergence of the series $B(z)$ is $1/4$, due to the asymptotics of $c_n$,
hence that of $q(z)$ is 
\eqn\radcon{\eqalign{ 
z^*~=~{1 \over 4} B({1  \over 4})~&=~ {1 \over 4} \sum_{n \geq 0} c_n^2~4^{-2n} \cr
&=~ {1 \over 4} C({x \over 4})C({1 \over 4x})\big\vert_{x^0}\cr
&=~ {1 \over 4} \oint {dx \over 2i \pi x}  C({x \over 4})C({1 \over 4x})\cr
&=~ \oint {dx \over 2i \pi x} (1-\sqrt{1-x})(1-\sqrt{1-1/x})\cr
&=~ \oint {dx \over 2i \pi x} (-1+\sqrt{2 -x -1/x})\cr
&=~ {1 \over 2 \pi} \int_0^{2 \pi} d \theta (-1+2 \sin {\theta \over 2})\cr
&=~ {4 - \pi \over \pi}~.\cr}}
Hence a leading behaviour for large $n$ \LZ\
\eqn\leadirtom{ q_n~\sim \left({\pi \over 4-\pi} \right)^{2n}~.}
This gives an upper bound on the leading behaviour of $M_n$
\eqn\upbom{ R ~\leq ~ \left({\pi \over 4-\pi} \right)^2~=~ 13.3923...}
A refined study of the equation \finlutms\ also yields the subleading power correction
\eqn\subleadirtom{ q_n~\sim {1 \over n^3} \left({\pi \over 4-\pi} \right)^{2n}~.}

Let us extend these considerations to the semi--meander case.
If ${\bar Q}_n^{(k)}$ denotes the number of $2$--irreducible semi--meanders of order $n$
with $k$ connected components, let us enumerate the ${\bar M}_n^{(k)}$
meanders of order $n$ with $k$ connected components, focussing on their leftmost
$2$--irreducible piece, say of order $p$ with $l$ connected components. 
\fig{A general semi--meander. The leftmost $2$--irreducible component is represented
in thick solid lines. The positions marked by $\#$ can be decorated with
any meanders to get the most general semi--meander with this leftmost $2$--irreducible
component. The last position, indicated by @, can be decorated with any semi--meander,
possibly winding around the source of the river ($*$). }{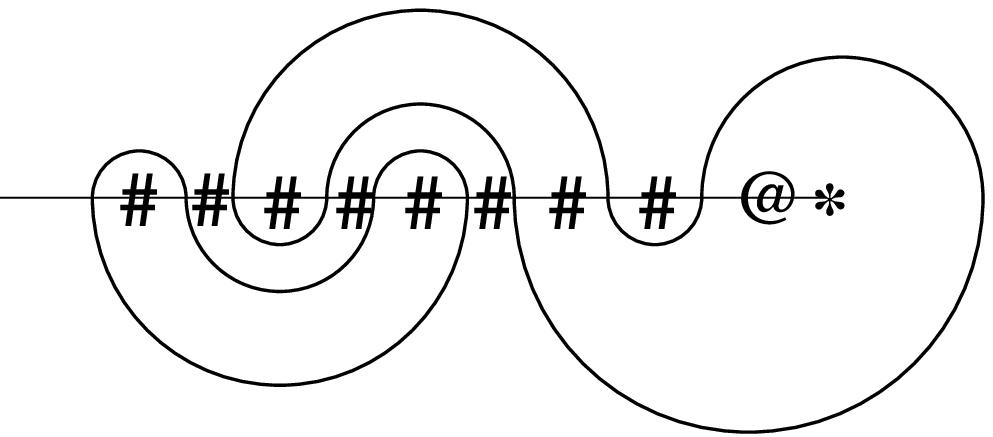}{8.truecm}
\figlabel\enusem
\noindent{} The most general semi--meander with given leftmost $2$--irreducible component
(as indicated by thick solid lines in Fig.\enusem) is obtained by decorating the segments
of river between any two consecutive bridges (there are $(p-1)$ such positions)
with meanders of respective orders $n_1$, $n_2$,... $n_{p-1}$, with
respectively
$k_1$, ...,$k_{p-1}$ connected components,
and the segment of river between
the source and the first bridge by any semi--meander of order $q$, 
with $r$ connected components,
such that
$p+q+2\sum n_i=n$ and $l+r+\sum k_i=k$. 
This amounts to the following relation
\eqn\relasemi{ {\bar M}_n^{(k)}~=~ \sum_{p+q+2(n_1+...+n_{p-1})=n \atop
l+r+k_1+...+k_{p-1}=k} {\bar Q}_p^{(l)} M_{n_1}^{(k_1)} ... M_{n_{p-1}}^{(k_{p-1})}
{\bar M}_q^{(r)}~,}
where the sum extends over non--negative values of $n_i$ and $k_i$,
with the convention that $M_0^{(0)}=1={\bar Q}_0^{(0)}={\bar M}_0^{(0)}$.
This translates into a relation between the generating functions
\eqn\genesemim{ \eqalign{
{\bar M}(z,x)~&=~ \sum_{n \geq k \geq 1} {\bar M}_n^{(k)}~ z^{n}~x^k\cr
{\bar Q}(z,x)~&=~  \sum_{n \geq k \geq 1} {\bar Q}_n^{(k)}~ z^{n}~x^k\cr}}
and $M(z,x)$ of eq.\genefuncme\ (note that in \genesemim\
we have given a weight $z$ per bridge in
the {\it semi--infinite river} framework, while there would be twice that number 
of bridges in the {\it rainbow closing} framework)
\eqn\finlutmsm{\eqalign{ 
{\bar M}(z,x)~&=~ {\bar Q}(z (1+M(z,x)),x) ~
{1+{\bar M}(z,x) \over 1+M(z,x)}~\cr
\Rightarrow {{\bar M}(z,x) \over 1+{\bar M}(z,x)} &(1+M(z,x))~=~
{\bar Q}(z(1+M(z,x)),x)~.\cr}}
In the special case $x=1$, with the generating function
\eqn\genesemimea{
{\bar q}(z)~=~1+{\bar Q}(z,1)~=~\sum_{n \geq 0} {\bar q}_n~z^n,}
where ${\bar q}_n=\sum_{k=1}^n{\bar Q}_n^{(k)}$ is the total number of 
$2$--irreducible multi--component
semi--meanders of order $n$,
and with $C(z)=1+{\bar M}(z,1)$, $B(z)=1+M(z,1)$ defined respectively 
in \catagen\ \defbz, the relation
\finlutmsm\ reduces to
\eqn\msmm{ 1+ zB(z)C(z)~=~{\bar q}(z B(z))~,}
where we used the fact that $(C(z)-1)/C(z)=zC(z)$ \algcat.
Reasoning as above, we find the convergence radius of the series ${\bar q}(z)$,
$z^*=B(1/4)/4=(4-\pi)/\pi$. Hence we have the asymptotics
\eqn\asyirsem{  {\bar q}_n~\sim ~ \left( {\pi \over 4- \pi }\right)^n~.}
This implies the upper bound on the leading behaviour of ${\bar M}_n$
\eqn\leadupbosem{ {\bar R}~\leq ~ {\pi \over 4 - \pi}~=3.659...}
This upper bound is below the mean field estimate \estimean\ which therefore
cannot be exact and needs to be improved.
Again, a more refined study of the relation \msmm\ gives the subleading power
law
\eqn\morefsem{\eqalign{ 
{\bar q}_n ~ &\sim ~ {1 \over n^{3 \over 2} }
\left( {\pi \over 4- \pi }\right)^n~ \cr
&\sim ~\sqrt{q_n}. \cr}}

\newsec{Conclusion}

In this paper we have studied the statistics of arch configurations, meanders and
semi--meanders.  This study emphasizes the role of exterior arches,
instrumental in the recursive generation of both semi--meanders and arch configurations, 
and whose average number is directly linked to the corresponding entropies.
A complete solution of these problems however requires the knowledge of the
correlation between these exterior arches and those of a given depth 
(see for instance eqs.\casekone\-\casenull).
Discarding these correlations leads to a rough mean field result.

Using the alternative formulation \luttefin\--\semimat\
in the framework of random matrix models, 
we are left with the computation of particular
Gaussian matrix averages of traces of words. Equation \limavtr\ provides 
a recursive way of computing any such average.  
As a by--product, we have derived an exact formula \purecom\ expressing 
the semi--meander numbers in terms of the characters of the symmetric group.

Finally, we have presented the exact solutions for simpler meandric problems such as the
closing of particular lower arch configurations (made of rereated motives), as well as
the $2$--irreducible meanders and semi--meanders.

In conclusion, we are still lacking of an efficient treatment of the main recursion
relation for semi--meanders.  Beyond the entropy problem, it would also be interesting
to determine values of critical exponents in this problem, such as the exponent governing
the subleading large order behaviour of the meander and semi--meander numbers \asympto.
The latter are nothing but the usual $\alpha$ and $\gamma$ configuration
exponents of polymer chains.  {}From this point of view, a fundamental issue is the
determination of the universality class of the (self--avoiding)
chain folding problem.

\noindent{\bf Acknowledgements}

We thank C. Itzykson, H. Orland, and J.-B. Zuber for stimulating discussions,
and M. Bauer for a careful and critical reading of the manuscript.

\appendix{A}{Combinatorial identities for arch statistics}

Many combinatorial identities can be proved directly by identifying the result
as a certain coefficient, say of $x^n$, in a (Laurent) series expansion
of some function $f(x)$ around $x=0$. In the following, we use the notation
\eqn\notcoeff{ f(x)\big\vert_{x^n} ~\equiv~ \oint 
{dx \over 2i \pi} {f(x) \over x^{n+1}}}
for the coefficient of $x^n$ in the small $x$ expansion of the function $f$.
 
\noindent{\bf Proof of eq.\idenpol.} We want to prove the identity \idenpol,
which may be rewritten as
\eqn\newidentpol{\eqalign{ 
\sum_{k=1}^n {k \over 2n-k} {2n-k \choose n} {k \choose l}~&=~
\sum_{m=0}^{n-1} {n-m \over n} {n+m-1 \choose m} {n-m \choose l}\cr
&=~ {2l+1 \over 2n+1} {2n+1 \choose n-l}~,\cr}}
where we changed the summation variable to $m=n-k$.
The above sum over $m$ can be expressed as
\eqn\calcint{ \sum_{r,m\geq 0 \atop r+m = n-l } 
{n+m-1 \choose m} \times {r+l \over n}{r+l \choose r}~,}
where we introduced a second summation over $r=n-l-m$.
Using the infinite series expansion
\eqn\infserex{ {1 \over (1-x)^\alpha} ~=~ \sum_{m=0}^\infty {m+\alpha -1 \choose m}
x^m ~,}
we identify the expression \calcint\ as the coefficient of $x^{n-l}$
of some series
\eqn\serexpan{\eqalign{ {1 \over (1-x)^n} \times &{1 \over n}(l + x {d \over dx})
{1 \over (1-x)^{l+1}} \bigg\vert_{x^{n-l}} \cr
&= ~ {l \over n (1-x)^{n+l+1}}\bigg\vert_{x^{n-l}} +{l+1 \over n (1-x)^{n+l+2}}
\bigg\vert_{x^{n-l-1}} \cr
&=~ {l \over n} { n+l +n-l \choose n-l} +{l+1 \over  n}{ n+l+1 +n-l-1  \choose n-l-1 }\cr
&=~{1 \over n}(l(n+l+1)+ (l+1)(n-l)){(2n)! \over (n+l+1)! (n-l)!} \cr
&=~ {2l+1 \over 2n+1} {2n+1 \choose n-l} ~.\cr}}
This proves the identity \idenpol.

\noindent{\bf Proof of the identity \compint.}
We want to prove the identity \compint, which reads
\eqn\recompint{\eqalign{
{1 \over n} \sum_{k=1}^n { k \choose l}&{n \choose k} {n \choose k-1} \cr
&=~{1 \over n} {n \choose l} \sum_{k=1}^n {n-l \choose k-l} {n \choose k-1} \cr
&=~{1 \over n} {n \choose l} {2n-l \choose n-l+1}~. \cr}}
The sum over $k$ in the second line of \recompint\ is nothing but the
coefficient of $x^{l-1}$ of some Laurent series, identified as
\eqn\laurid{\eqalign{
\sum_{k=1}^n {n-l \choose k-l} {n \choose k-1}~&=~
(1+x)^{n-l} (1+{1 \over x})^n \bigg\vert_{x^{1-l}} \cr
&=~ (1+x)^{2n-l} \big\vert_{x^{n-l+1}}={2n-l \choose n-l+1}~.\cr}}
This completes the proof of the identity \compint.

\listrefs

\listtoc

\bye